\documentclass[aps,prx,reprint,preprintnumbers,superscriptaddress,nofootinbib,longbibliography,floatfix]{revtex4-2}
\pdfoutput=1
\usepackage{rotating}
\usepackage{array}
\usepackage{amsmath}
\usepackage[normalem]{ulem}
\usepackage{slashed}
\usepackage{booktabs}
\usepackage[pdftex,table]{xcolor}
\usepackage{units}
\usepackage{xfrac}
\usepackage{mathtools}
\usepackage{empheq}
\usepackage[]{units}
\usepackage{multirow}
\usepackage{amssymb}
\usepackage{url}
\usepackage{comment}
\usepackage{physics}
\usepackage{color,soul}
\usepackage{bbm}
\usepackage[caption=false]{subfig}
\usepackage{xspace}
\usepackage{makecell}

\usepackage{hyperref}
\hypersetup{
  colorlinks=true,
  citecolor=blue,
  linkcolor=blue,
  urlcolor=blue
}

\newlength{\myfigurewidth}
\newlength{\myfigureskipeqh}
\newcommand{\threefigeqh}[3]{\resizebox{\myfigurewidth}{!}{\includegraphics[height=1cm]{#1}\hspace{\myfigureskipeqh}\includegraphics[height=1cm]{#2}\hspace{\myfigureskipeqh}\includegraphics[height=1cm]{#3}}}

\newcommand{\geant}{\texttt{Geant4}\xspace}
\newcommand{\challenge}{\textit{CaloChallenge}\xspace}
\newcommand{\caloscore}{\texttt{CaloScore}\xspace}
\newcommand{\caloscorev}{\texttt{CaloScore v2}\xspace}
\newcommand{\caloflow}{\texttt{CaloFlow}\xspace}
\newcommand{\icaloflow}{\texttt{iCaloFlow}\xspace}
\newcommand{\fastcalogan}{\texttt{FastCaloGAN}\xspace}
\newcommand{\jetnet}{\texttt{JETNET}\xspace}
\newcommand{\diffu}{\texttt{CaloDiffusion}\xspace}
\newcommand{\glam}{\texttt{GLaM}\xspace}

\newcolumntype{A}{r@{\hspace{0em}}l}
\newcolumntype{B}{A@{$\,\pm\,$}A}

\begin{document} 

\preprint{FERMILAB-PUB-23-384-CSAID-PPD}

\title{Denoising diffusion models with geometry adaptation for high fidelity calorimeter simulation}

\author{Oz Amram}
\email{oamram@fnal.gov}

\author{Kevin Pedro}
\email{pedrok@fnal.gov}
\affiliation{Fermi National Accelerator Laboratory, Batavia, IL 60510, USA}

\begin{abstract}
    
Simulation is crucial for all aspects of collider data analysis, but the available computing budget in the High Luminosity LHC era will be severely constrained.
Generative machine learning models may act as surrogates to replace physics-based full simulation of particle detectors, and diffusion models have recently emerged as the state of the art for other generative tasks.
We introduce \diffu, a denoising diffusion model trained on the public \challenge datasets to generate calorimeter showers.
Our algorithm employs 3D cylindrical convolutions, which take advantage of symmetries of the underlying data representation.
To handle irregular detector geometries, we augment the diffusion model with a new geometry latent mapping (\glam) layer to learn forward and reverse transformations to a regular geometry that is suitable for cylindrical convolutions.
The showers generated by our approach are nearly indistinguishable from the full simulation, as measured by several different metrics.
\end{abstract}

\maketitle
\flushbottom

\section{Introduction}
\label{sec:intro}

High quality simulation plays a crucial role in modern particle physics experiments.
Most experiments rely on the \geant~\cite{Geant1,Geant2,Geant3} toolkit to simulate interactions of particles with their detector.
Achieving accurate results requires simulating the interactions of both the primary particle incident on the detector and the numerous secondary particles produced through interactions with the detector material.
For this reason, simulations of calorimeters, which are designed to capture the energy produced by the shower of secondary particles, usually requires the most computational resources.
Simulating calorimeters currently consumes a significant fraction of the computing resources of modern collider experiments~\cite{HEPSoftwareFoundation:2018fmg}.
The problem will be exacerbated at the High Luminosity LHC, which will feature larger data volumes, more complex detectors~\cite{CMS:2017jpq}, and a higher pileup environment.
Future high granularity detectors will require more computational resources to simulate because of their more complex geometries and higher levels of precision~\cite{Pedro:2020kbk}.
At the same time, reconstruction will require a larger fraction of the computing budget because of the expected superlinear scaling of important algorithms with increasing pileup~\cite{Ju:2021ayy}.

These resource constraints mean that full, detailed detector simulation using \geant will not be possible for every simulated event.
Instead, `fast simulation' methods that approximate the output of \geant using less computation will be employed. 
Most major experiments have developed fast simulation frameworks based on parametric approximations manually tuned to \geant~\cite{Abdullin:2011zz,Giammanco:2014bza,Sekmen:2016iql,ATLAS:2010bfa,Lukas:2012kua,ATLFast3}.
These parametric models generally suffer from deficiencies in modeling detailed observables of calorimeter showers, limiting their usage in physics analysis.

In order to overcome these challenges, machine learning (ML) models are increasing in popularity as fast surrogate models for \geant~\cite{CaloGAN,Chekalina:2018hxi,FastCaloGAN,Buhmann:2020pmy,Diefenbacher:2020rna,CaloFlow,ATLFast3,CaloFlow2,Buhmann:2020pmy,Buhmann:2021lxj,ATLAS:2022jhk,Buhmann:2021caf,CaloScore,Buhmann:2023pmh,L2LFlows,Hashemi:2023ruu,Diefenbacher:2023prl,Mikuni:2023dvk,CaloClouds,iCaloFlow,Mikuni:2023tqg} 
(see Ref.~\cite{Adelmann:2022ozp} for an overview and Ref.~\cite{Butter:2022rso} for a recent review).
These techniques borrow from the growing field of ML-based generative modeling, which has made significant advances in recent years.

In high energy physics (HEP), the first class of generative models proposed for this purpose were generative adversarial networks (GANs)~\cite{CaloGAN}. 
GANs are trained by iterating between a `generator network' that learns to produce artificial samples and a `discriminator network' which attempts to distinguish the artificial samples from true ones. 
GANs are able to generate high quality showers orders of magnitude faster than \geant. 
The ATLAS experiment has now employed calorimeter GANs in their fast simulation framework~\cite{ATLFast3}.
GANs are also used for fast simulation by the LHCb~\cite{Barbetti:2023bvi} experiment and 
are being explored for emulating the high granularity, 7.5M channel, pixel detector of Belle-II \cite{Hashemi:2023ruu}.
However, GAN training does not reliably converge because the two competing objectives create a saddle point in the loss space rather than a minimum.
Additionally, GANs are known to suffer from `mode collapse', in which the generator network only learns to produce samples from a subset of the full data space. 

Variational Autoencoders (VAEs) have also been proposed for calorimeter simulation \cite{Buhmann:2020pmy, Buhmann:2021lxj, ATLAS:2022jhk}. 
A VAE consists of an encoder, which maps the input data to a smaller latent space, and a decoder, which maps the latent space back to the original data. 
A VAE is distinguished from a regular autoencoder by forcing the latent space to follow a multivariate Gaussian distribution via an additional term in the training loss. 
New samples can then be generated by drawing random samples from a multivariate Gaussian in the latent space and applying the decoder model. 
However, VAEs on their own do not seem have the expressive power of GANs and other state-of-the-art models and generally achieve worse quality on complex high-dimensional data such as calorimeter showers. 
Refs.~\cite{Buhmann:2020pmy,Buhmann:2021lxj} instead use a bounded information bottleneck AE (BIB-AE), which is a novel combination of the VAE and GAN architecture. 

Normalizing flows (NFs) have also been proposed for calorimeter simulations~\cite{CaloFlow,iCaloFlow,L2LFlows}. 
NFs are based on a series of invertible transformations that convert the input distributions to multivariate Gaussians.
Once trained, new samples can be generated by sampling the Gaussian space and applying the inverse transformations to convert to the data space. 
However, as the dimensionality of the data has to be preserved in each stage of the flow, it can be difficult to scale NFs to very high-dimensional data. 

Recently, a new class of models has become dominant in ML image generation tasks: denoising diffusion models~\cite{DDPM,LatentDiffusion,elucidating_design}. 
In this work, we explore the use of denoising diffusion models to generate calorimeter showers. 
Diffusion models are based on a `noising process' that continuously perturbs an image until it is degraded to pure noise. 
A `denoising model' is then trained to invert the diffusion process.
New samples can be generated by constructing a sample in the noise space and repeatedly `denoising' it back to the original space.
The use of diffusion models in image generation has proliferated because of their straightforward training procedure, high quality results, straightforward scaling to high-dimensional data, and manageable computational requirements. 
Diffusion models were first used for calorimeter simulation in \caloscore~\cite{CaloScore,Mikuni:2023tqg} with promising results.
\caloscore is a score-based diffusion model, which is similar but distinct from the denoising diffusion model employed in this paper. 
Recent work has combined diffusion with point clouds~\cite{CaloClouds,point_cloud_comparison} and demonstrated distillation of diffusion models to improve generation time of jet particle clouds~\cite{Mikuni:2023dvk,PCDroid}.
Several other works apply diffusion to HEP in other contexts~\cite{Shmakov:2023kjj,Butter:2023fov,Mikuni:2023tok}. 

Our approach, dubbed \diffu, is a denoising diffusion model for calorimeter simulation and employs several novel optimizations to make use of the geometric structure of the data.
In contrast to other recent works~\cite{CaloClouds,point_cloud_comparison}, which have advocated for point cloud representations of calorimeter showers, \diffu uses voxelized image-like representations of the calorimeter data. 
The use of this voxelized representation retains the geometric information of the data, allowing for several optimizations that exploit the cylindrical structure and scale well for high-dimensional datasets. 
We additionally introduce a new geometry latent mapping (\glam) component, which is able to map irregular detector geometries into a regular structure suitable for symmetry-preserving operations such as convolutions.

We test our approach on the public datasets provided as part of the Fast Calorimeter Simulation Challenge (\challenge)~\cite{CaloChallenge}. 
The challenge released three datasets of showers simulated with \geant in calorimeters with increasing granularity.
We find that \diffu is able to generate very quality showers that are difficult to distinguish from \geant for all datasets of the \challenge. 
Based on quantitative metrics, we demonstrate significant gains over previous state-of-the-art methods, particularly for the high-dimensional datasets of the \challenge.

\section{Diffusion Models}
\label{sec:background}

Diffusion models are defined in terms of a `noising process', which is a Markov chain that starts from data points $x_0$ (following a probability distribution
$q(x_0)$) and iteratively adds Gaussian noise.
The data points $x_{t}$ at time $t$ are generated from data points at the previous time step $x_{t-1}$ by adding Gaussian noise $\epsilon$. 
At the final time step $T$, the probability distribution of data points $q(x_T | x_0)$ can then be computed based on the original $x_0$ via a product of Gaussian likelihoods.
This is summarized in the following equations:
\begin{align}
        x_t  &= \sqrt{1 - \beta_t} x_{t-1} + \beta_t \epsilon, \\ 
        q(x_t | x_{t-1}) &= \mathcal{N}(x_t | \sqrt{1 - \beta_t}x_{t-1}, \beta_t), \\
        q(x_T | x_0) &= \prod\limits_{t=1}^{T} q(x_t | x_{t-1}),
\end{align}
where we denote Gaussian likelihoods as $\mathcal{N}(x | \mu, \sigma^2)$, $\epsilon \sim N(0,\mathcal{I})$, and $\beta_t$ is a `variance schedule' that controls how much Gaussian noise is added at each time step.

For a sufficiently large $T$ (the total number of diffusion steps), the Gaussian noise will overwhelm the original data and $x_T$ will follow a multivariate Gaussian distribution.
Therefore, a new sample $x_0$ could be generated by sampling $x_T$ from a multivariate Gaussian and inverting the diffusion process in order to produce $x_0 \sim q(x_0)$.
An exact inversion of the diffusion process requires knowing the reverse distribution $p(x_{t-1} | x_t)$, which encodes how likely a particular data point $x_{t-1}$ is given the noisier version $x_t$. 
Direct calculation of $p(x_{t-1} | x_t)$ could be done via Bayes' rule $p(x_{t-1} | x_t) = \sfrac{q(x_t | x_{t-1}) q(x_{t-1})}{q(x_t)}$, but this is intractable because evaluating $q(x_t) = \int dx_0 q(x_0) \prod\limits_{t=1}^{T} q(x_t | x_{t-1})$ requires an integral over the entire data distribution $q(x_0)$. 
We therefore approximate $p(x_{t-1} | x_t)$ as:
\begin{equation}
    p(x_{t-1} | x_t) = \mathcal{N}(x_{t-1} | \mu_{\theta}(x_t,t,z), \beta_t \mathcal{I}),
\end{equation}
where the estimated mean $\mu_{\theta}$ is modeled by a neural network with parameters $\theta$, conditioned on $t$ and additional information $z$. 
There are multiple ways to parameterize $\mu_{\theta}(x_t,t,z)$ and we employ two different approaches as discussed in Section~\ref{sec:implementation}.

Because sums of Gaussians also follow a Gaussian distribution, $x_t$ can be directly sampled from $x_0$ in a single step:
\begin{align}
    q(x_T | x_0) &= \mathcal{N}(x_t | \sqrt{\overline{\alpha_t}}x_0, (1 - \overline{\alpha}_t) \mathcal{I}) \\
    x_t &= \sqrt{\overline{\alpha}_t}x_0 + \sqrt{1-\overline{\alpha}_t}\epsilon
    \label{eq:noising}
\end{align}
where $\alpha_t \equiv 1 - \beta_t$ and $\overline{\alpha}_t = \prod_{\tau=1}^{t} \alpha_\tau$. 
The variance of the noise for time step $t$ is therefore $1-\overline{\alpha}_t$,
which can be used to define the noise schedule as an alternative to $\beta_t$.
This property is convenient because efficiently computing $x_t$ from $x_0$ allows $t$ to be randomly sampled during training.

Training a denoising diffusion model proceeds via the following steps: sampling a batch of images $x^{\prime}$ from the training set; choosing random time steps $t^{\prime}$;
producing a set of noised images $x^{\prime}_{t^{\prime}}$ based on Eq.~\eqref{eq:noising}; and comparing the model's prediction for $\mu_{\theta}$ to the true value to compute the loss.
Once the model has been trained, new samples can be generated by first sampling $x_T \sim \mathcal{N}(0, \mathcal{I})$,
then repeatedly evaluating $p(x_{t-1} | x_t)$ from the trained model until $x_0$ is reached. 

The denoising diffusion approach employed here shares many features with score-based diffusion, or score-matching models, such as \caloscore~\cite{CaloScore,Mikuni:2023tqg}. 
The score-based approach defines a stochastic differential equation (SDE) that continuously corrupts the data into a known distribution.
Rather than directly learning to invert the denoising process, the neural network is trained to evaluate the score of the data, $\nabla_x \log q(x)$, which can then be used to reverse the SDE in order to generate new samples. 
There are different ways to parameterize the SDE, based on the choices of the  `diffusion' and `drift' functions. 
However, the `variance preserving' formulation is deeply tied to the denoising diffusion approach employed here:
the optimal score-matching network is identical to the optimal denoising network (see Appendix B of Ref.~\cite{elucidating_design} for a short derivation). 
Both score-based and denoising-based diffusion models are being actively explored in the ML literature~\cite{elucidating_design}.
We focus on the denoising variant here because of its conceptual simplicity. 

\section{Datasets}
\label{sec:datasets}

To facilitate a comparison with other work, we test our methods on the datasets of the \challenge.
The first dataset from the \challenge consists of voxelized showers from single particles, $\gamma$ or $\pi^{\pm}$, interacting with the ATLAS detector in the $\eta$ range [0.2, 0.25]~\cite{ATLAS_showers}.
15 different incident particle energies, spanning the range 256 MeV up to 4 TeV in powers of 2, are included. 
10,000 events per incident energy are provided, except for the highest energies, which have fewer events and therefore higher statistical uncertainty.
In total, 242000 (241600) events are provided for the photon (pion) dataset.
These datasets were used by ATLAS to train the \fastcalogan~\cite{FastCaloGAN} model used in \texttt{AltFast3}~\cite{ATLFast3}. 
The voxelized representations have 5 and 7 layers with 368 and 533 voxels, respectively, for the $\gamma$ and $\pi^{\pm}$ showers.
There are different numbers of angular and radial bins within each layer to reflect the varying granularity of the ATLAS calorimeter.
For the photon (pion) datasets, layers 1 and 2 (1, 2, 12, and 13) have 10 angular bins and the rest have only a single angular bin. 
Each layer has a unique binning in the radial direction.
For example, the first layer of the pion dataset has 8 variable-width bins covering a radial distance up to 600 cm, while the last layer has 10 variable-width bins covering up to 2000 cm.
Because of the unique binning in each layer, only two bins from the first layer exactly align with a bin from the last layer.
`There are a total of 30 (23) unique radial bin edges for the photon (pion) dataset. 

Datasets 2 and 3 of the \challenge each consist of 200,000 showers from an electron incident on a cylindrical sampling calorimeter with 45 layers, each with an active (silicon) and passive (tungsten) component.
The electron energy spans the range of 1 GeV to 1 TeV with a log-uniform distribution.
Each layer in dataset 2 has 9 radial bins and 16 angular bins, leading to a total of $45\times16\times9 = 6480$ voxels in each shower.
Dataset 3 features a much higher granularity; each layer has 18 radial and 50 angular bins, leading to a total of $45\times50\times18 = 40500$ voxels in each shower.

Following the specifications of the \challenge, we split the available events evenly between training and evaluation for all datasets.
The resulting size of the training sample, only $O(100\text{K})$ showers, is relatively limited, especially for very high-dimensional data such as dataset 3.
It is likely generating additional showers for training would lead to improved performance.
However, if this limited sample is taken to represent only a small portion of a real particle detector geometry, it may be a realistic estimate of the practically achievable training sample size, given restrictions on available computing resources.
For example, the approach employed by ATLAS for \fastcalogan involves training a separate model for each of 100 different $\eta$ regions of the detector and thus 
can only generate a limited number of events for each $\eta$ region. 

\section{Methods}
\label{sec:methods}

\subsection{Preprocessing}

We apply several stages of preprocessing to the showers before the diffusion process.
First, the energy in each voxel is divided by the incident particle energy, yielding the normalized energy $E_i$ in voxel $i$. 
As in previous work~\cite{CaloFlow,CaloScore}, a `logit' transformation is then applied to the voxel energies:
\begin{equation}
    u_i = \log\left(\frac{x}{1-x}\right), \quad x = \delta  + (1 - 2*\delta) * E_i,
\end{equation}
where $\delta = 10^{-6}$ avoids discontinuities at $x = 0$ and $x = 1$.
We then subtract the mean and divide by the standard deviation of the transformed voxel energy distribution $u_i$:
\begin{equation}
    u^{\prime}_i = \frac{u_i - \overline{u}}{\sigma_{u}}.
\end{equation}
The distribution of preprocessed voxel energies $u^{\prime}$ has zero mean and unit variance, which is important to ensure the signal-to-noise ratio during the diffusion process has the appropriate magnitude.

The incident particle energy is used as a conditioning input to the model.
We first apply a logarithm to the energy and then scale the resulting values to fall in the range 0 to 1.

\subsection{Diffusion Specifics}\label{sec:implementation}

We train our model based on a diffusion process with 400 noising steps.
We follow Ref.~\cite{improved_diffu} and use a `cosine' noise schedule, defined as: 
\begin{equation}
\overline{\alpha}_t = \cos\left(\frac{ \frac{t}{T} + s} {1 + s} \cdot \frac{\pi}{2}\right)
\end{equation}
with $s = 0.008$. 
This noise schedule adds noise more slowly during the intermediate steps of the diffusion process than the simple linear schedule originally used in Ref.~\cite{DDPM}. 
This preserves information for longer during the process, and
we find it reduces the number of diffusion steps needed to maintain high quality.

As mentioned in Section~\ref{sec:background}, there are different choices for the parameterization of the training objective of the model.
The most obvious approach is to predict the denoised image $x_0$ directly.
Ref.~\cite{DDPM} suggests predicting the normalized noise component, $\epsilon$, and then computing $\mu_{\theta}$ as: 
\begin{equation}
\mu_{\theta}(x_t,t,z) = \frac{1}{\sqrt{\overline{\alpha_t}}}\left(x_t - \frac{1 - \alpha_t}{\sqrt{1 - \overline{\alpha}_t}} \epsilon_{\theta}(x_t, t)\right).
\end{equation}
In this case, training proceeds by minimizing the loss:
\begin{equation}
\mathcal{L} = \mathbb{E}_{t,\epsilon}\left[|| \epsilon_{\theta}(x_t,t,z) - \epsilon ||^{2} \right] .
\end{equation}
The argument in favor of predicting the normalized noise component as the training objective is that it allows the model output to stay in a consistent range, so the model learns to make
subtle refinements when the noise levels are small.
However, when the noise levels are large, small inaccuracies in the model prediction can lead to large changes to the image in the sampling process.
This can be somewhat mitigated by skipping the first steps of the diffusion process during generation in order to avoid this divergent behavior.
We find this parameterization works well for datasets 1 and 2. 

For dataset 3, we find the training objective suggested by Ref.~\cite{elucidating_design}, where the model predicts a weighted average of the noise component and the denoised image, yields better results:
\begin{equation}
\begin{aligned}
\mathcal{L} = \mathbb{E}_{t,\epsilon}\left[w(t)\left|\left| \vphantom{\frac{1}{c_\text{out}(t)}}  \right.\right.\right. &F_{\theta}(x_t, t, z) - \vphantom{x}\\
&\left.\left.\left.\frac{1}{c_\text{out}(t)}\left(x_0 - c_\text{skip}(t) \cdot (x_t) \right) \right|\right|^{2}\right] .
\end{aligned}
\end{equation}
The different weighting functions are chosen to be proportional to the standard deviation of the total amount of noise at each step $t$, $\sigma(t) = \sqrt{1 - \overline{\alpha}_t}$.
Specifically, $w(t) = 1 + \sfrac{1}{\sigma(t)^2}$, $c_\text{skip}(t) = \sfrac{1}{(\sigma(t)^2 + 1)}$, and $c_\text{out}(t) = \sfrac{1}{(1 + \sfrac{1}{\sigma(t)^2})}$.
With this combination of terms, the model trades off between predicting the noise component when the noise is small, and predicting the denoised image when the noise is large. 
For $t \to 0$, $\sigma(t) \to 0$ and $c_\text{skip} \to 1$, so the training objective of the model is roughly proportional to $\epsilon$.
But for $t \to T$, $\sigma(t) \to 1$ and $c_\text{skip} \to \sfrac{1}{2}$, and the training objective is a weighted average of the denoised image $x_0$ and the noise component $\epsilon$.
This scheme makes the model less sensitive to inaccurate predictions at high noise levels during the sampling process.
This effect is more important for dataset 3 because of its higher sparsity, which leads to a longer tail in the voxel energy distribution.
When using this training objective, skipping the first iterations of the diffusion process when sampling is no longer required.

We follow the stochastic sampling algorithm proposed in Ref.~\cite{DDPM}, in which a small amount of additional noise is added back to the sample after each denoising step:
\begin{equation}
    x_{t-1} = \frac{1}{\sqrt{\overline{\alpha_t}}}\left(x_t - \frac{1 - \alpha_t}{\sqrt{1 - \overline{\alpha}_t}} \epsilon_{\theta}(x_t, t)\right) + \sigma_t\epsilon^{\prime}
\end{equation}
for $\epsilon^{\prime} \sim \mathcal{N}(0, \mathbb{I})$ and $\sigma_t = \beta_t (\sfrac{1-\overline{\alpha}_{t-1}}{1-\overline{\alpha}_t})$.

As long as the model is conditioned on the noise level, the number of diffusion steps in the sampling need not be the same as the number of steps in the training.
Decreasing the number of diffusion steps will linearly improve the computational time needed to generate samples, but may produce samples of lower quality.
This provides significant flexibility in trading between sample quality and computation time for a trained model.
As the optimal balance between sample quality and computation time will be application-specific, in this work we primarily focus on sample quality.
For datasets 1-photon, 2 and 3 we choose 200 diffusion steps for sampling because we find it does not significantly degrade sample quality compared to 400 steps, but further reductions do.
For dataset-pions we find that 200 diffusion steps noticeably reduces the sample quality and therefore report results using 400 steps.
Additionally, for datasets 1 and 2, we find that skipping the first two denoising steps (i.e. starting from $x_{T-2}$ rather than $x_{T}$) 
avoids instabilities caused by imperfect estimates of $\epsilon$ at the highest noise levels.

After generation, we apply a cutoff on the minimum voxel energy to match the minimum value in the \challenge datasets.  
This corresponds to a value of 10 MeV for dataset 1, and 15 keV for datasets 2 and 3. 
Voxels below this value are set to zero. 
We note that the threshold for datasets 2 and 3 is likely unrealistically low for a real detector operating in the energy range considered, however we use these values to maintain consistency with the \challenge. 

\subsection{Network Architecture}

The primary input to the network is the noisy representation of the shower. However, additional, conditional information is provided as well.
The conditional information consists of the scaled logarithm of the incident energy of the particle and the noise level of the current diffusion step ($\sqrt{1 - \overline{\alpha_t}}$).
This conditional information is encoded into a 128 dimensional vector via a two-layer fully connected network.

\begin{figure*}[!ht]
    \centering
    \raisebox{-0.5\height}{\includegraphics[width=0.58\textwidth]{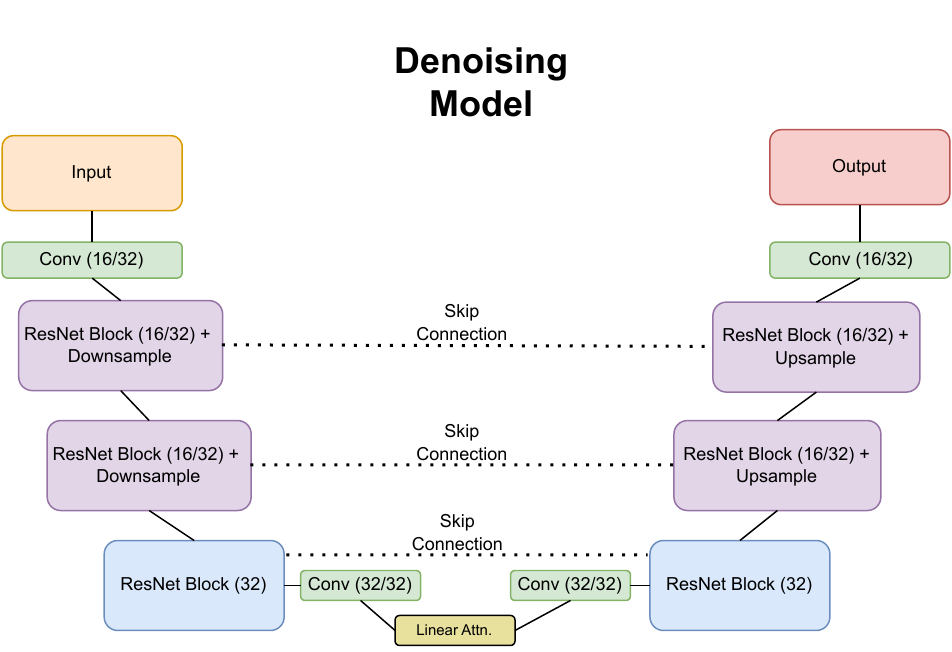}}
    \raisebox{-0.5\height}{\includegraphics[width=0.4\textwidth]{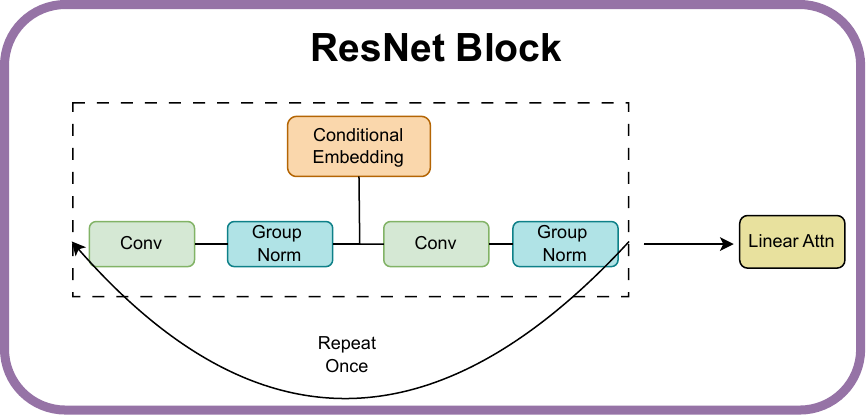}}
    \caption{Left: a schematic of the network architecture. The numbers in parentheses for each module indicate the number of filters used in that module for the network for datasets 1 and 2 / dataset 3. Right: A detailed view of the operations in a ResNet block.}
    \label{fig:model_arch}
\end{figure*}

The denoising model uses a U-net~\cite{Unet} architecture, which is commonly employed in diffusion tasks.
U-net architectures resemble an encoder-decoder pattern, where the input is gradually compressed to a smaller space, but unlike an autoencoder, skip connections are used so that there is no information bottleneck.
Our U-net has an initial convolution followed by a series of ResNet blocks~\cite{RESNET}. 
Conditional convolutions are created by adding the conditional information as an additional bias term after the first convolutional layer of each ResNet block.
For datasets 1 and 2 (3) we use three ResNet blocks for the encoder with 16/16/32 (32/32/32) filters.
Convolutional layers with a stride of 2 and appropriate padding are used to reduce the data size by a factor of two in each dimension after each of the first two ResNet blocks. 
Linear self-attention layers~\cite{linear_attention} are applied after each ResNet block.
The architecture is then mirrored, with three more ResNet blocks with the same filter sizes.
Convolutional transpose layers are used to upsample by a factor of two after each ResNet block to return to the original data dimension.
A schematic of the network architecture is shown in Fig.~\ref{fig:model_arch}.

In total, the models for datasets 1 and 2 (dataset 3) consists of ${\sim} 520\text{K}$ (${\sim} 1.2\text{M}$) parameters.
The model architectures were not extensively optimized, and it is likely the performance could be further improved with a dedicated optimization procedure.

\section{Geometric Innovations}

\subsection{Optimizations for Cylindrical Geometry}

Regular convolutions achieve their power by exploiting the underlying symmetry of the data: translation invariance along each of the coordinate dimensions.
When a convolutional layer is applied to an image, the filters perform the same local operation across the whole input image.
This allows for expressive, parameter-efficient operations on high-dimensional images. 
However, while calorimeter showers represented in a voxelized cylindrical geometry have a regular structure, they are not inherently translation invariant. 
The distribution of energy deposited in each layer encodes important information about the shower, which would be spoiled by translating the shower in either direction along the layer axis.
Likewise, the distribution of energy in the radial direction encodes important information about the transverse spread of the shower and falls rapidly as a function of the distance away from the shower center.
Additionally, in a realistic detector, sensors in different layers may have different sizes or be made of different materials.
The one coordinate dimension that may be translation invariant is the angular dimension. 
However, this dimension has a periodic topology that will not be respected by regular convolutions. 
We therefore design several novel optimizations of the convolution operation tailored to cylindrical data that improve the output fidelity.

In order to respect the periodicity of the angular dimension in cylindrical calorimeters, our denoising network uses cylindrical convolutions rather than standard Cartesian ones.
The angular dimension is represented in a linear array, so neighboring values with coordinates near the extrema of the angular range are far apart in the array representation.
Before each cylindrical convolution operation, a circular padding is added in the angular dimension, such that both ends of the linear array are extended with the values from the opposite end.
This ensures that when a 3D convolution is applied, the voxels close to the ends of the linear array properly interact with their angular neighbors on the opposite end.
This padding is only applied to the angular dimension; the radial and $z$ dimensions remain unchanged.

To allow our convolutional operations to violate translation invariance, we devise a novel scheme for location-conditional convolutions. 
This is implemented by augmenting the shower image with additional input channels that encode the position of each voxel.
We construct one `layer image' in which the value of each voxel corresponds to the layer number of that voxel, normalized to the range 0 to 1.
We similarly construct a `radial image' that encodes the radial distance of each voxel, also normalized from 0 to 1.
For dataset 1, we observe slight non-uniformities in the energy distribution as a function of the angular bin, and find slight performance gains from including an `angular image' as well.
These additional images are concatenated to the per-voxel shower energy as additional input channels.
This allows the filters in the convolutional operations to produce different results in different parts of the geometry. 
The output of the denoising network is still a single channel corresponding to the energy in each voxel.
As these images are the same for every input, in principle they do not supply any additional information to the model.
Therefore, one would expect that they would be unnecessary for a sufficiently large and expressive model.
However, in practice, with the models employed in this work, we have found this technique makes it easier for the network to learn the non-uniformities of the underlying data.

\subsection{\texorpdfstring{\glam}{GLaM}: Geometry Latent Mapping}\label{sec:glam}

Though datasets 2 and 3 feature significantly larger numbers of voxels than dataset 1, their regular binning allows convolutional operations to be readily applied.
In contrast, the irregular binning in dataset 1 poses a challenge for fully utilizing the geometric structure of the data.
Overcoming this challenge is important for the application of these techniques to real detectors, which often do not have perfectly regular geometries. 
Previous approaches have either used fully connected networks~\cite{FastCaloGAN} or 1D convolutions with a very large network size~\cite{CaloScore}.
Point clouds approaches have also gained some recent support as a way around this problem~\cite{CaloClouds,point_cloud_comparison}.

We instead employ a new method called Geometry Latent Mapping or \glam.
\glam learns a mapping from the data geometry to a perfectly regular geometric structure that is similar to the actual irregular geometry.
This embeds the data in a regular space so that computationally efficient operations, such as cylindrical convolutions, 
can be used to accomplish the primary task of the ML algorithm (here, the denoising task of the diffusion model).
The reverse transformation to bring the results of the primary task back to the original space is also learned by \glam.

\begin{figure*}[!ht]
\centering
\includegraphics[width=0.8\textwidth]{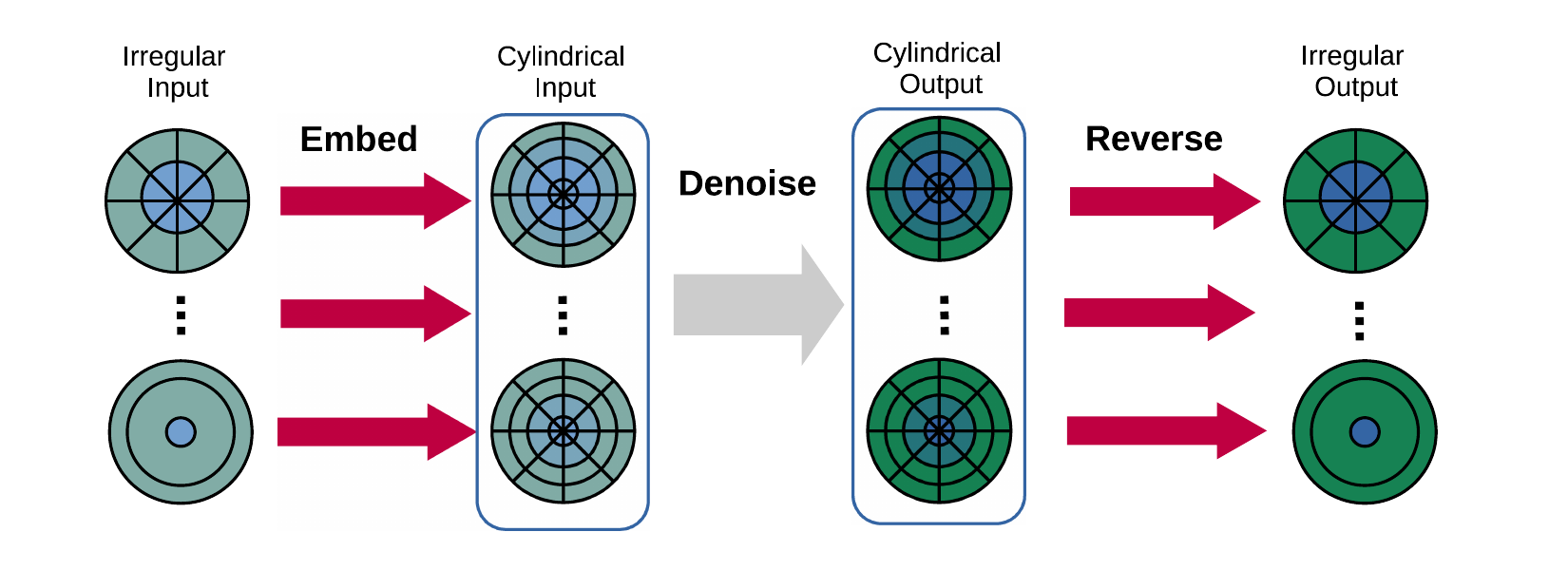}
\caption{A diagram of the Geometry Latent Mapping (\glam) approach.}
\label{fig:GLaM}
\end{figure*}

Separate mappings can be learned for different regions of the detector geometry. (A practical example is discussed below.)
The embedding for a particular region is therefore only based on local information from that region.
This ensures that the size of the embedding matrices remains small and that the embedded space reflects the inherent locality of the geometric structure.
\glam is philosophically similar to the approach of Latent Diffusion~\cite{LatentDiffusion}, which encodes data into a latent space learned by an autoencoder using a perceptual loss~\cite{PerceptualLoss} prior to the generative task.
However, with \glam, the embedded space can be larger than the input space, has a direct geometric interpretation, and is learned simultaneously with the generative task. 
A schematic of the \glam approach is shown in Fig.~\ref{fig:GLaM}.

We apply \glam to dataset 1 to learn a mapping of the input data to a regular cylindrical structure. 
During the diffusion training, the noise is still added in the original, irregular data structure, so that the embedding acts as just a part of the denoising model.\footnote{Adding the noise to the regular geometry results in a weak training signal for the embedding map, and therefore is more applicable to a situation in which the embedding is fixed or has been learned by some other means.}
We choose the radial and angular binning of this regular structure to be the superset of all the bin boundaries of the individual layers. 
This results in 10 angular bins and 30 radial bins for the photon dataset and 10 angular bins and 23 radial bins for the pion dataset. 
A separate mapping for each layer is then learned from the original binning in that layer to this regular structure.
The mapping along the radial dimension for layer $\ell$ is accomplished via a single matrix $C^\ell$, of size $c^\ell \times c^{\prime}$, where $c^\ell$ is the number of radial bins in the original geometry and $c^{\prime}$ is the number of bins in the regular geometry. 
The mapping back to the original space is likewise accomplished via a single matrix, $D^\ell$, of size $c^{\prime} \times c^\ell$.
The values of the $C^\ell$ matrix are trainable parameters, but initialized to values reflecting the geometric overlap of the original and regular binning scheme:
\begin{equation}
    C^\ell_{j,k} = 
\begin{cases}
    \frac{r^2_k - r^2_{k+1}}{r^2_j - r^2_{j+1}} + \kappa_{j,k} & \parbox{0.3\columnwidth}{if $r_k \geq r_j$\\ and $r_{k+1} \leq r_{j+1}$}\\
    \kappa_{j,k}              & \text{otherwise}.
\end{cases}
\end{equation}
Here, the values $r_j$ denote the bin boundaries in the original geometry, $r_k$ denote the bin boundaries in the regular geometry, and $\kappa$
is a tensor of Gaussian noise with mean zero and standard deviation $10^{-5}$.

Because $C^\ell$ generally maps between spaces of different dimensionality, the matrix is rectangular and does not have an analytic inverse. 
$D^\ell$ is instead initialized to the Moore-Penrose pseudo-inverse of $C^\ell$, so that initially $C^\ell D^\ell = \mathcal{I}$.
However, during training it is computationally difficult to backpropagate through the Moore-Penrose pseudoinverse function, so we instead allow the values of the inverse mapping $D^\ell_{k,j}$ to be independently trainable.

In dataset 1, layers either have a single angular bin or the same 10 angular bins.
We therefore take these 10 bins to be the regular structure and evenly divide the energy of layers with only a single angular bin among these 10 bins.
We found that learning a mapping for the angular dimension, similar to the one used for the radial dimension, provided no performance improvements beyond the simple energy splitting.

This is quite a simple ansatz, with the embedding being fully specified by 3180 (3404) parameters for the photon (pion) dataset.
We nevertheless find it works quite well in combination with cylindrical convolutions.
We find that a single embedding matrix with a geometrically-informed initialization yields significantly better results
than fully connected neural network layers initialized with standard techniques.

\section{Results}

We compare the showers generated with \diffu to those from \geant for all datasets from the \challenge.

\begin{figure*}[!ht]
    \centering
    \includegraphics[width=0.98\textwidth]{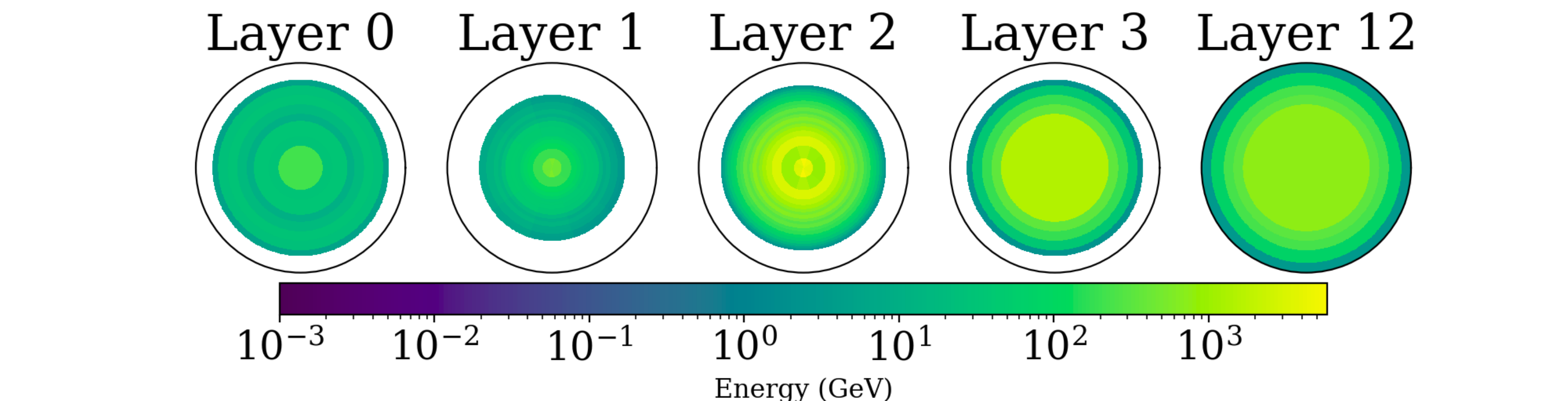}
    \includegraphics[width=0.98\textwidth]{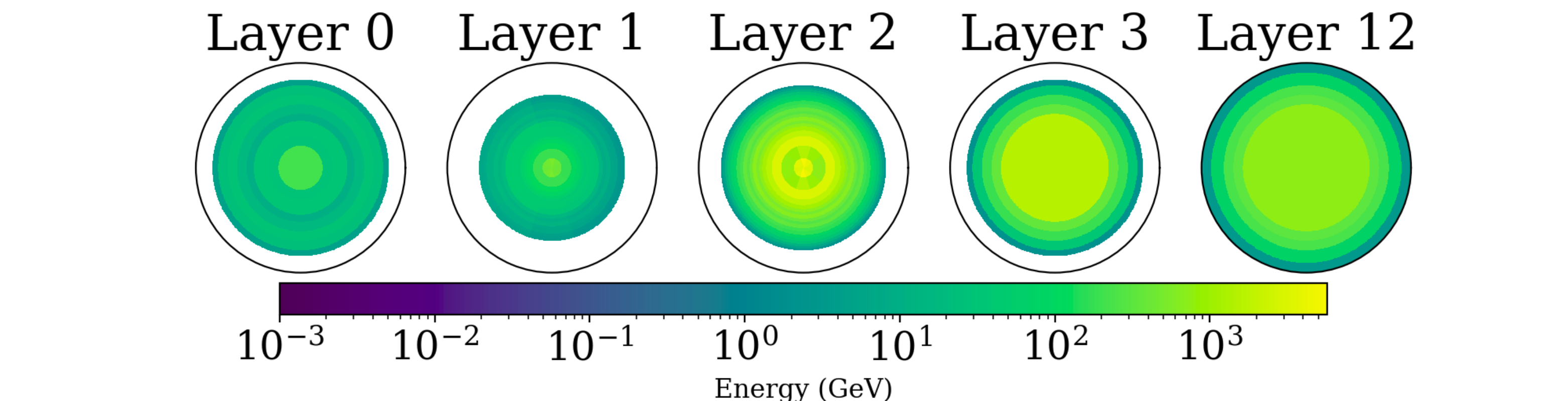}
    \caption{A comparison of the average showers produced by \geant (top) and \diffu (bottom) for the photon sample of dataset 1.}
    \label{fig:avg_showers}
\end{figure*}

\begin{figure*}[!ht]
    \centering
    \includegraphics[width=0.24\textwidth]{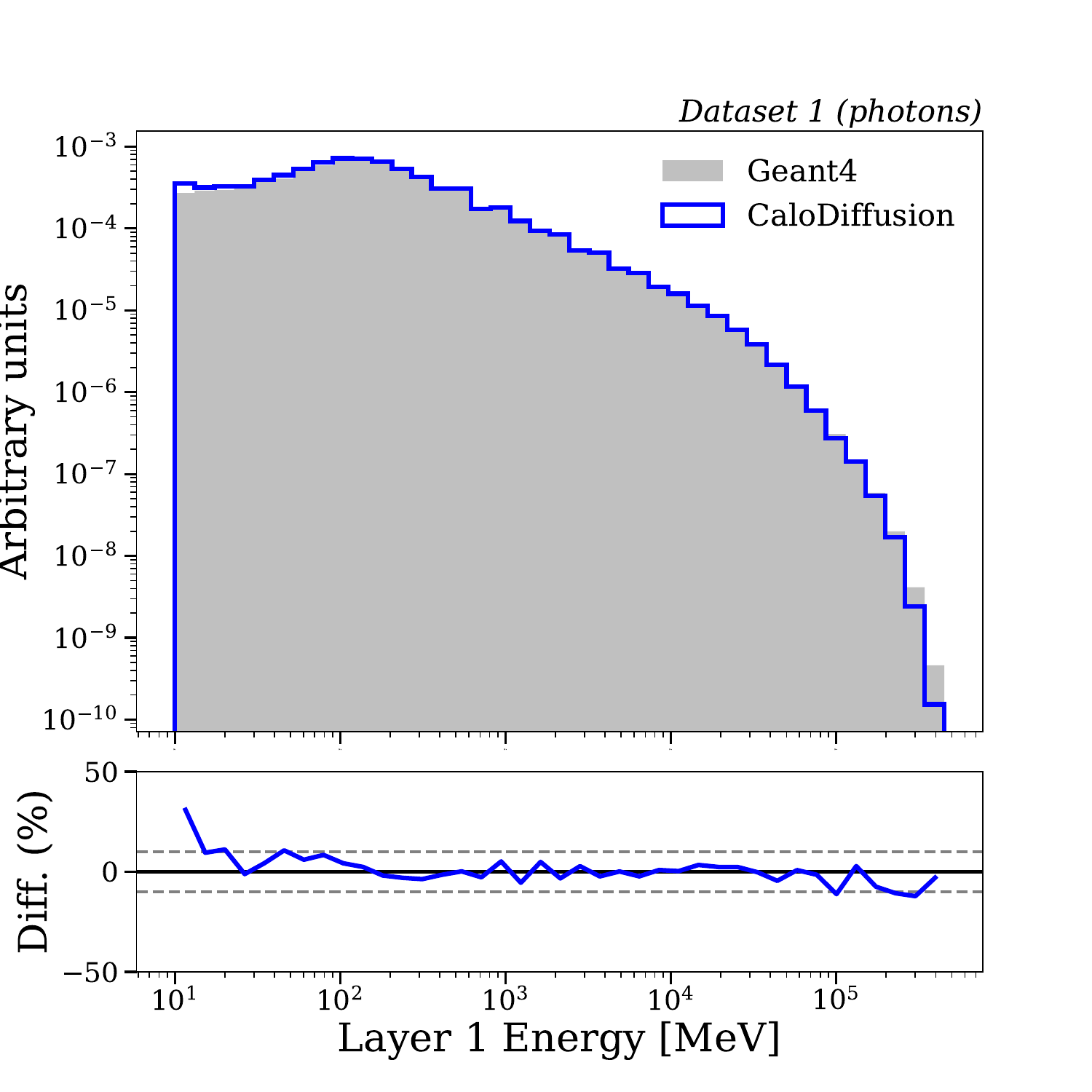}
    \includegraphics[width=0.24\textwidth]{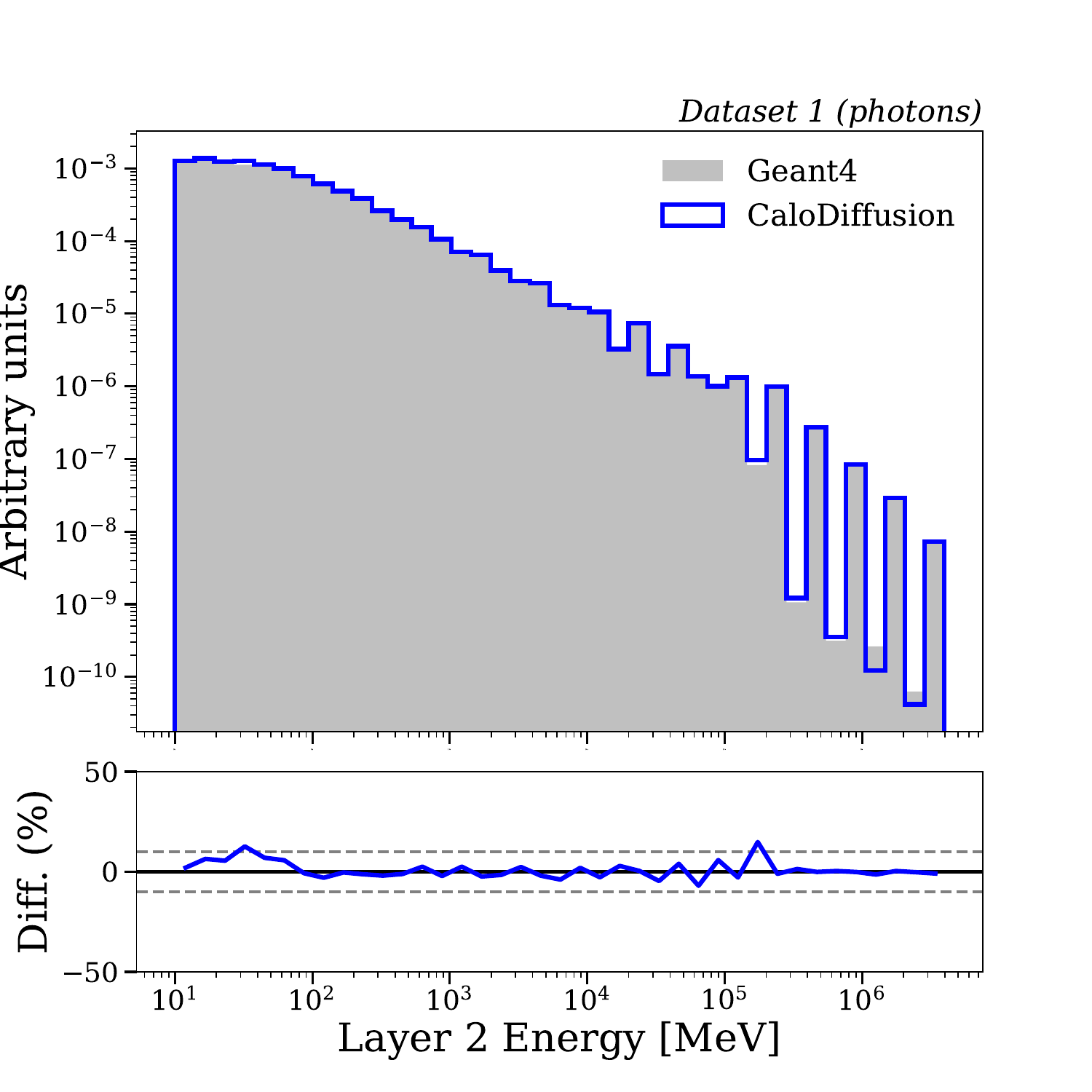}
    \includegraphics[width=0.24\textwidth]{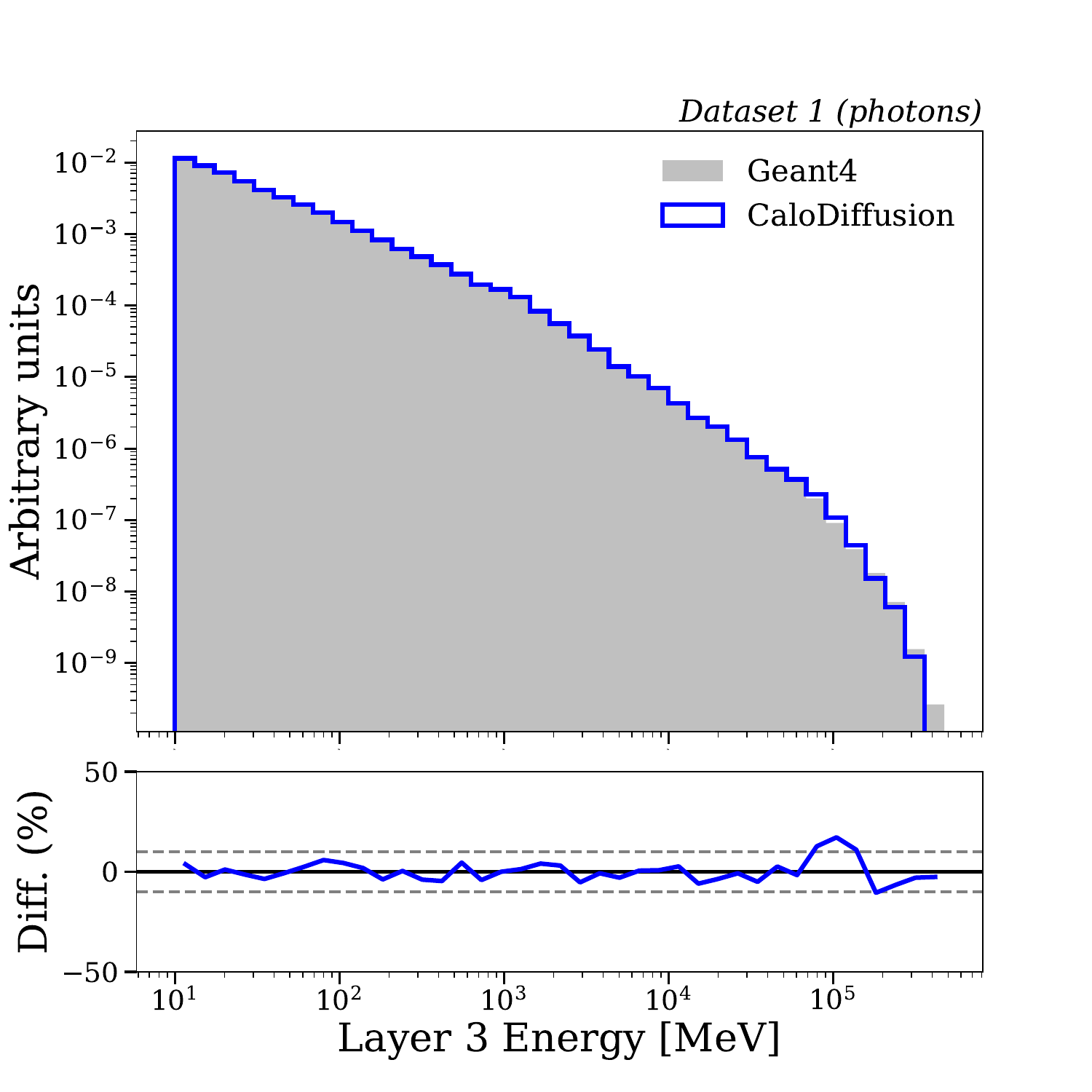}
    \includegraphics[width=0.24\textwidth]{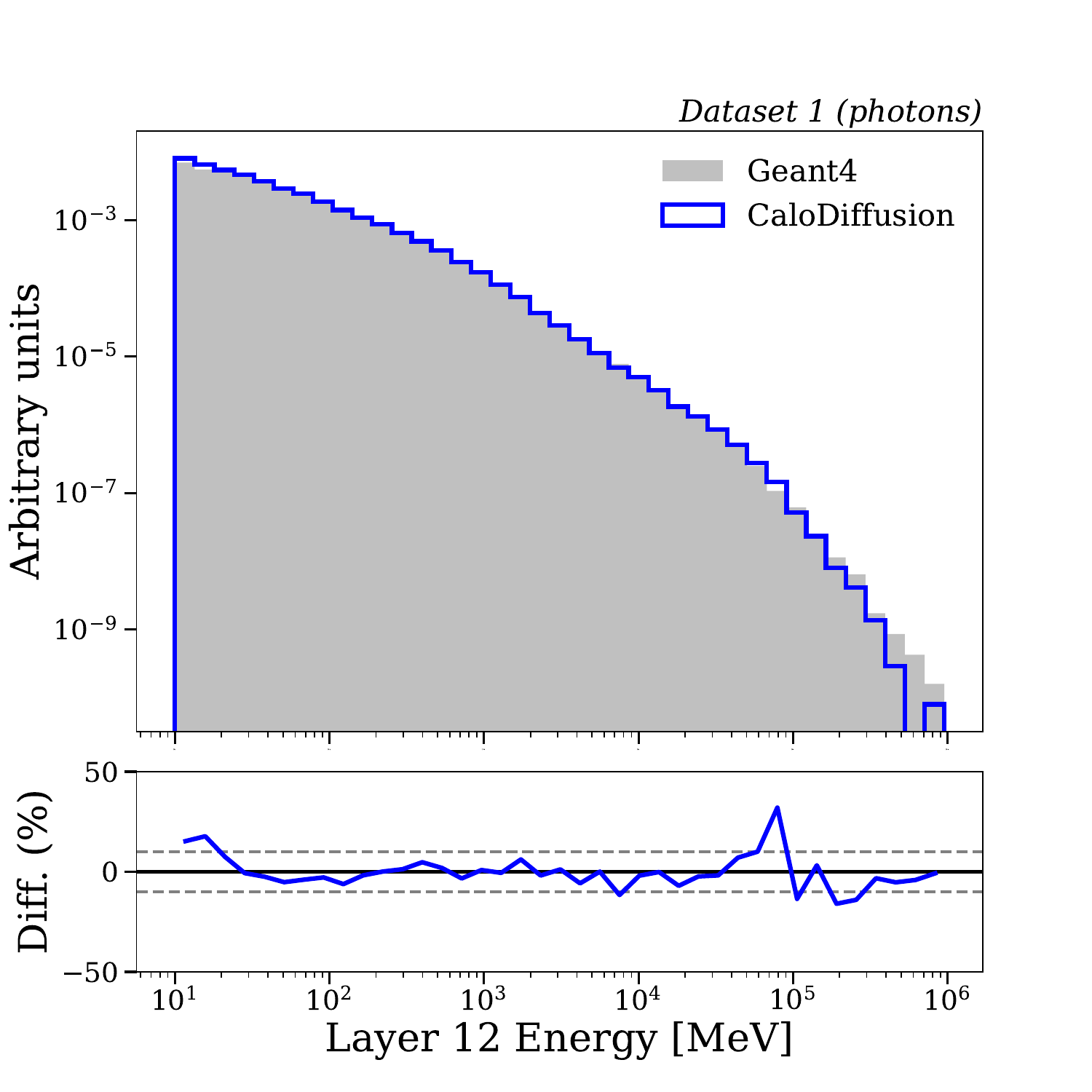}\\

    \includegraphics[width=0.24\textwidth]{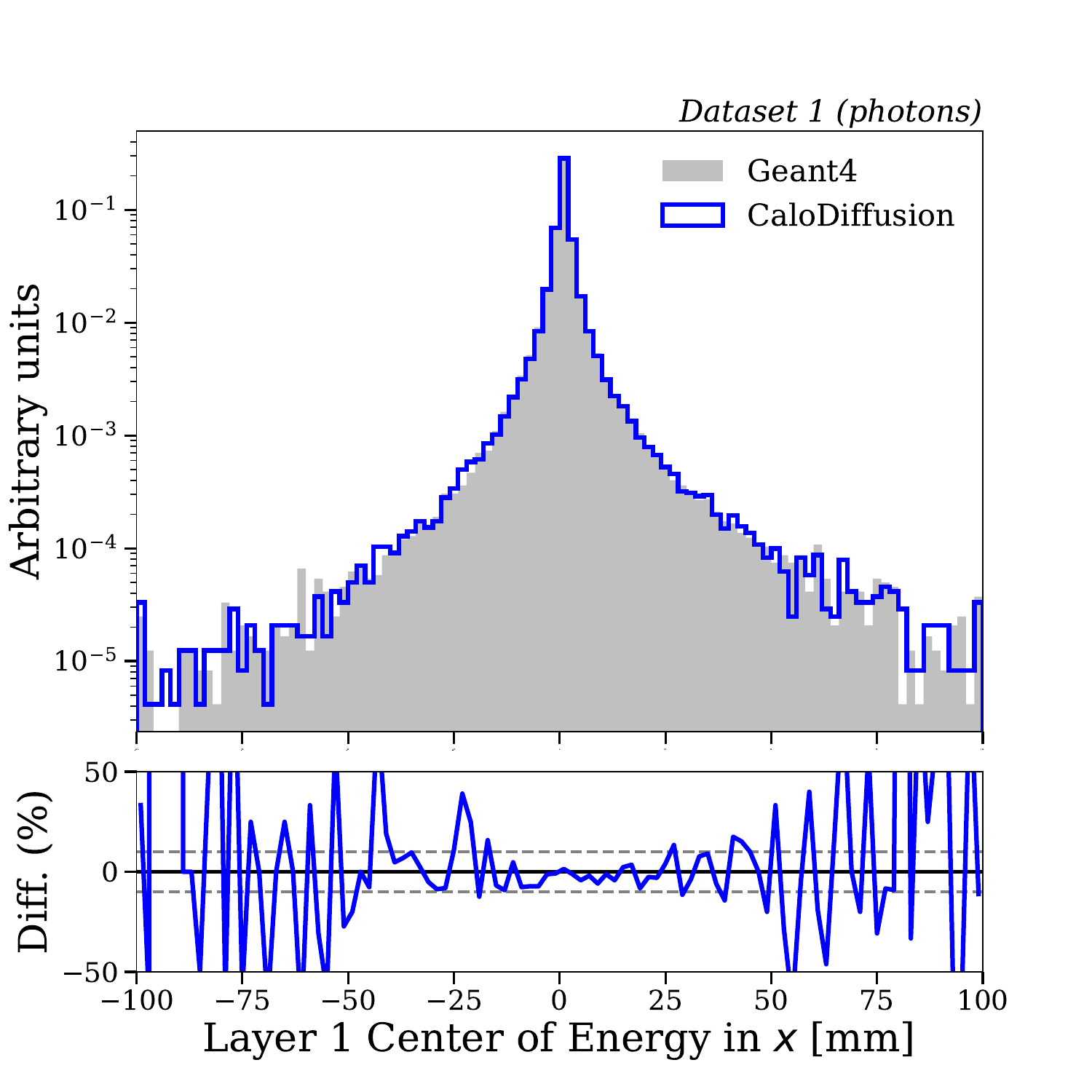}
    \includegraphics[width=0.24\textwidth]{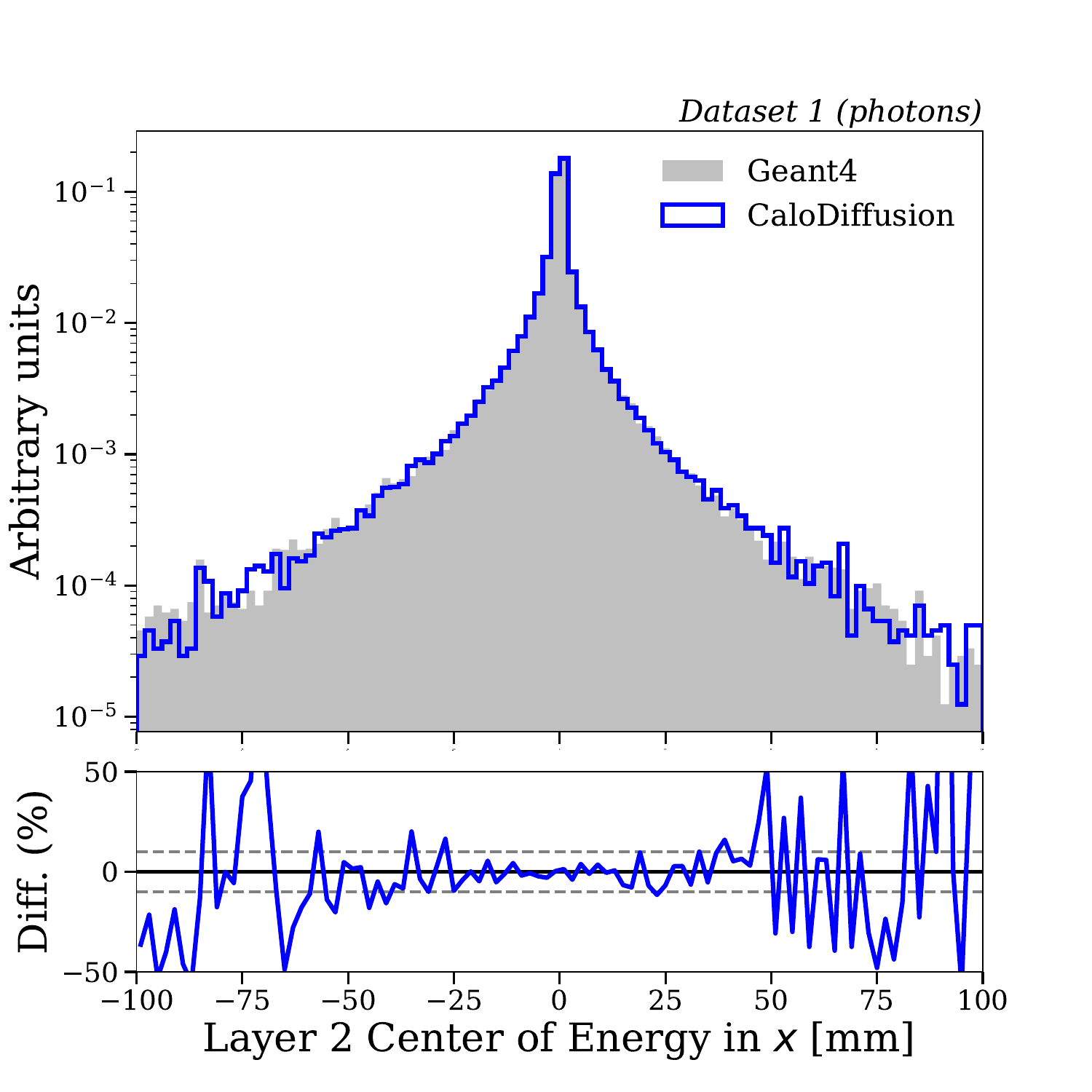}
    \includegraphics[width=0.24\textwidth]{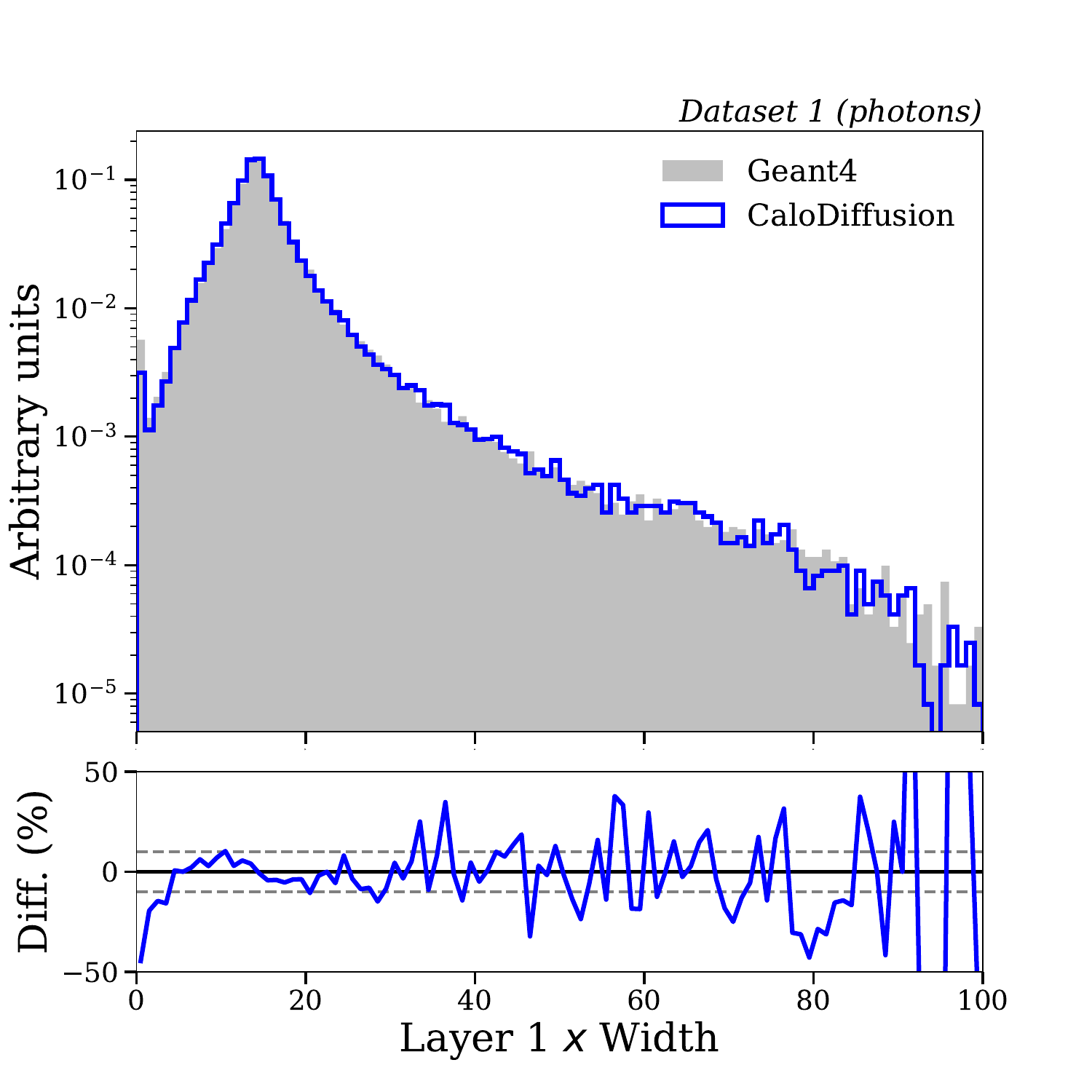}
    \includegraphics[width=0.24\textwidth]{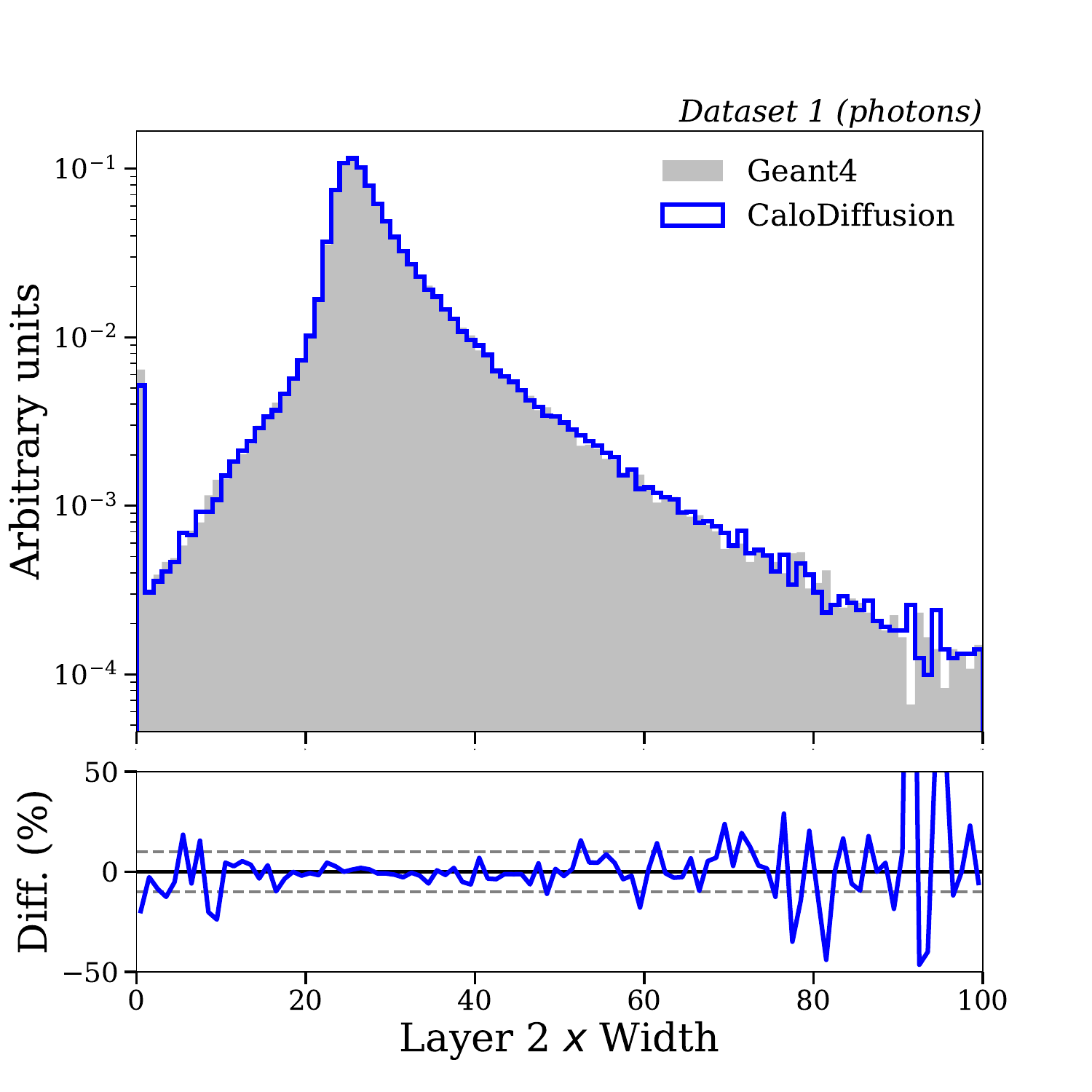}\\

    \threefigeqh{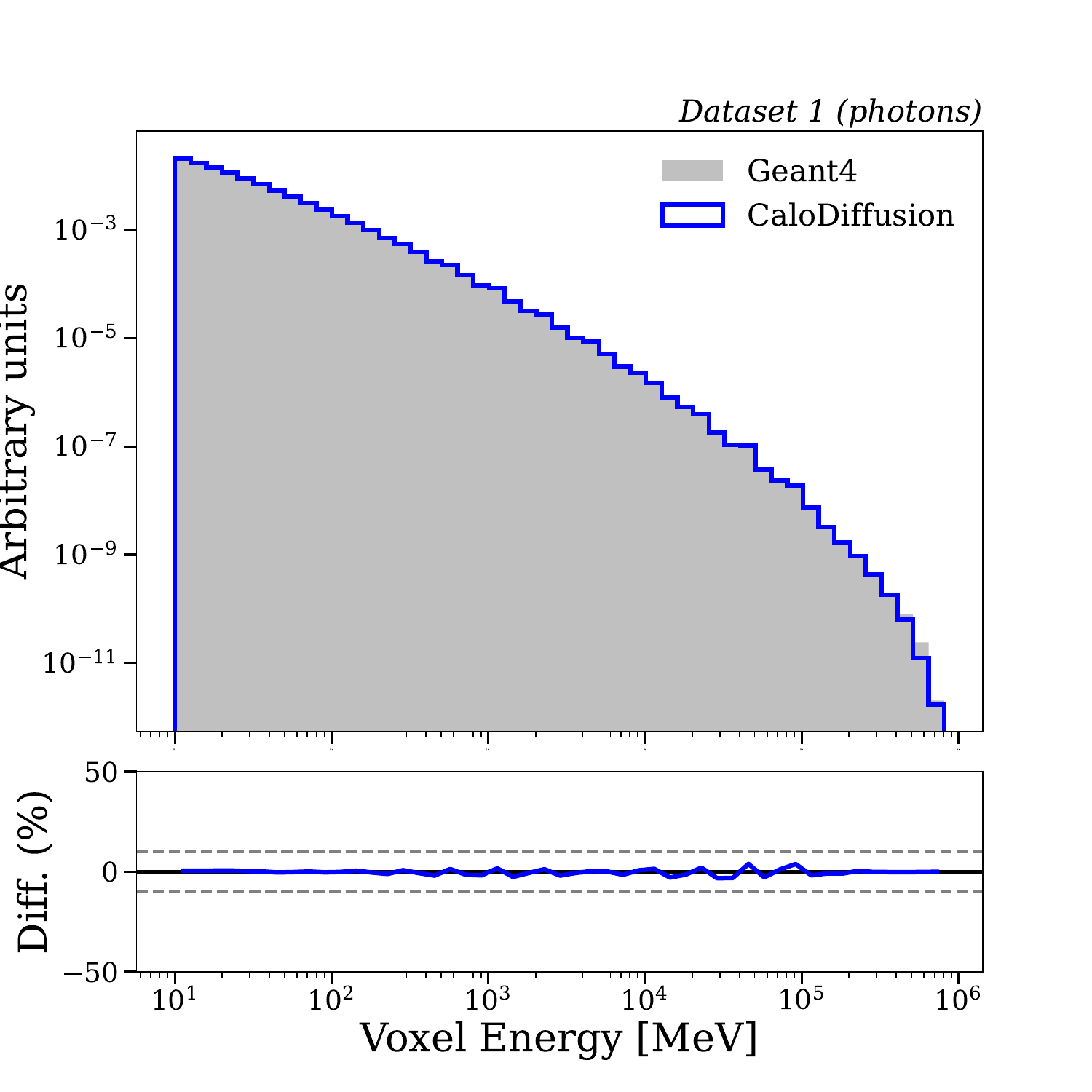}{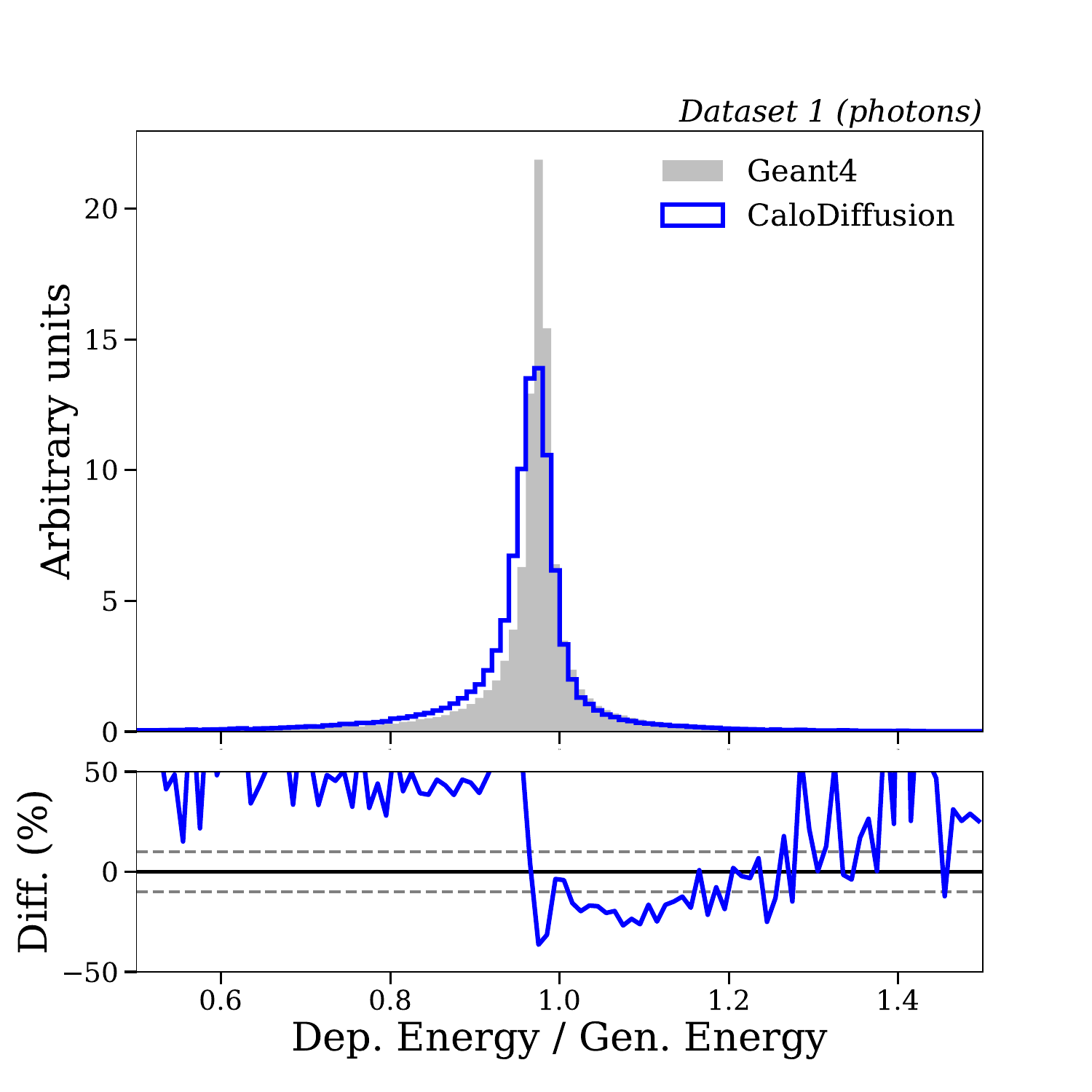}{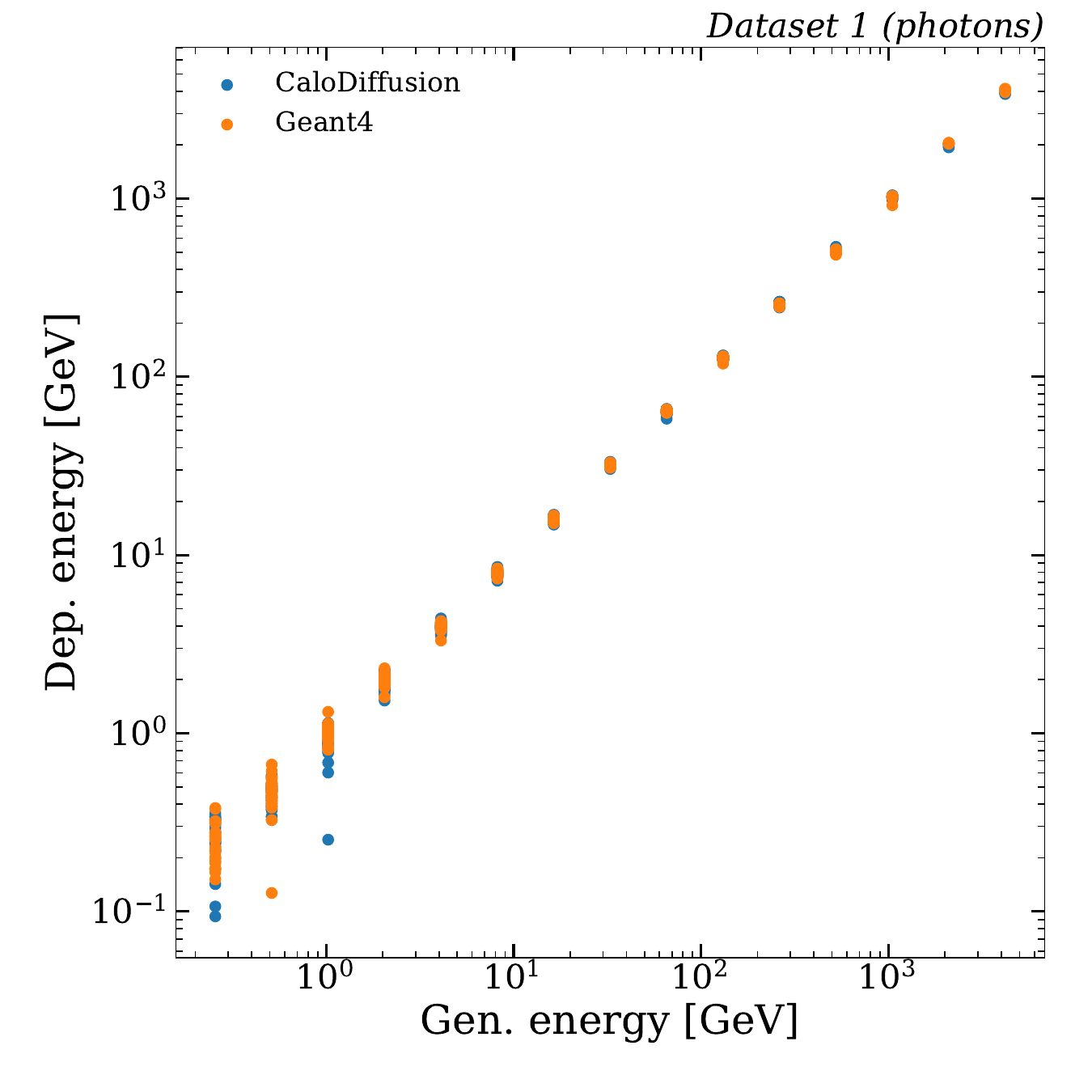}

    \caption{A comparison between \geant and \diffu showers across a variety of observables for the photon sample of dataset 1. The top row shows the distribution of energy in different layers of the calorimeter. The middle row shows the distribution of the center and width of the energy spread in two reference layers. The bottom row shows the distribution of voxel energies, the distribution of total shower energy divided by the incident energy, and a scatter plot of deposited energy versus incident energy. }
    \label{fig:dataset1_phot}
\end{figure*}

\begin{figure*}[!ht]
    \centering
    \includegraphics[width=0.24\textwidth]{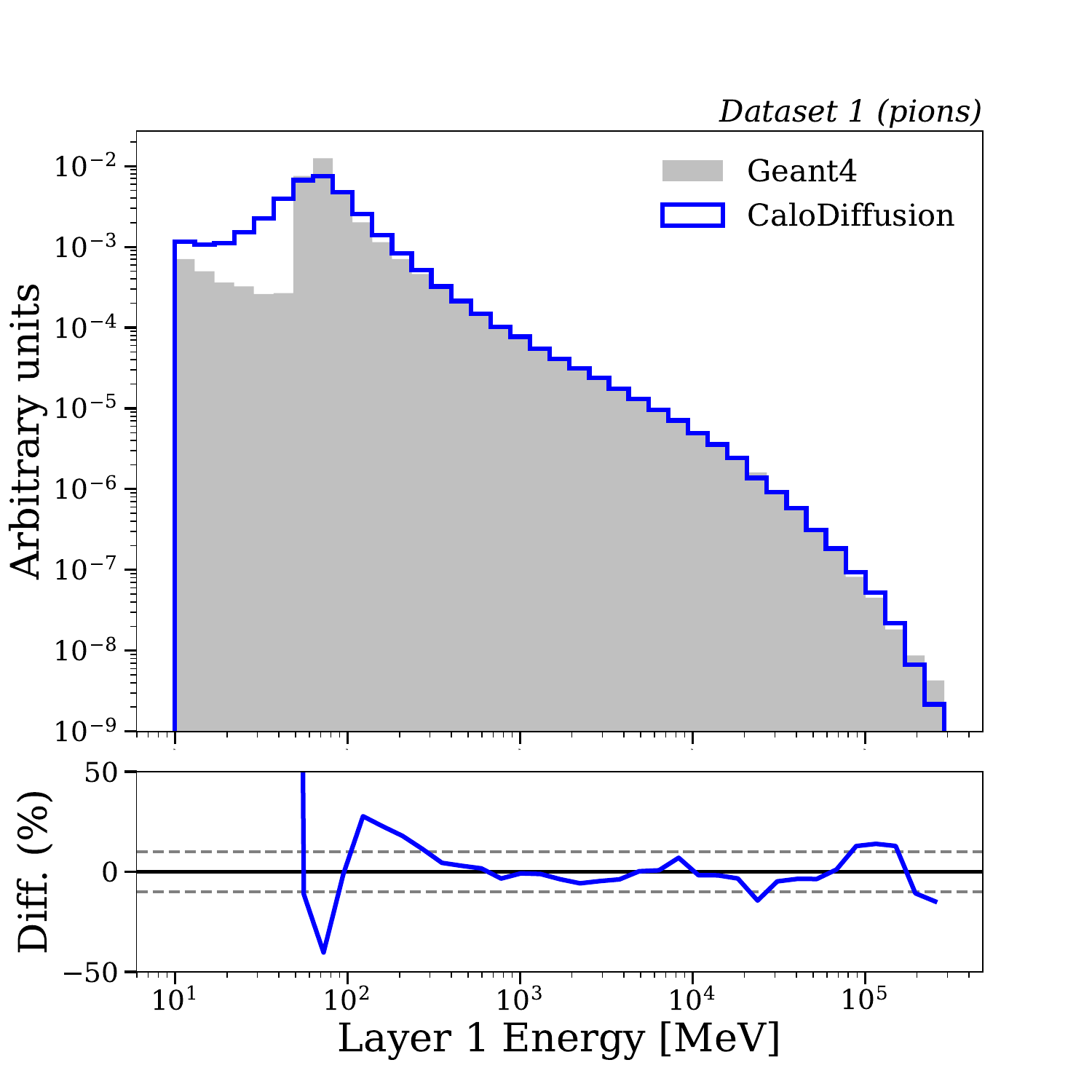}
    \includegraphics[width=0.24\textwidth]{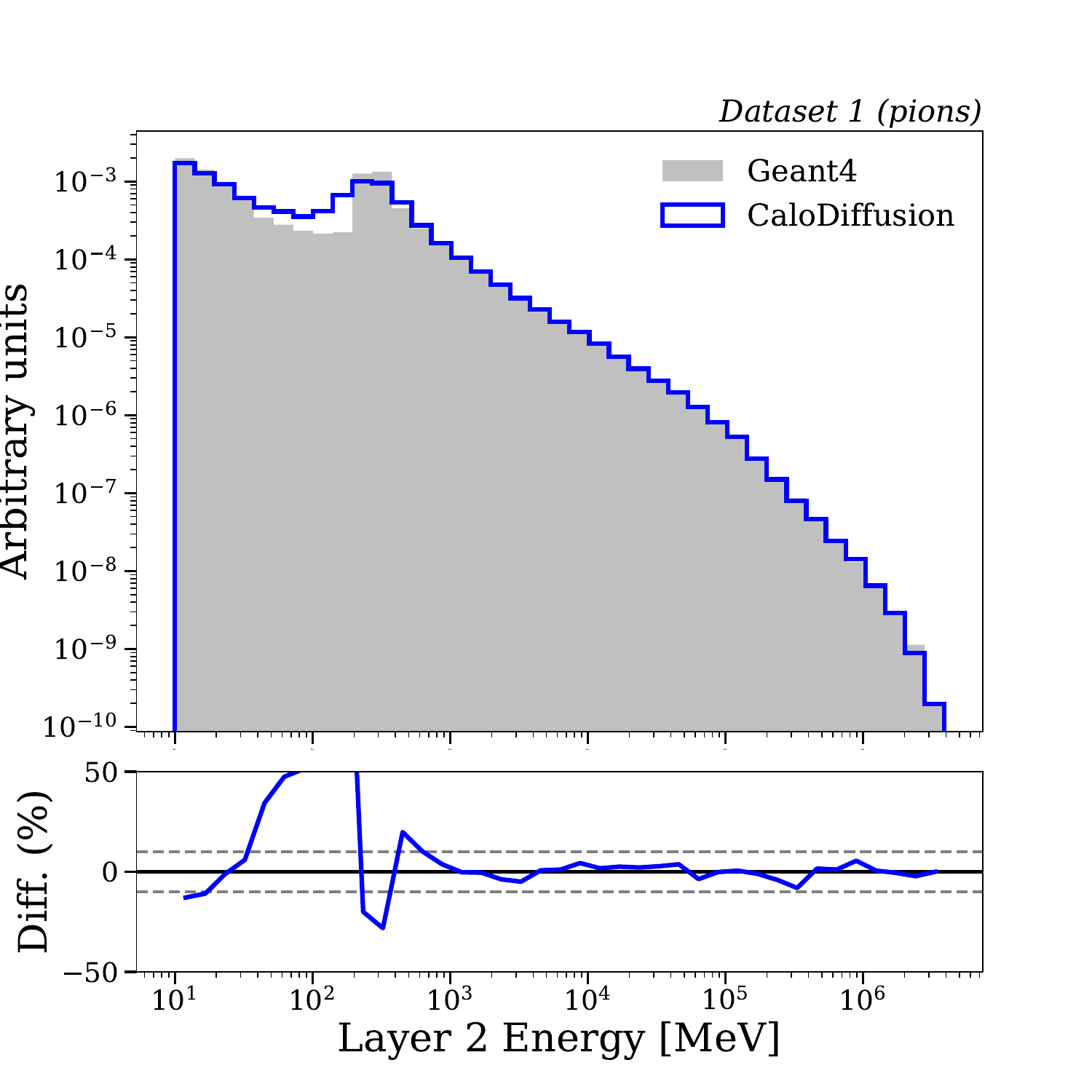}
    \includegraphics[width=0.24\textwidth]{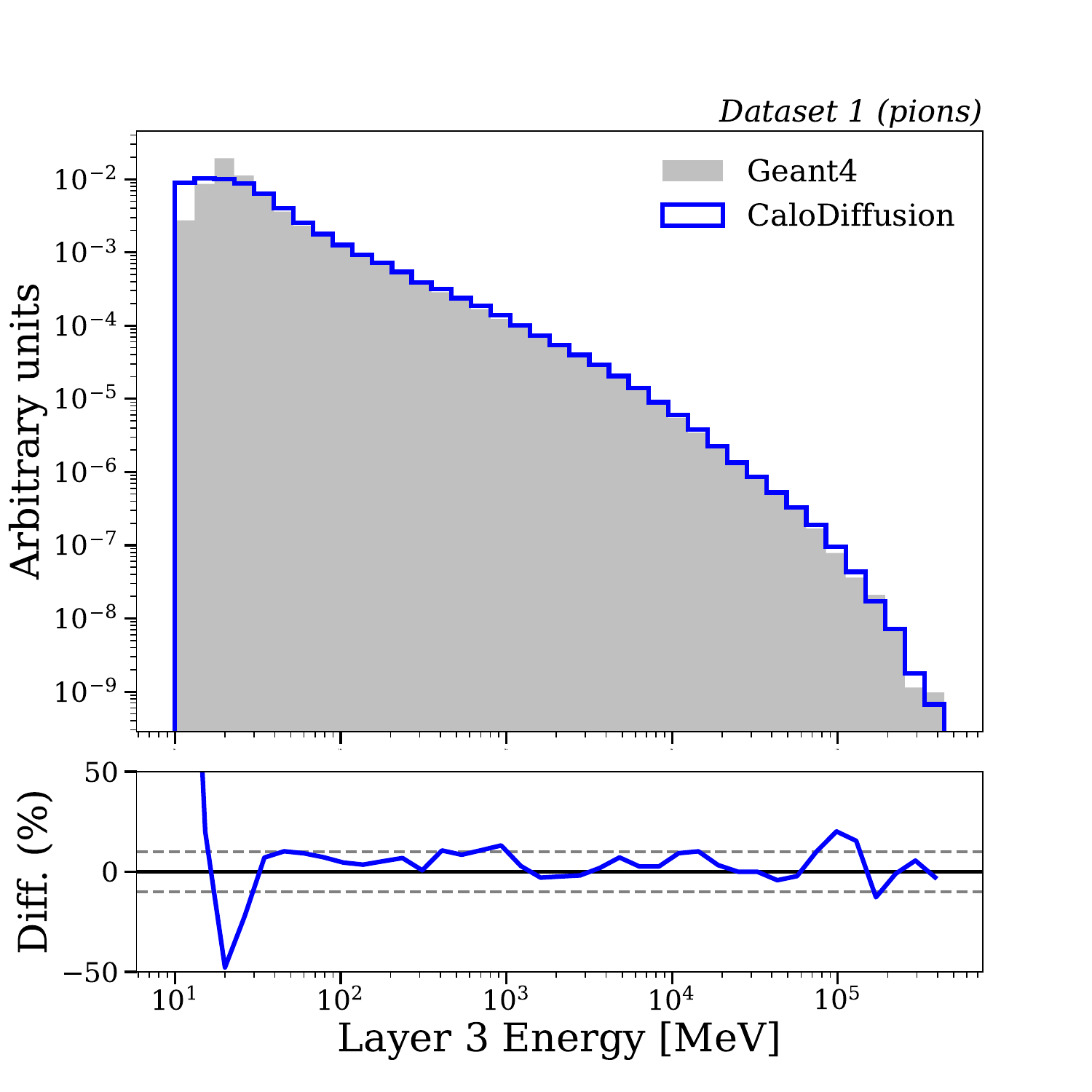}
    \includegraphics[width=0.24\textwidth]{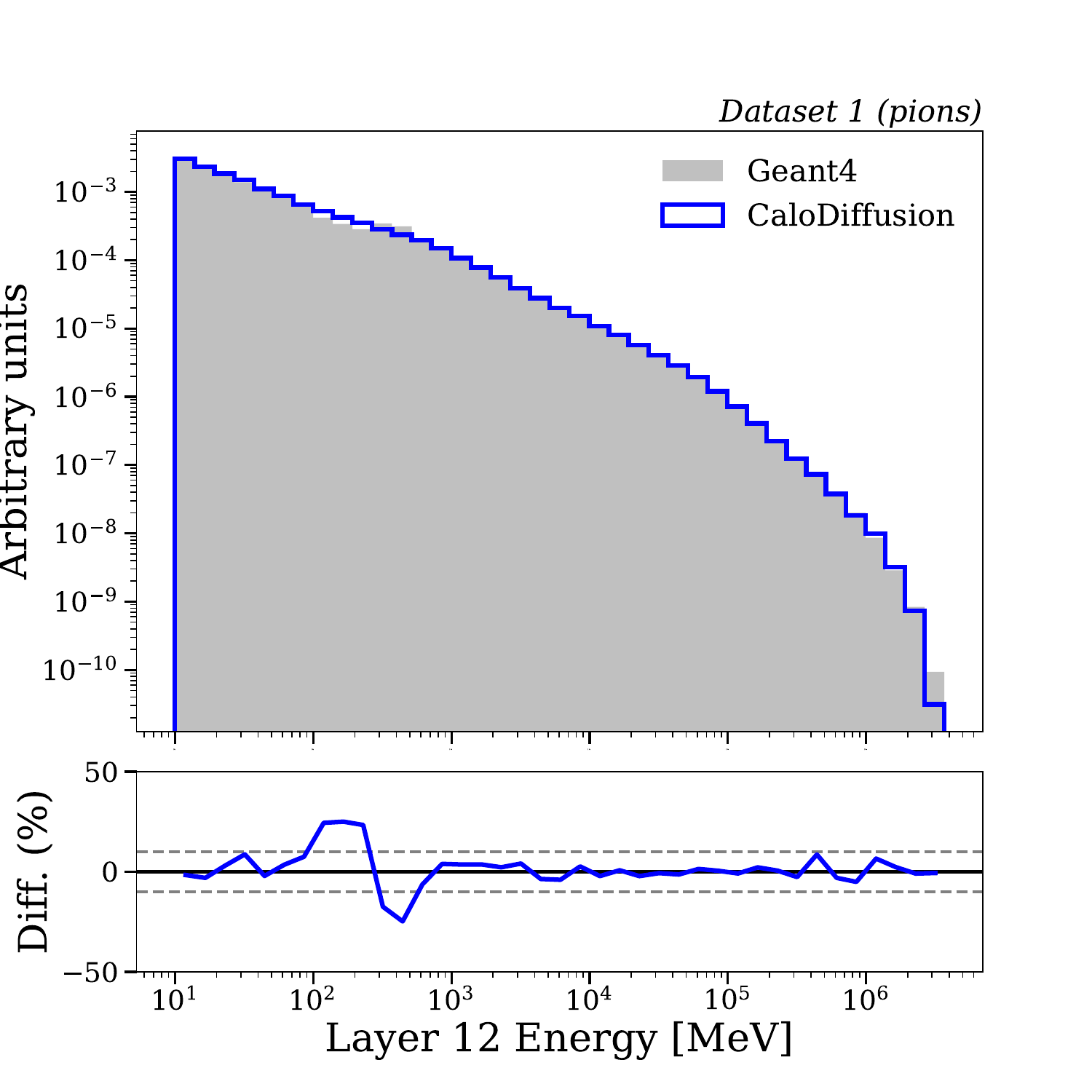}\\

    \includegraphics[width=0.24\textwidth]{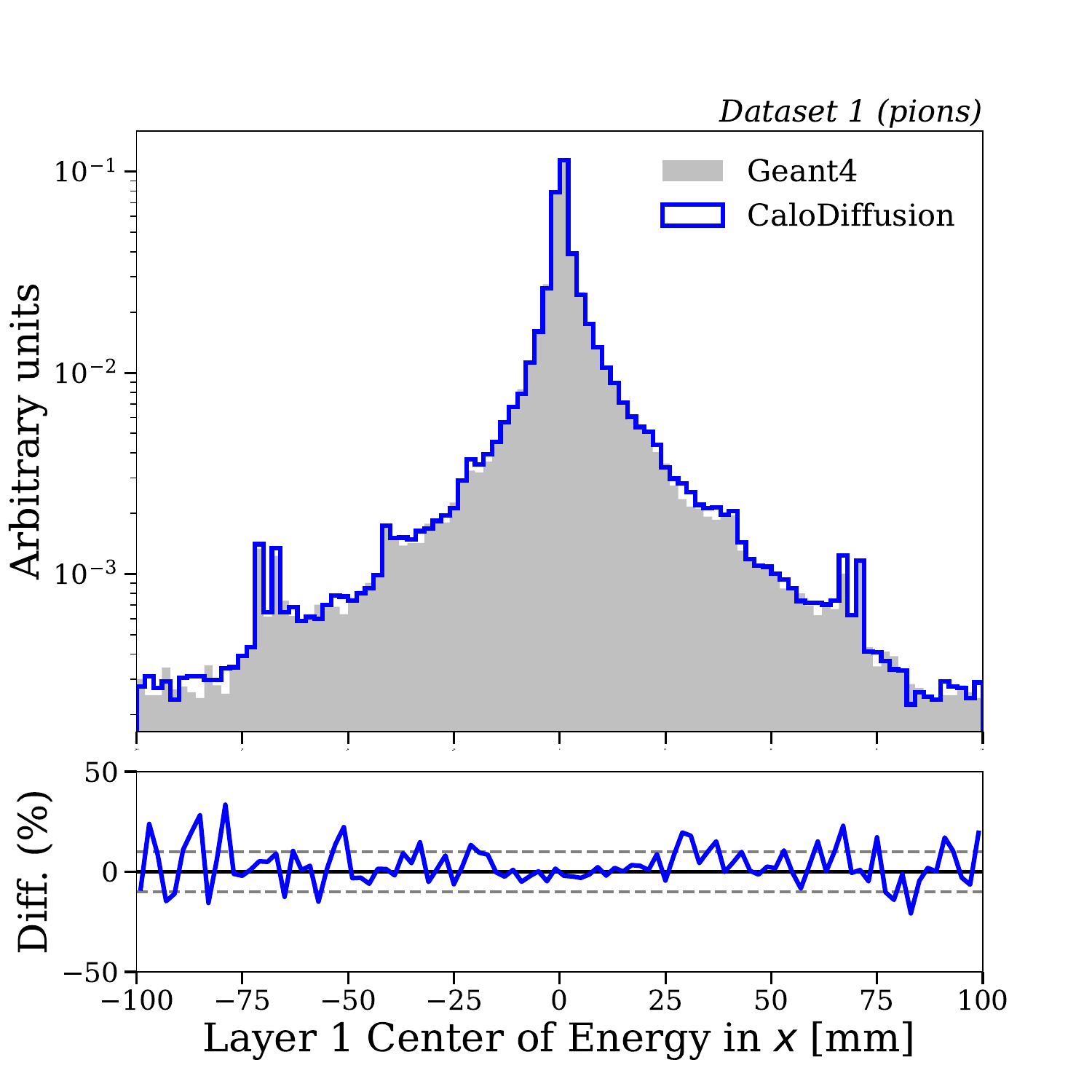}
    \includegraphics[width=0.24\textwidth]{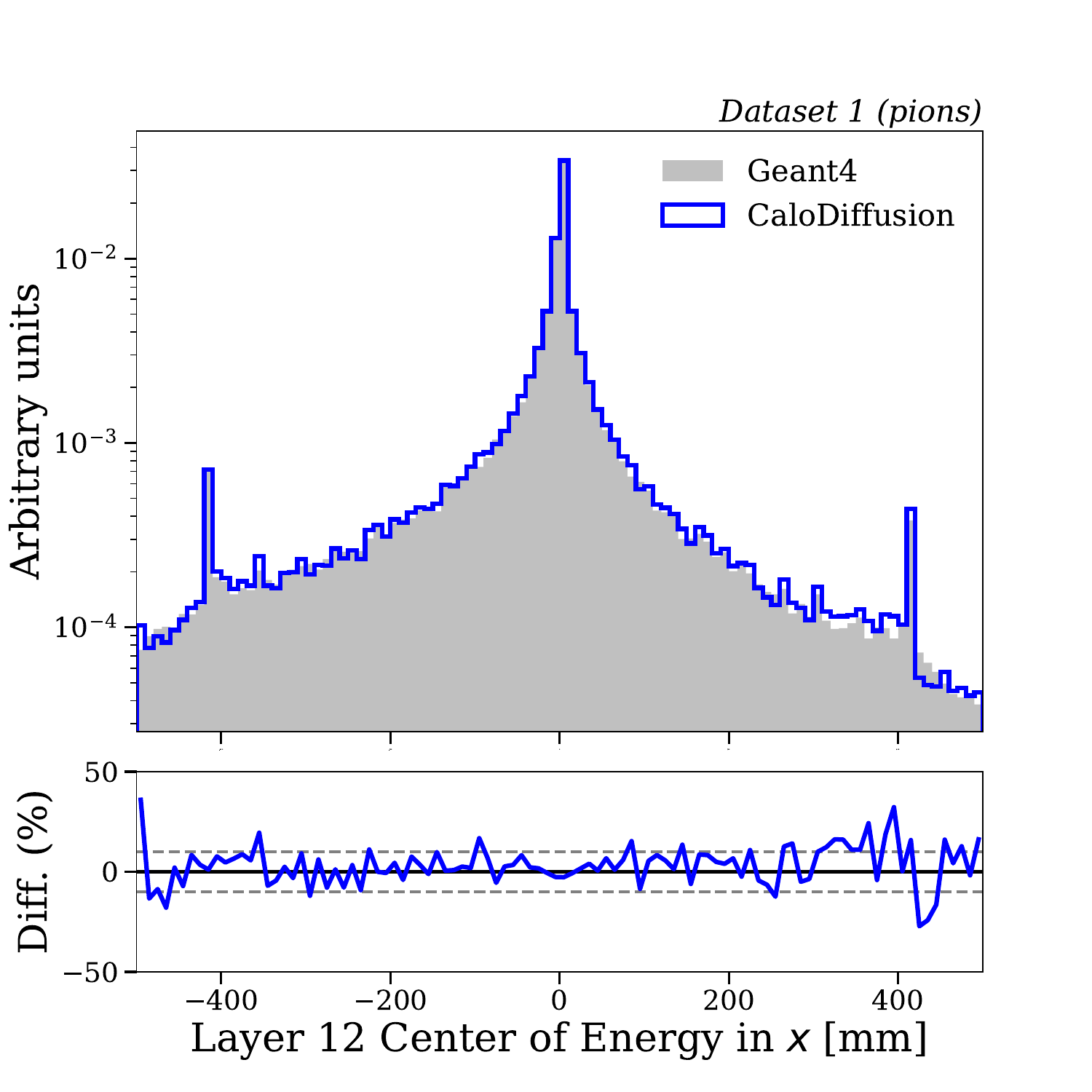}
    \includegraphics[width=0.24\textwidth]{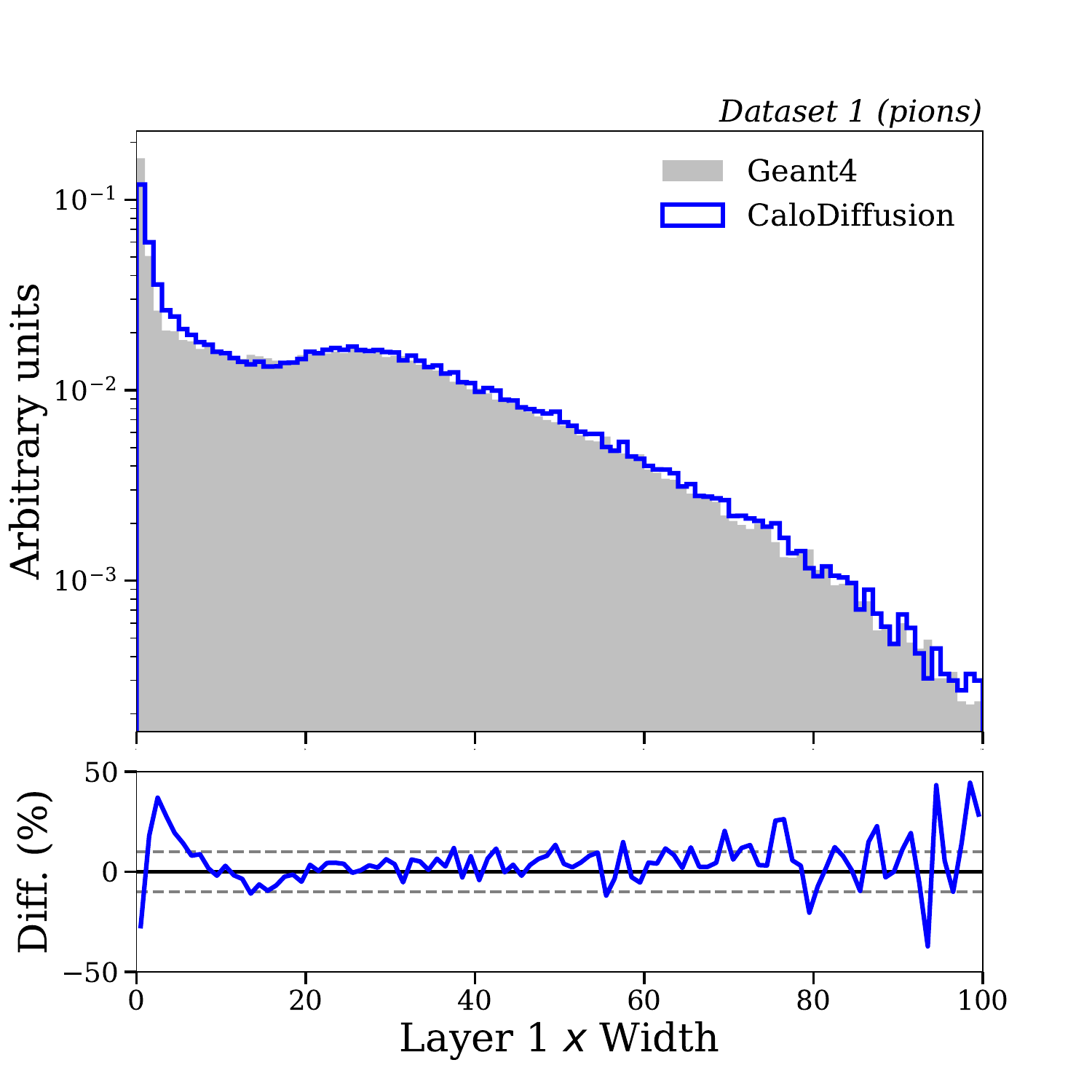}
    \includegraphics[width=0.24\textwidth]{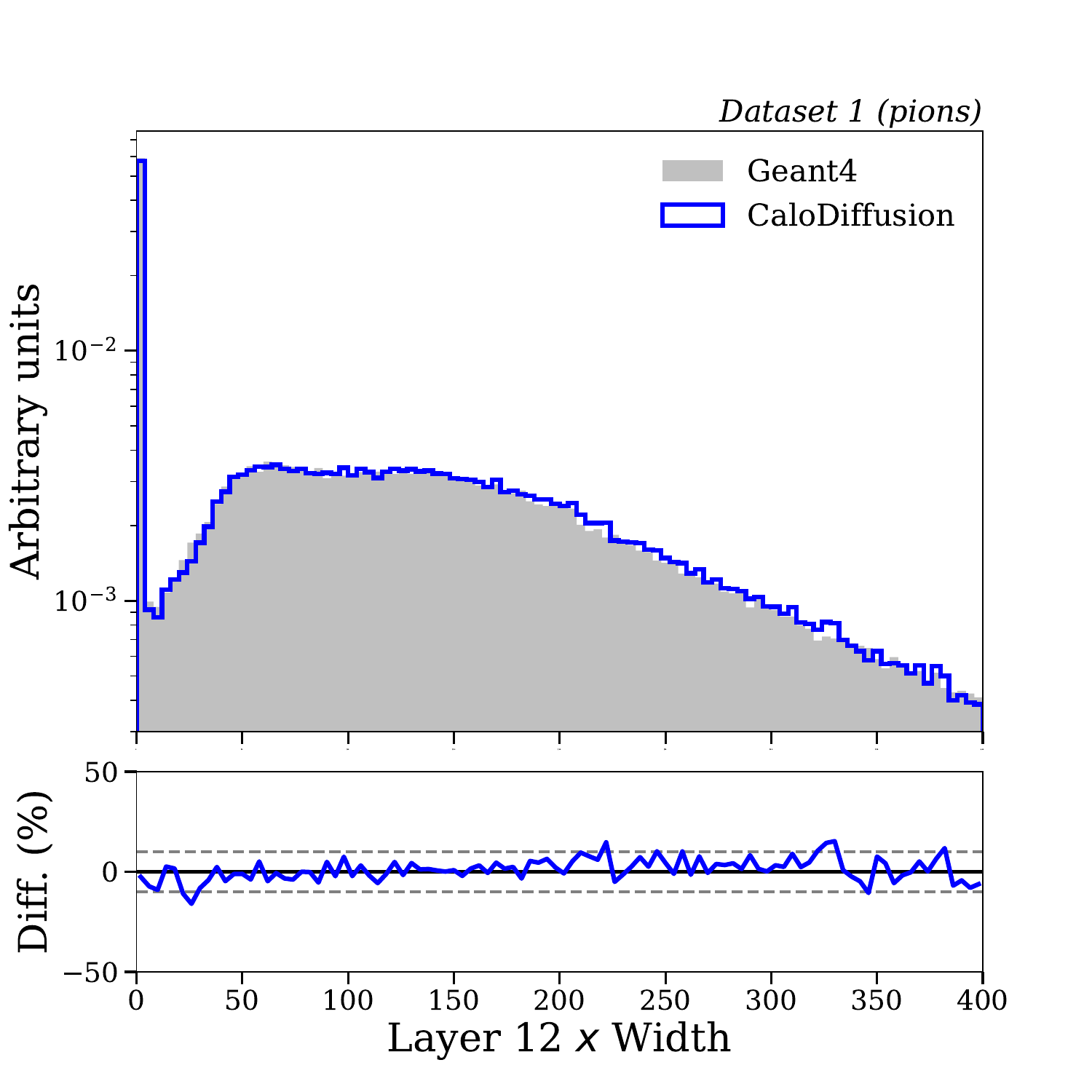}\\

    \threefigeqh{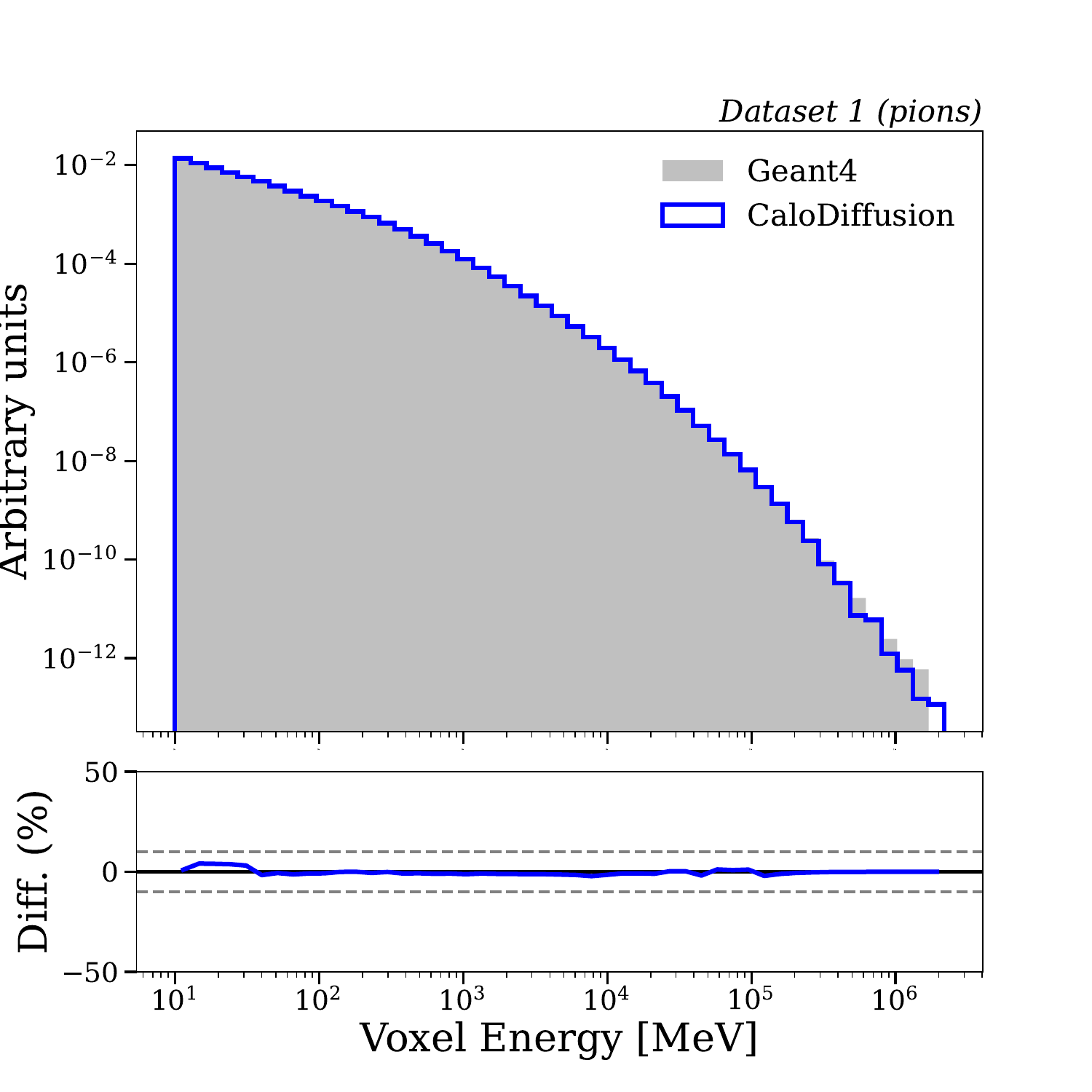}{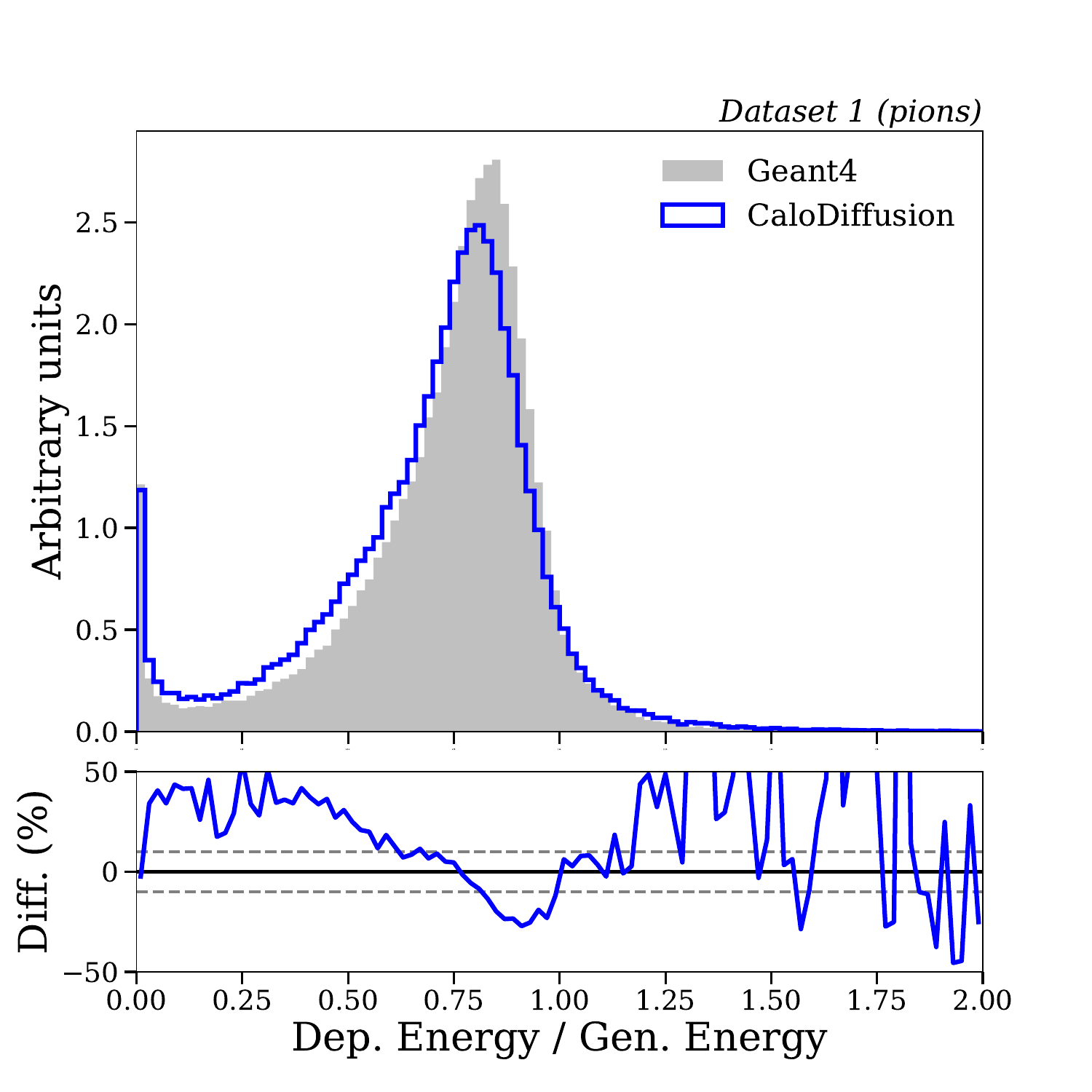}{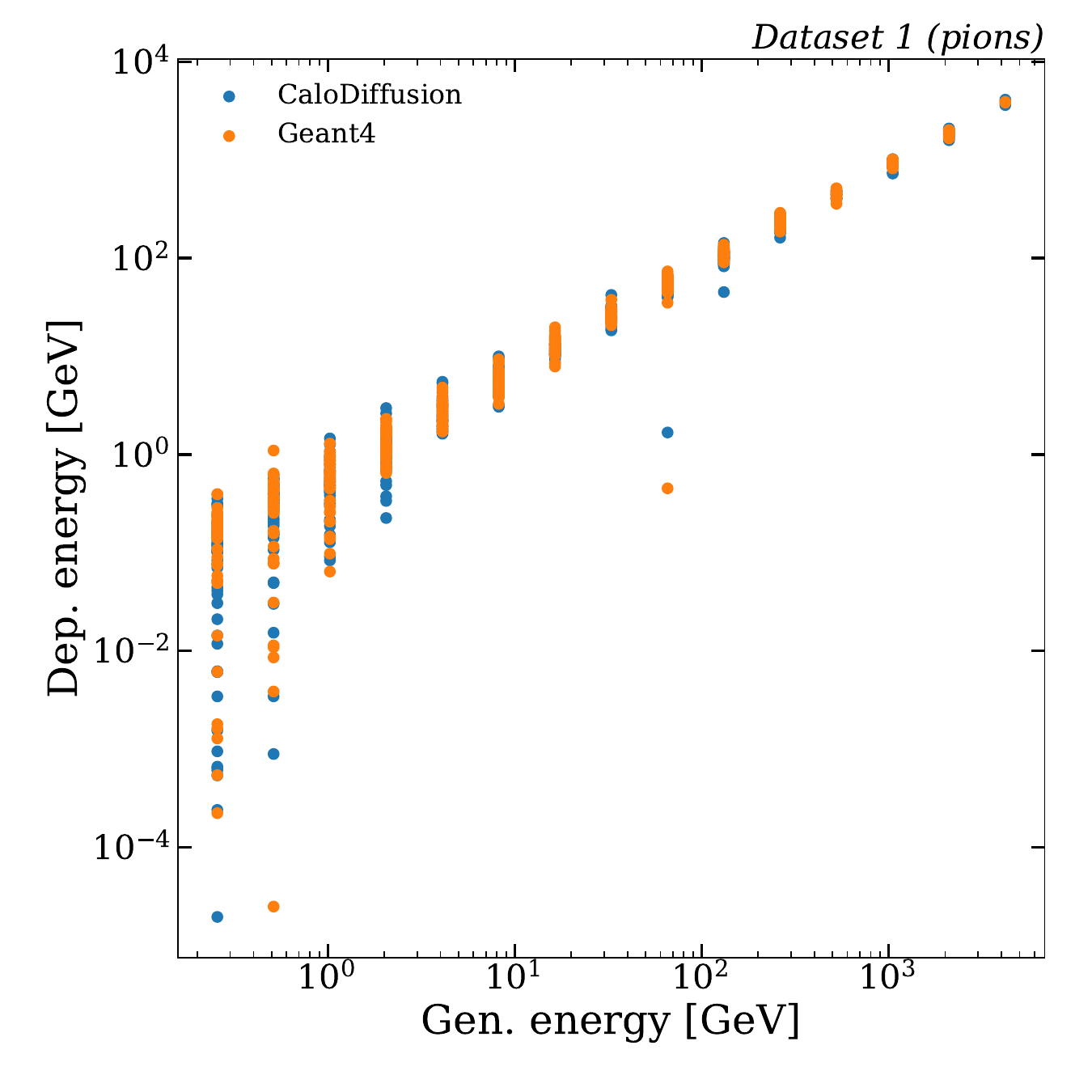}
    
    \caption{A comparison between \geant and \diffu showers across a variety of observables for the pion sample of dataset 1. The top row shows the distribution of energy in different layers of the calorimeter. The middle row shows the distribution of the center and width of the energy spread in two reference layers. The bottom row shows the distribution of voxel energies, the distribution of total shower energy divided by the incident energy, and a scatter plot of deposited energy versus incident energy. }
    \label{fig:dataset1_pion}
\end{figure*}

A comparison of the average showers produced by \geant and \diffu, with the \glam embedding approach, for the photon sample of dataset 1 is shown in Fig.~\ref{fig:avg_showers}.
A comparison of various energy distributions for the photon and pion samples of dataset 1 used in the evaluation of the \challenge are shown in Figs.~\ref{fig:dataset1_phot} and~\ref{fig:dataset1_pion}, respectively.
The spatial properties of the shower are characterized by the Cartesian center energy of the shower, defined as $\overline{x} = \frac{\langle x_i E_i \rangle}{\sum E_i}$ for cell location $x_i$ and energy $E_i$; and the shower width, defined as $\sqrt{\frac{\langle x_i^2 E_i \rangle}{\sum E_i} - \overline{x}^2}$. 

\begin{figure*}[!ht]
    \centering
        \includegraphics[width=0.32\textwidth]{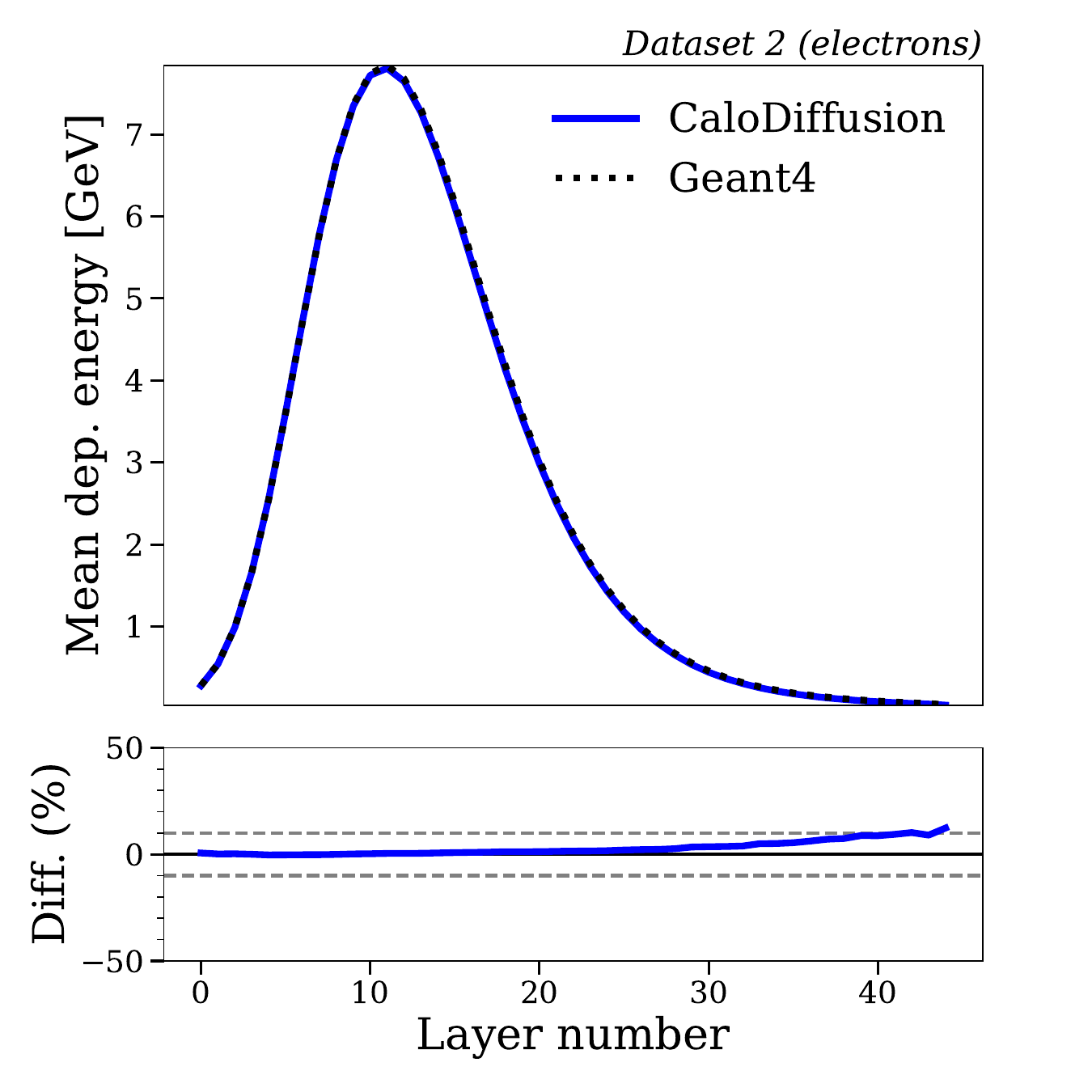}
        \includegraphics[width=0.32\textwidth]{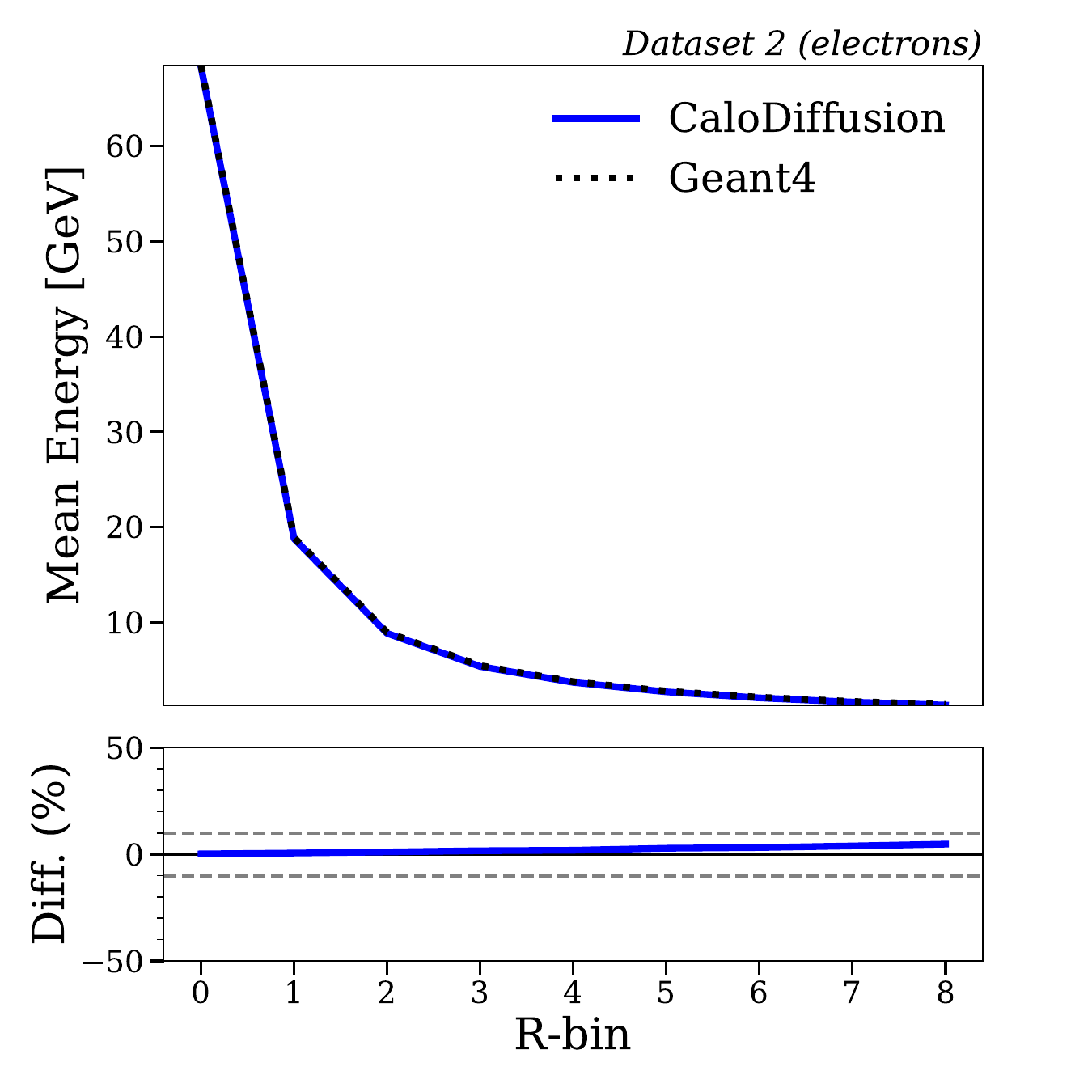}
        \includegraphics[width=0.32\textwidth]{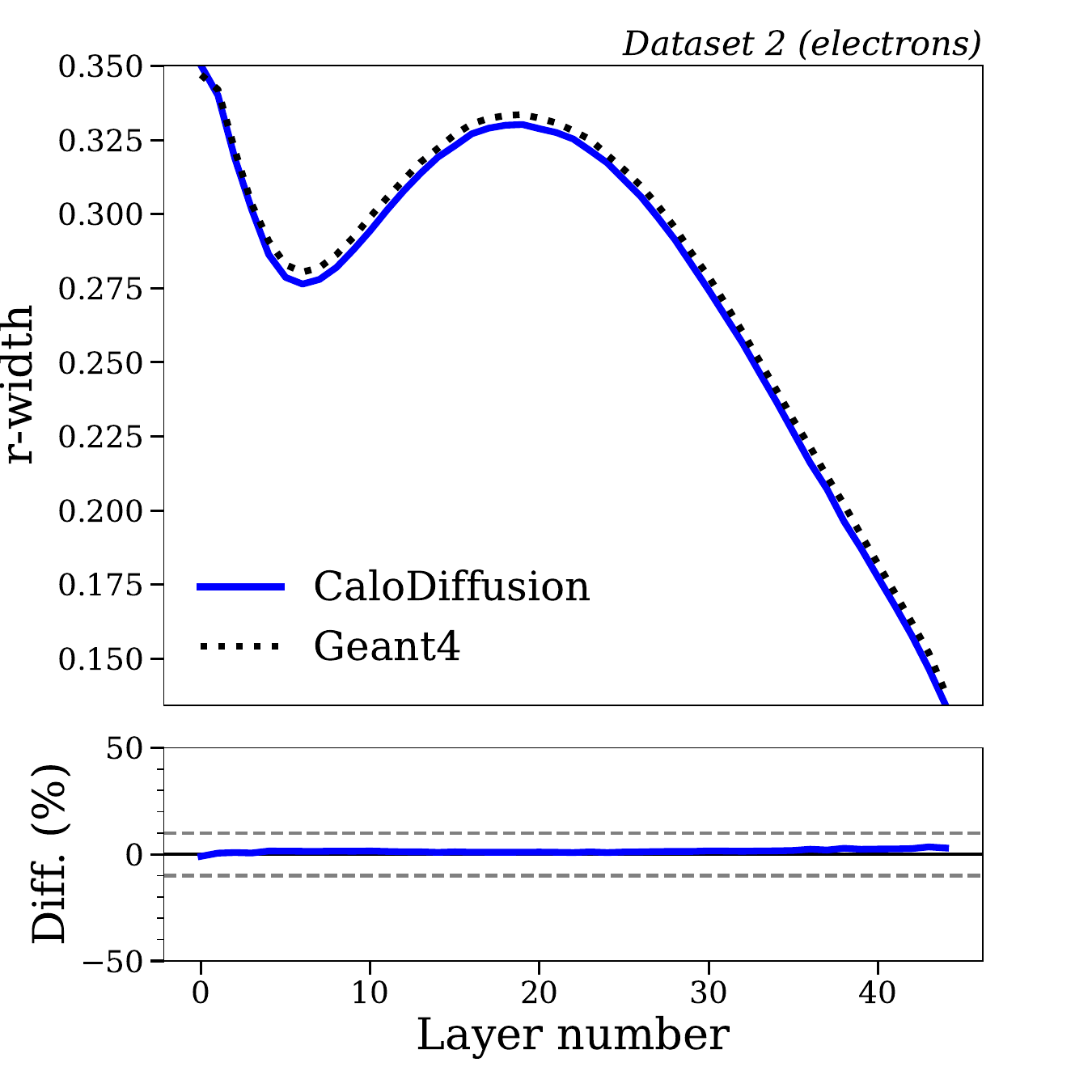}

        \includegraphics[width=0.32\textwidth]{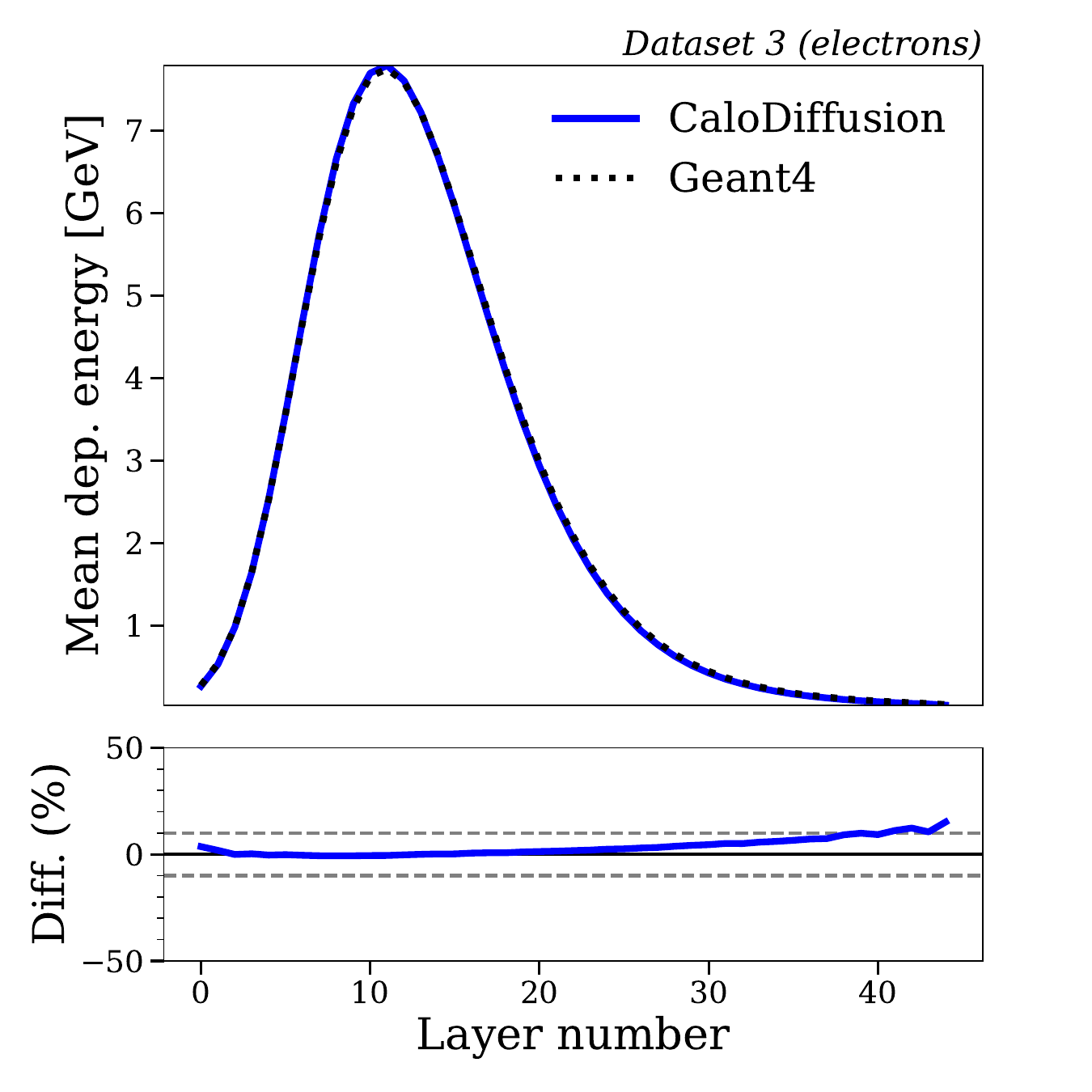}
        \includegraphics[width=0.32\textwidth]{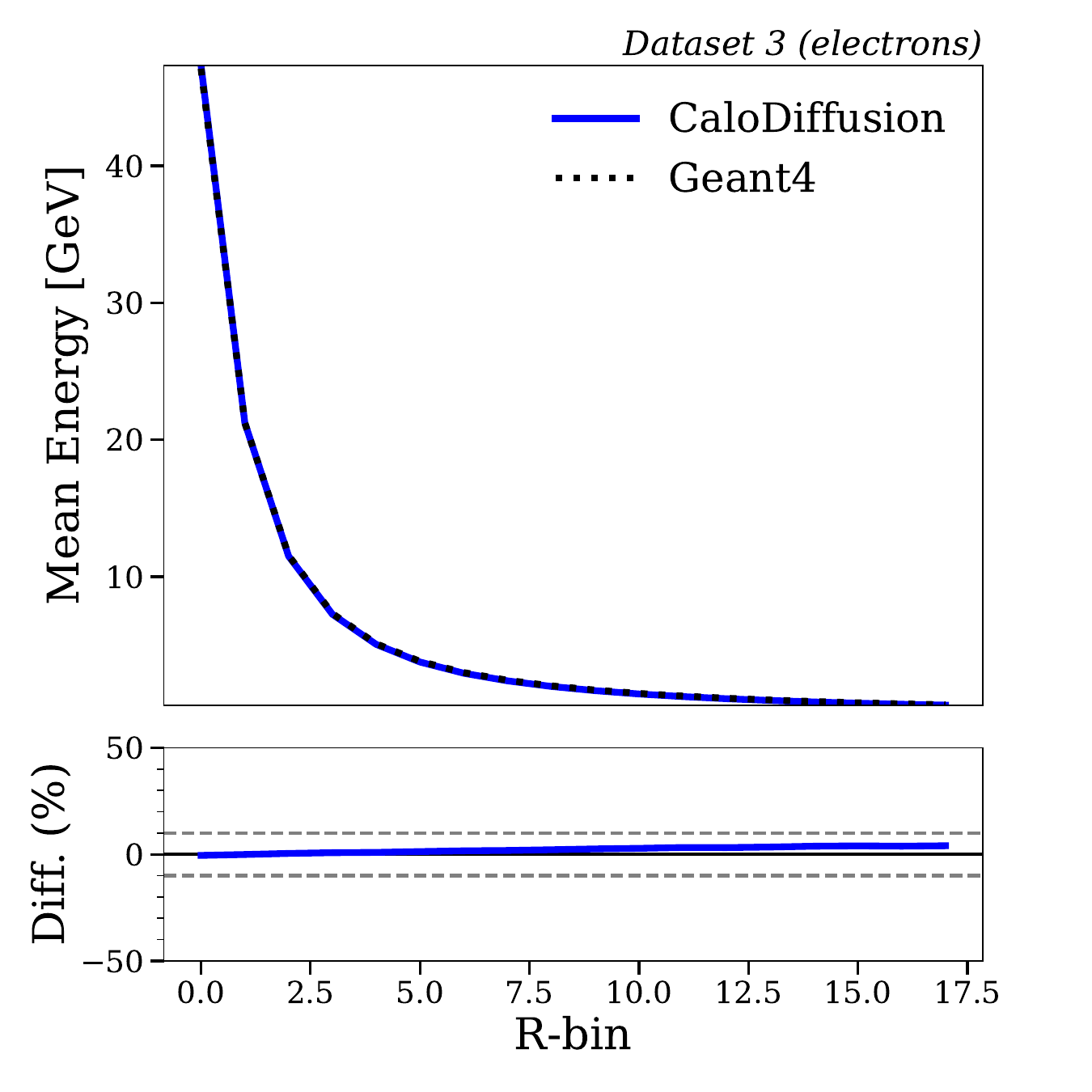}
        \includegraphics[width=0.32\textwidth]{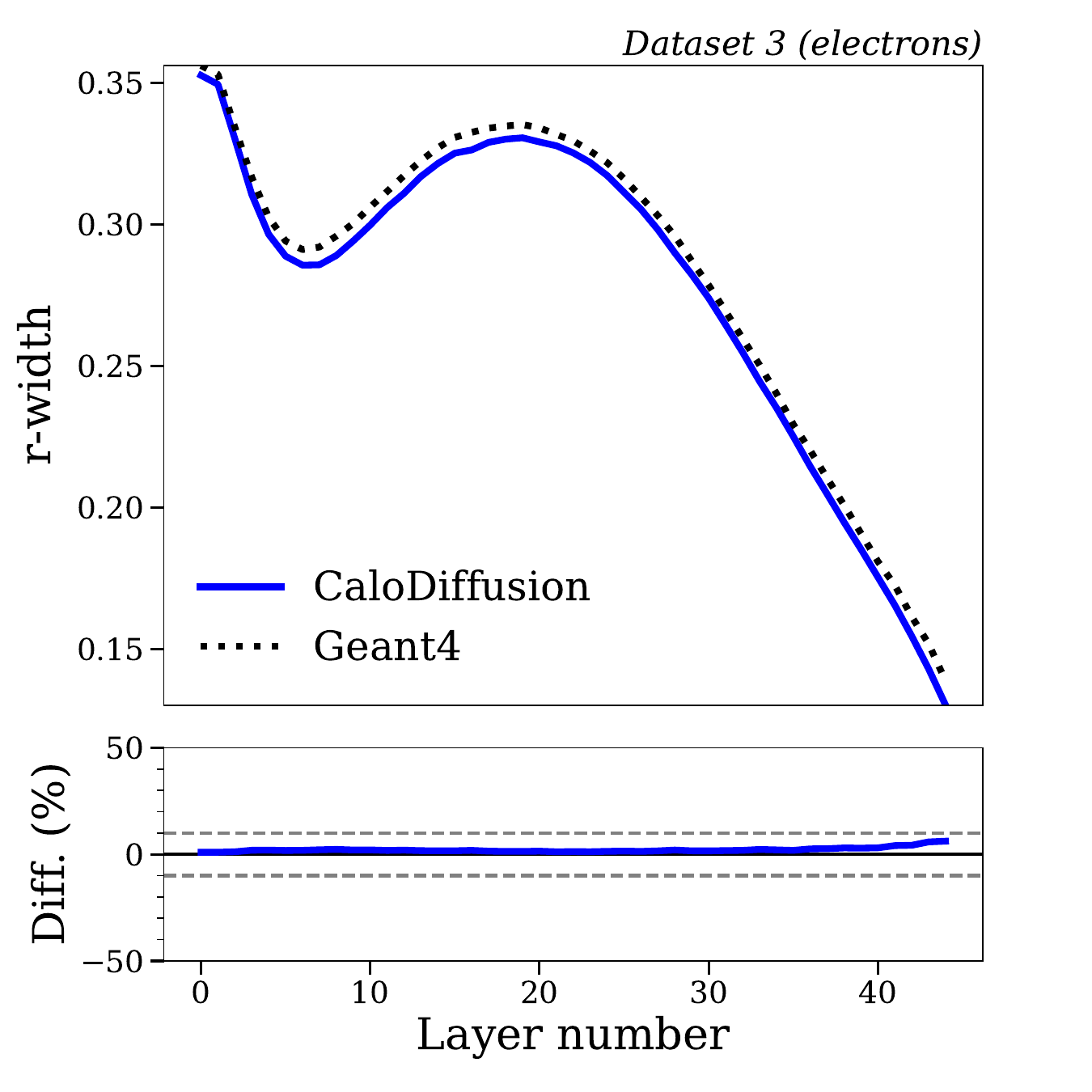}

    \caption{A comparison between \geant and \diffu showers on datasets 2 (top row) and 3 (bottom row).
    The average shower energy is shown as a function of layer (left) and as a function of radial bin (center). 
    The width of the shower in the radial direction is also shown (right).}
    \label{fig:spatial_distribution}
\end{figure*}

\begin{figure*}[!ht]
    \centering
        \threefigeqh{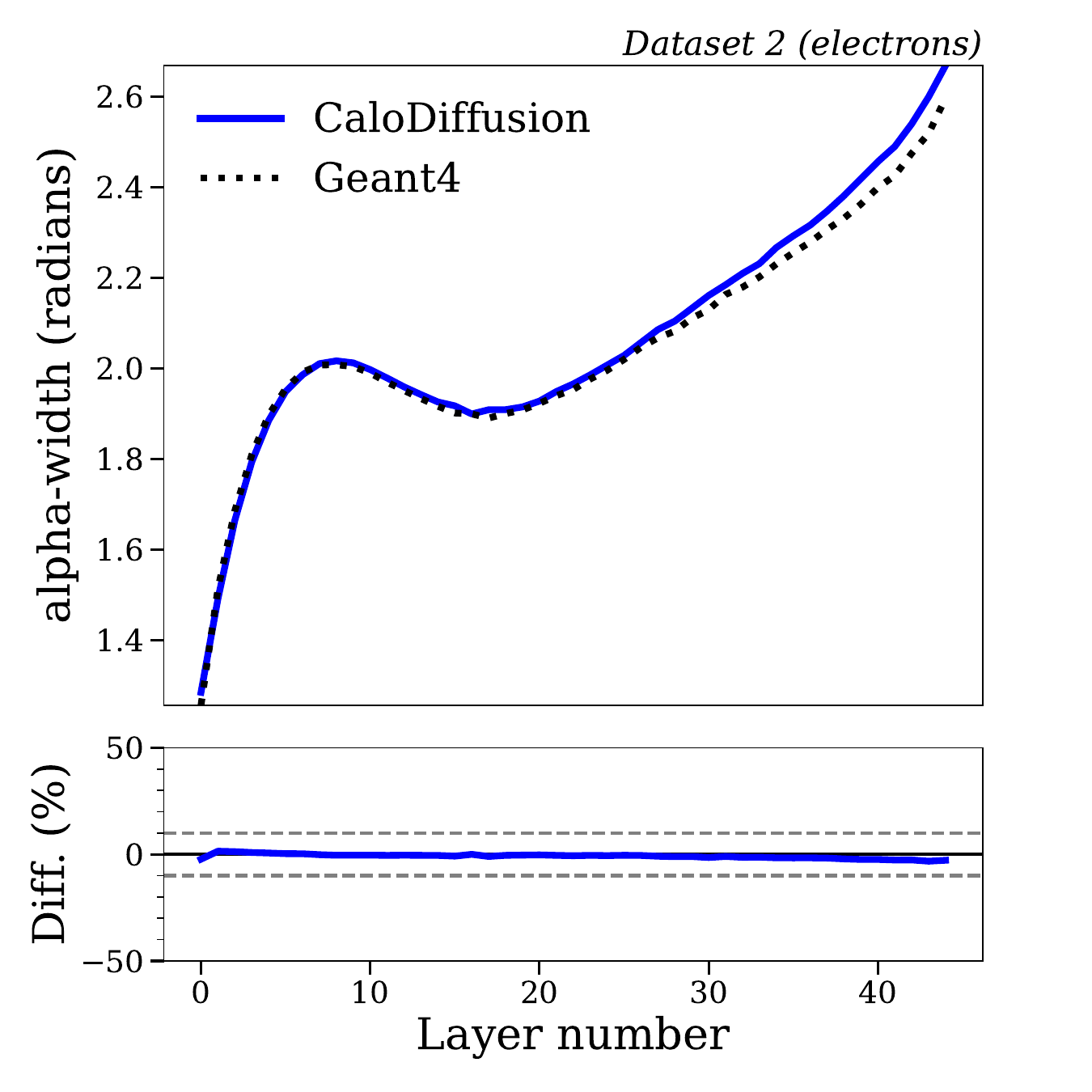}{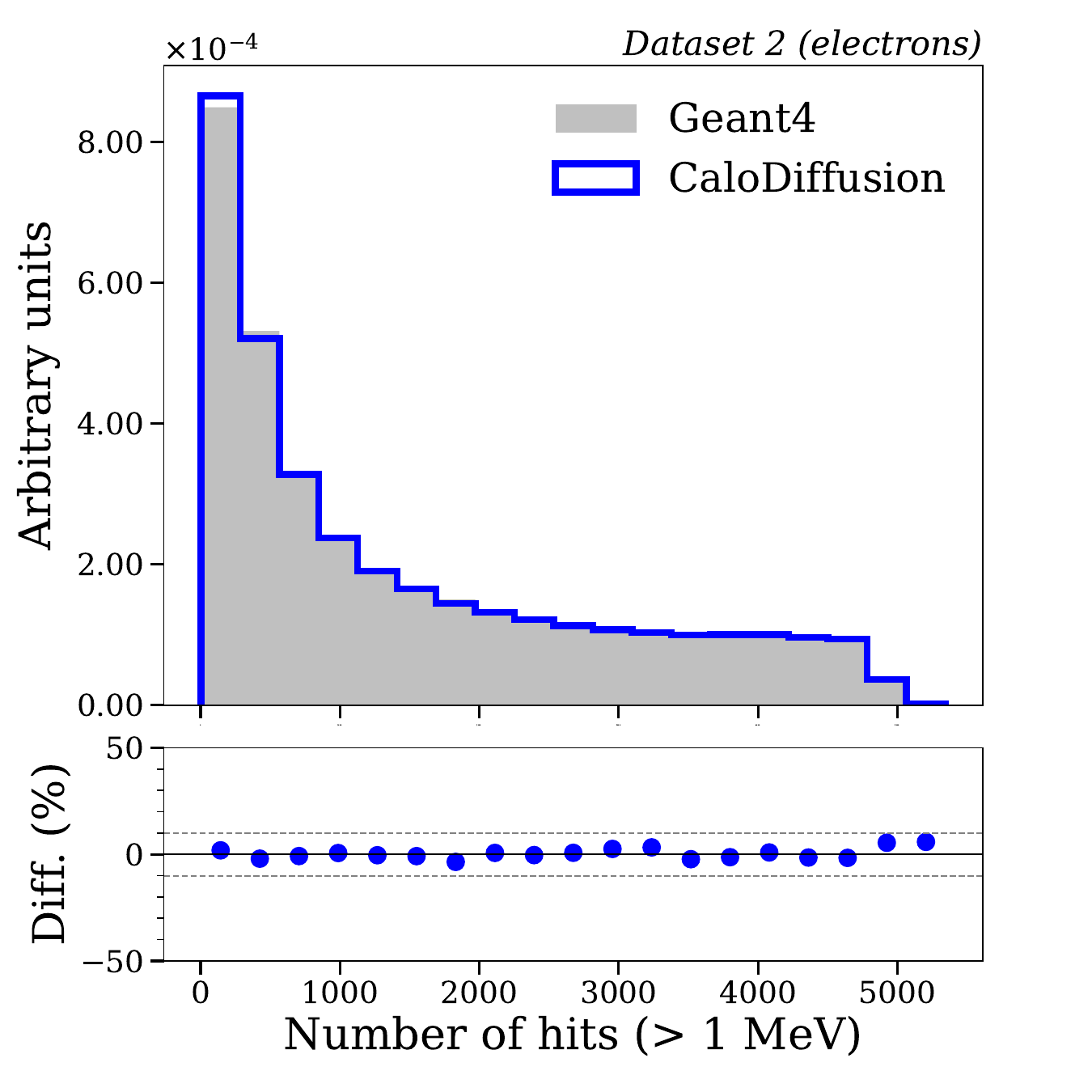}{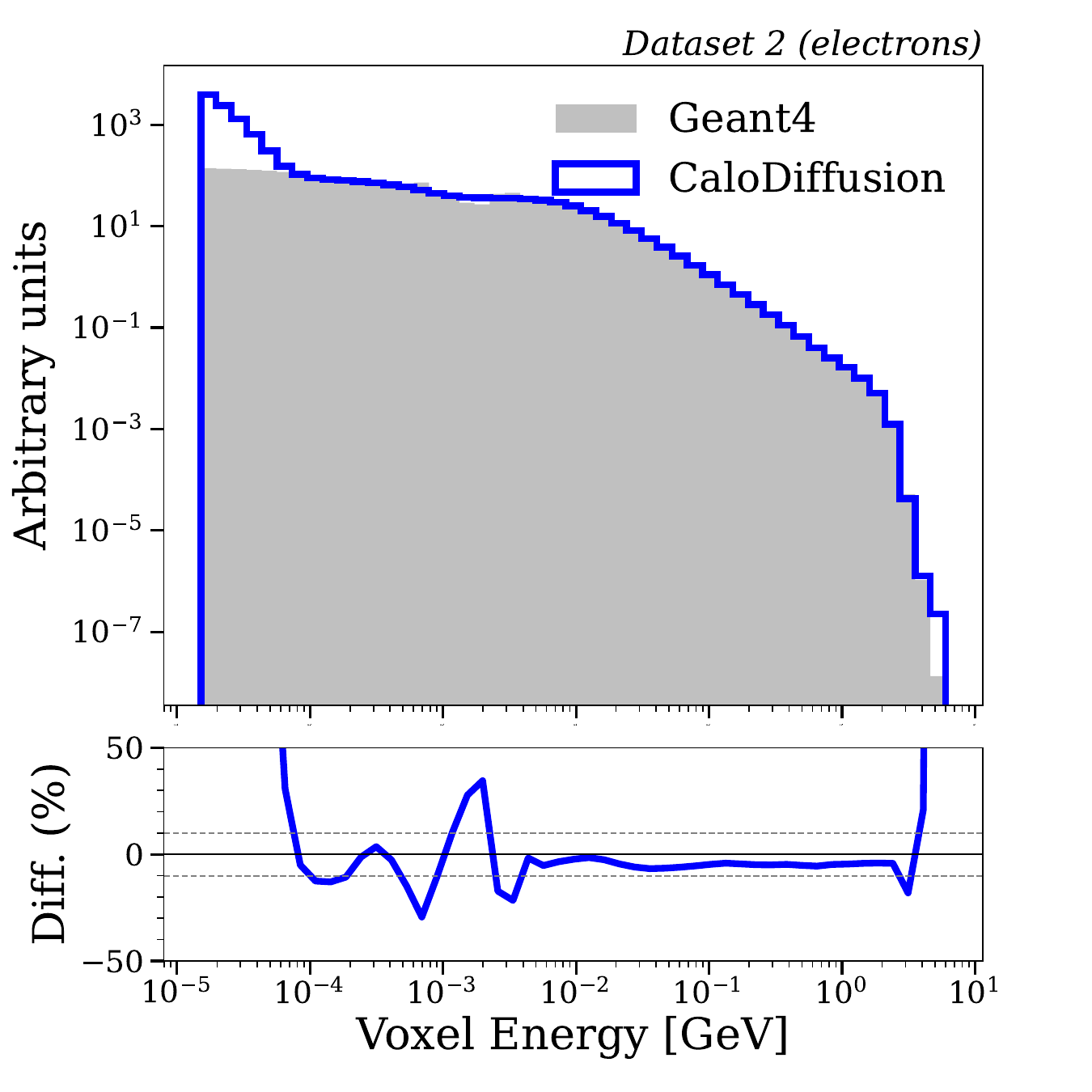}

        \threefigeqh{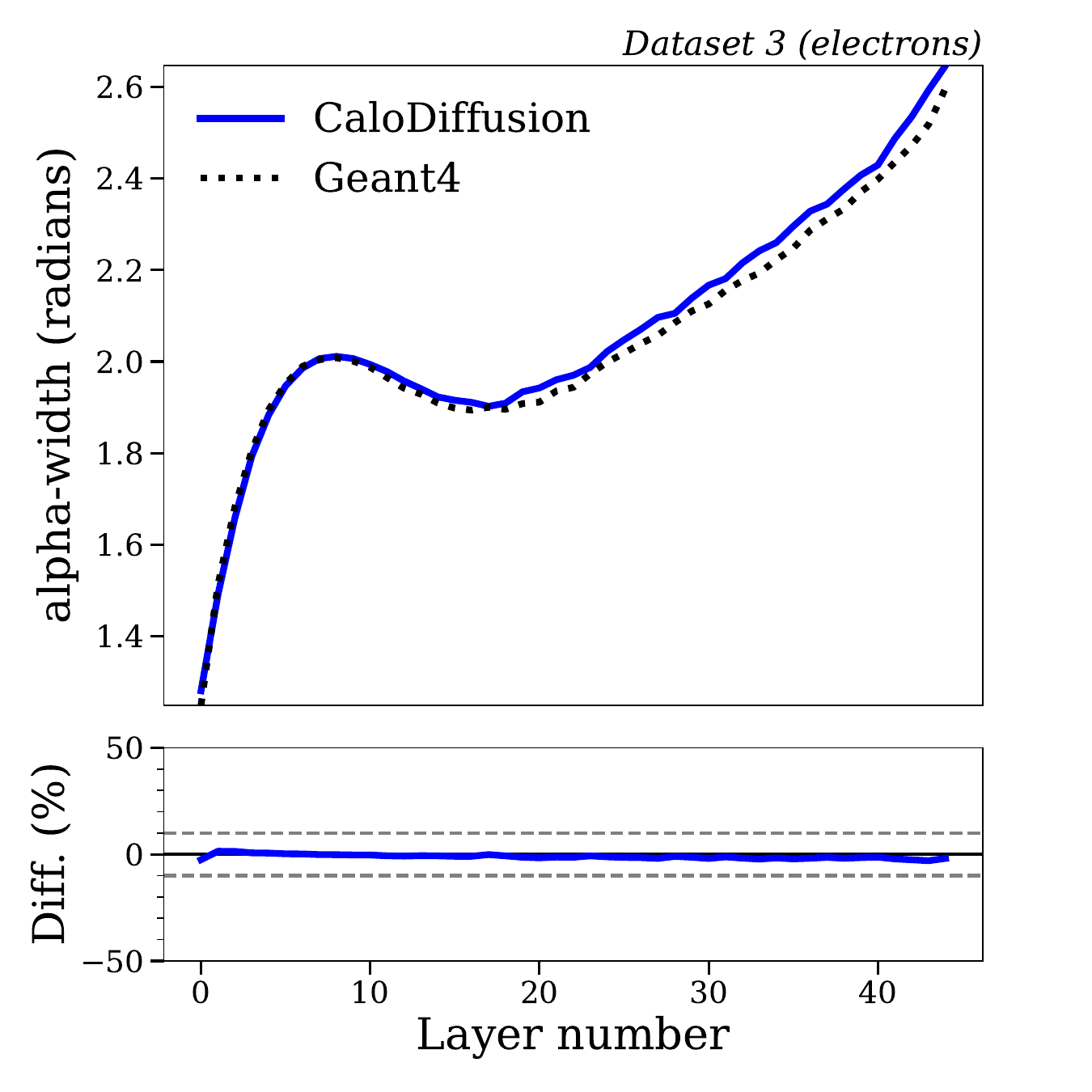}{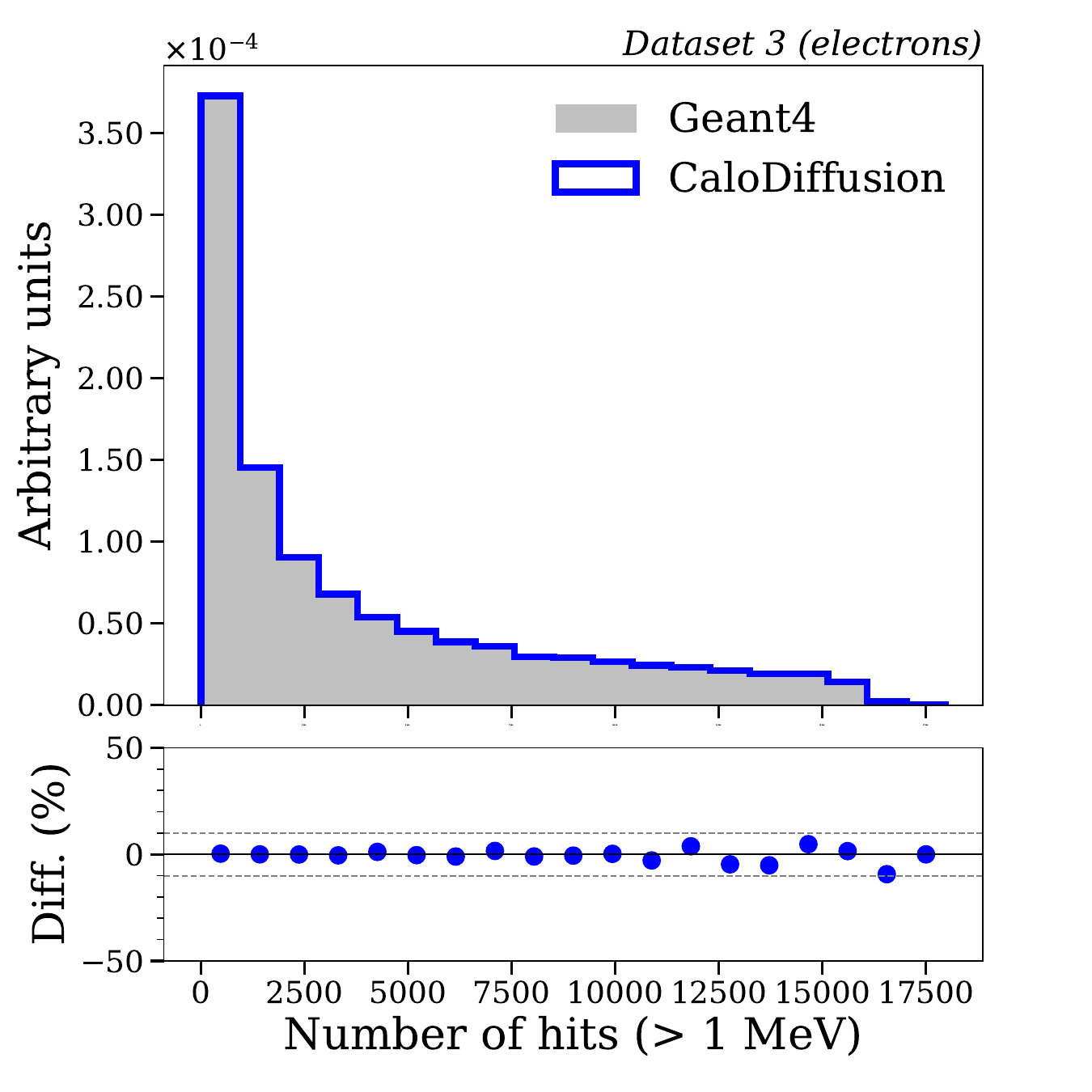}{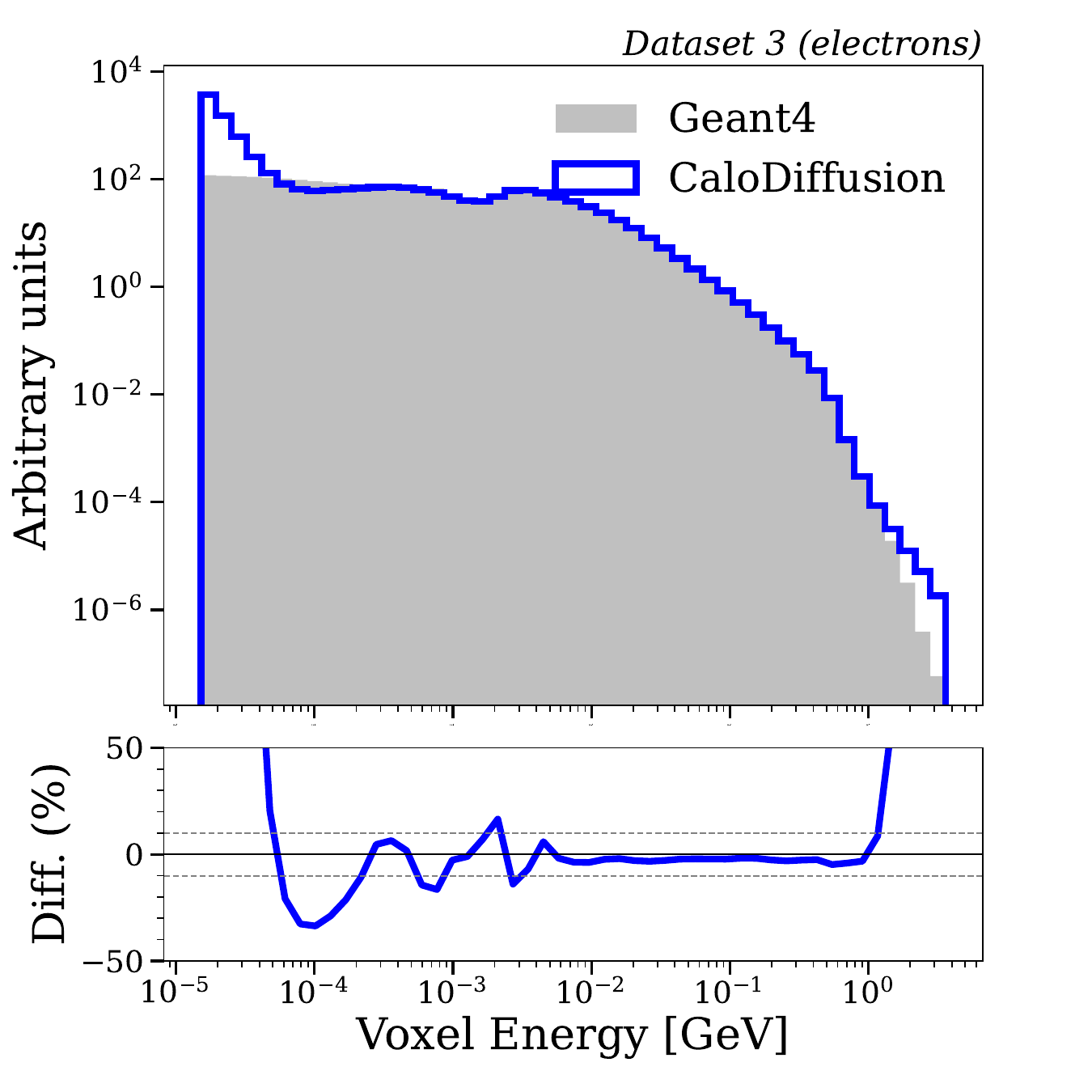}

    \caption{A comparison between \geant and \diffu showers on datasets 2 (top row) and 3 (bottom row).
    The quantities shown are the width of the shower in the angular direction (left), the distribution of total number of non-zero voxels (center), and the energy per voxel (right).}
    \label{fig:hits_distribution}
\end{figure*}

\begin{figure*}[!ht]
    \centering

        \threefigeqh{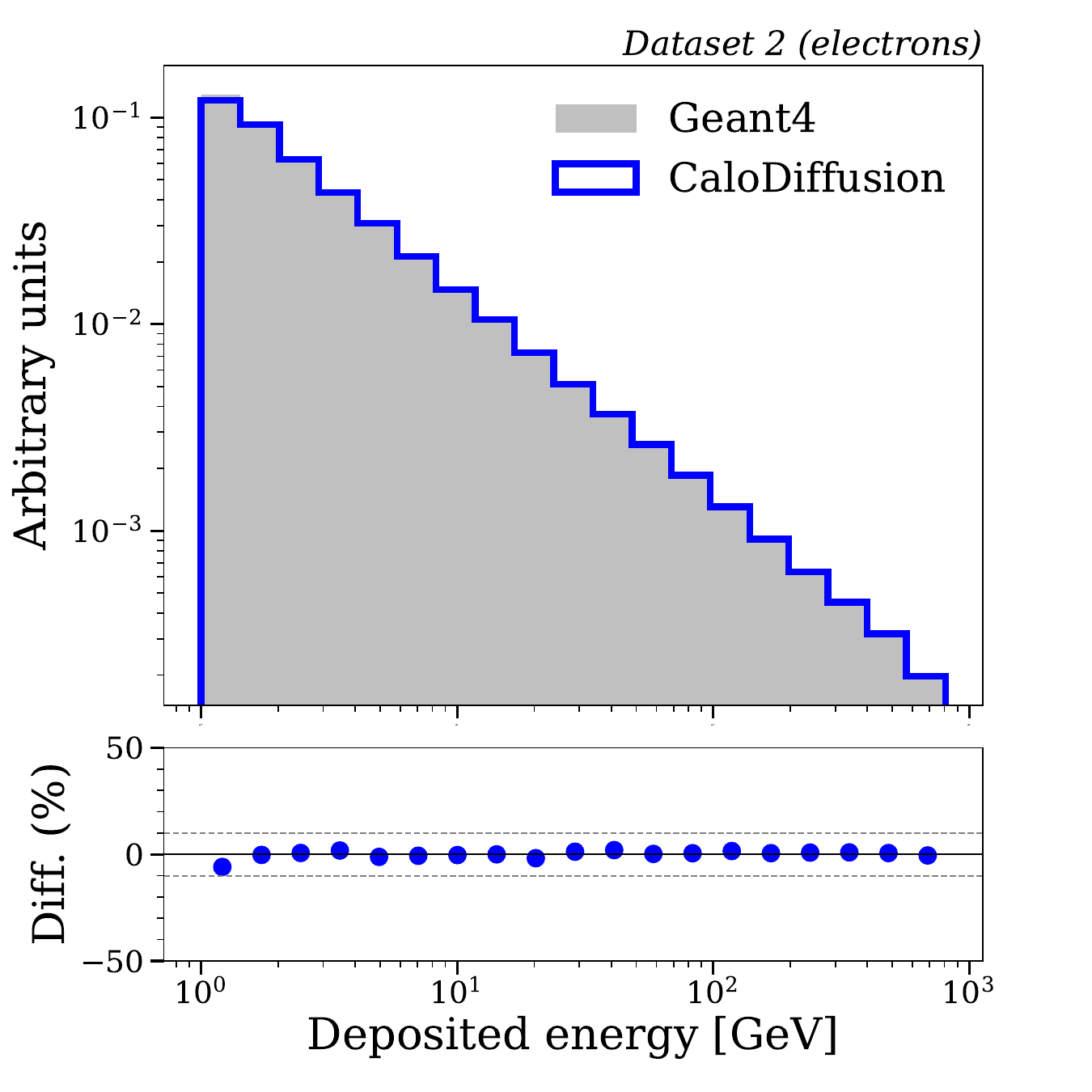}{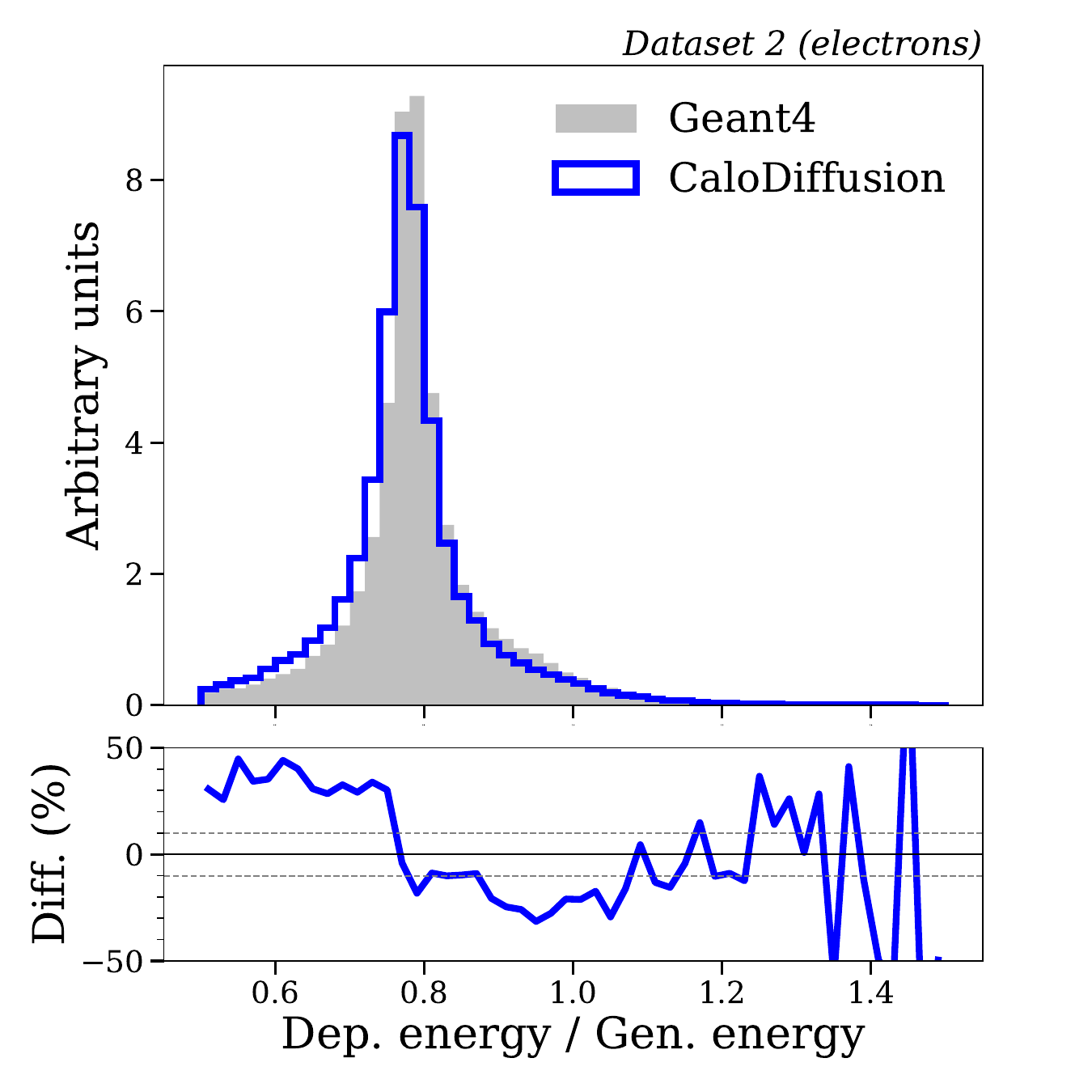}{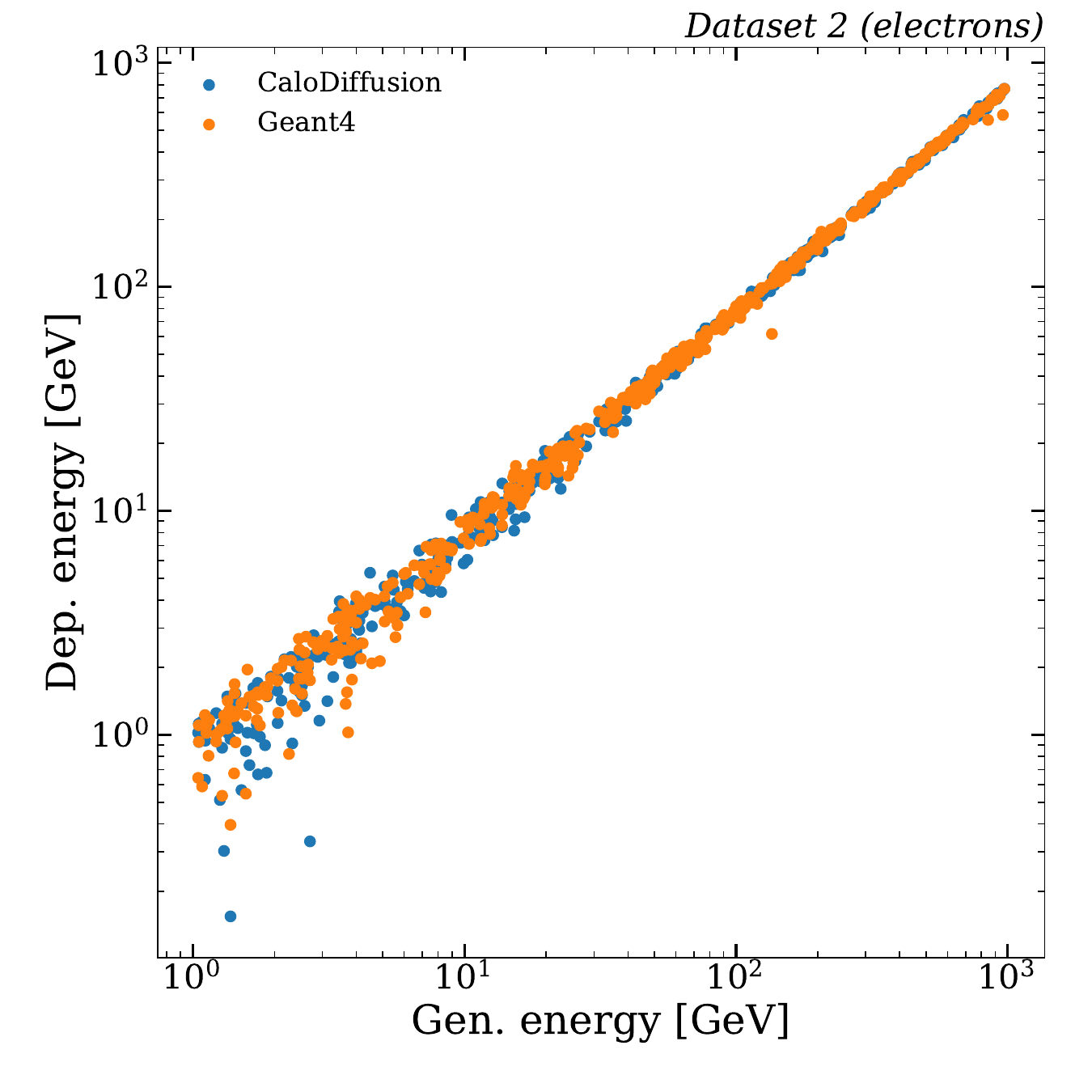}
        \threefigeqh{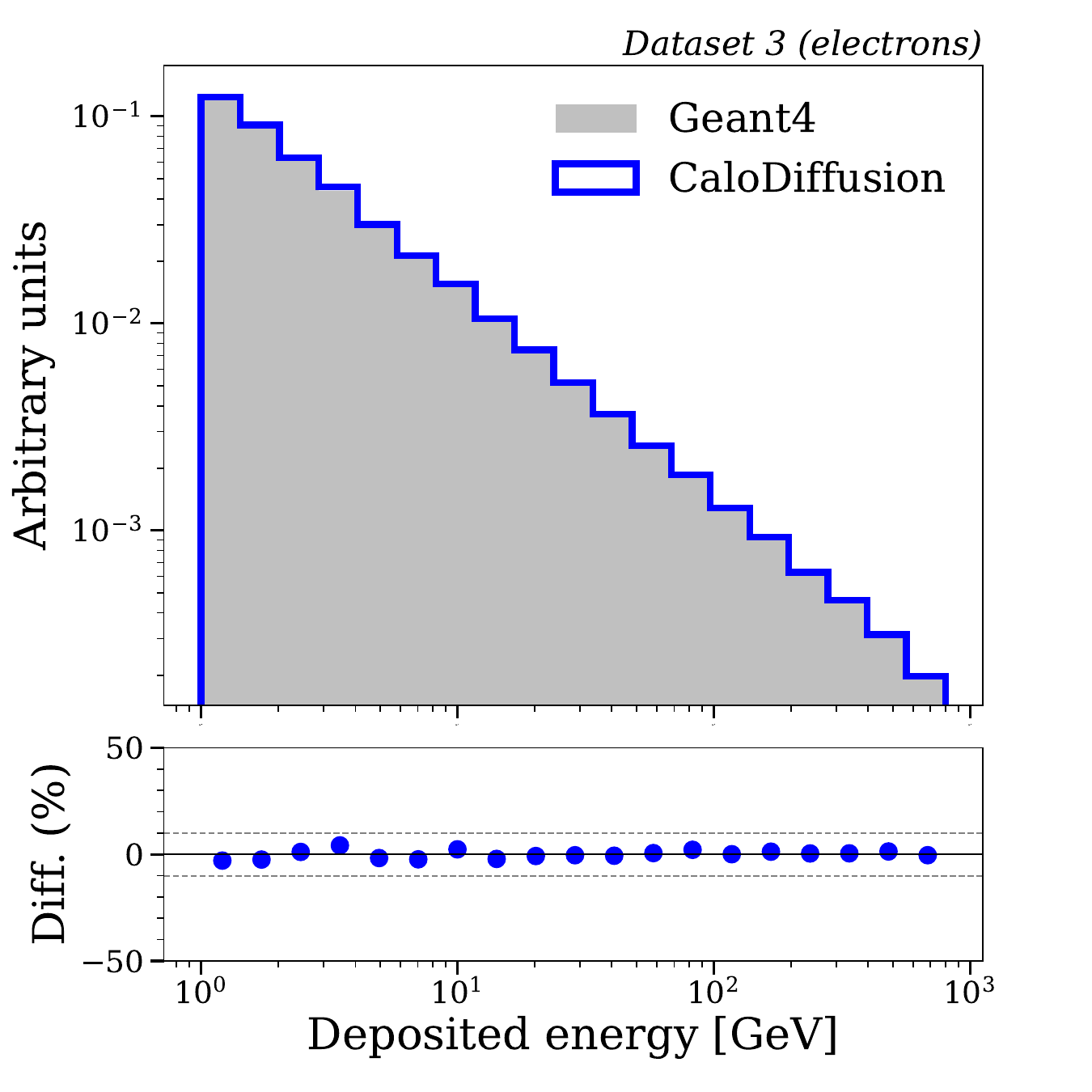}{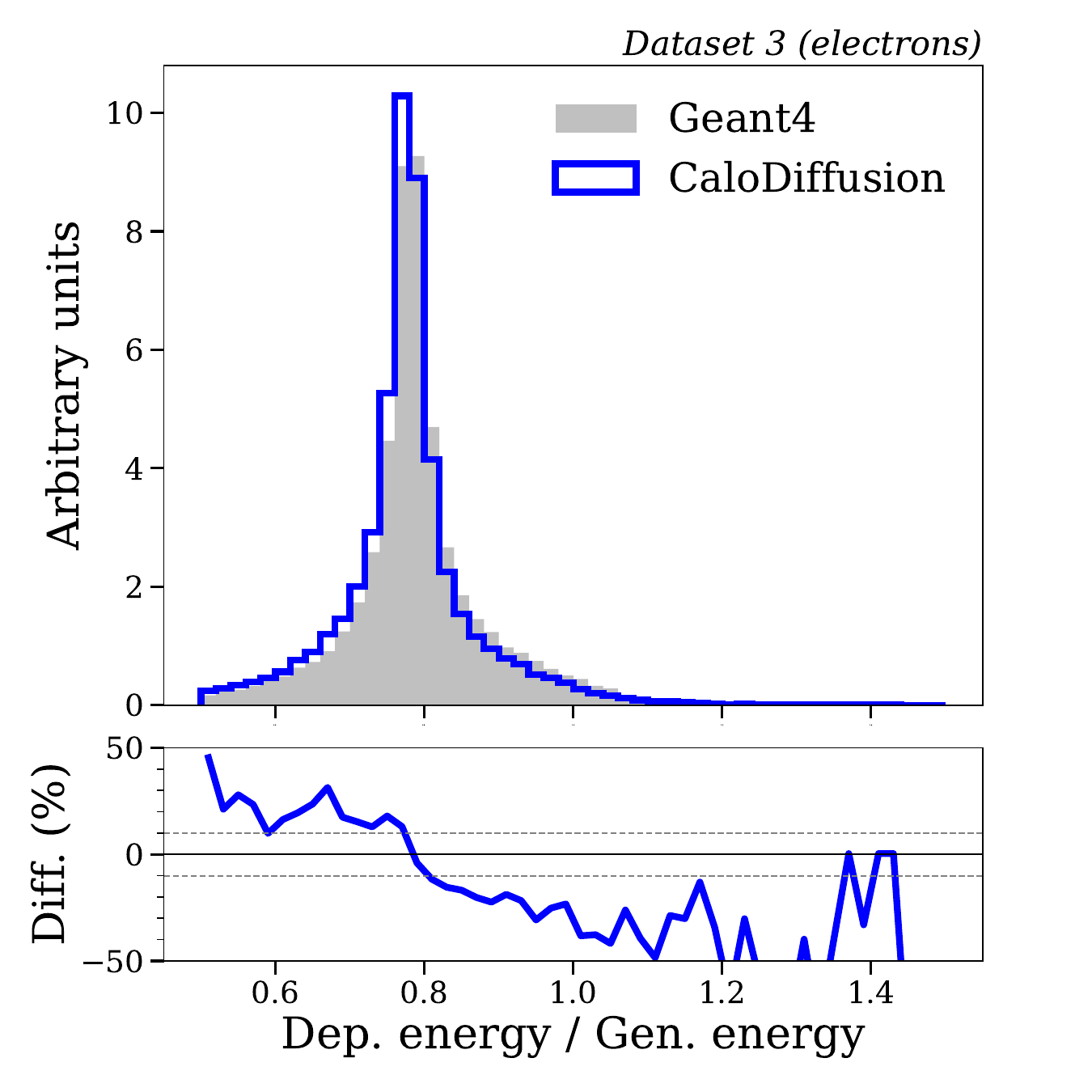}{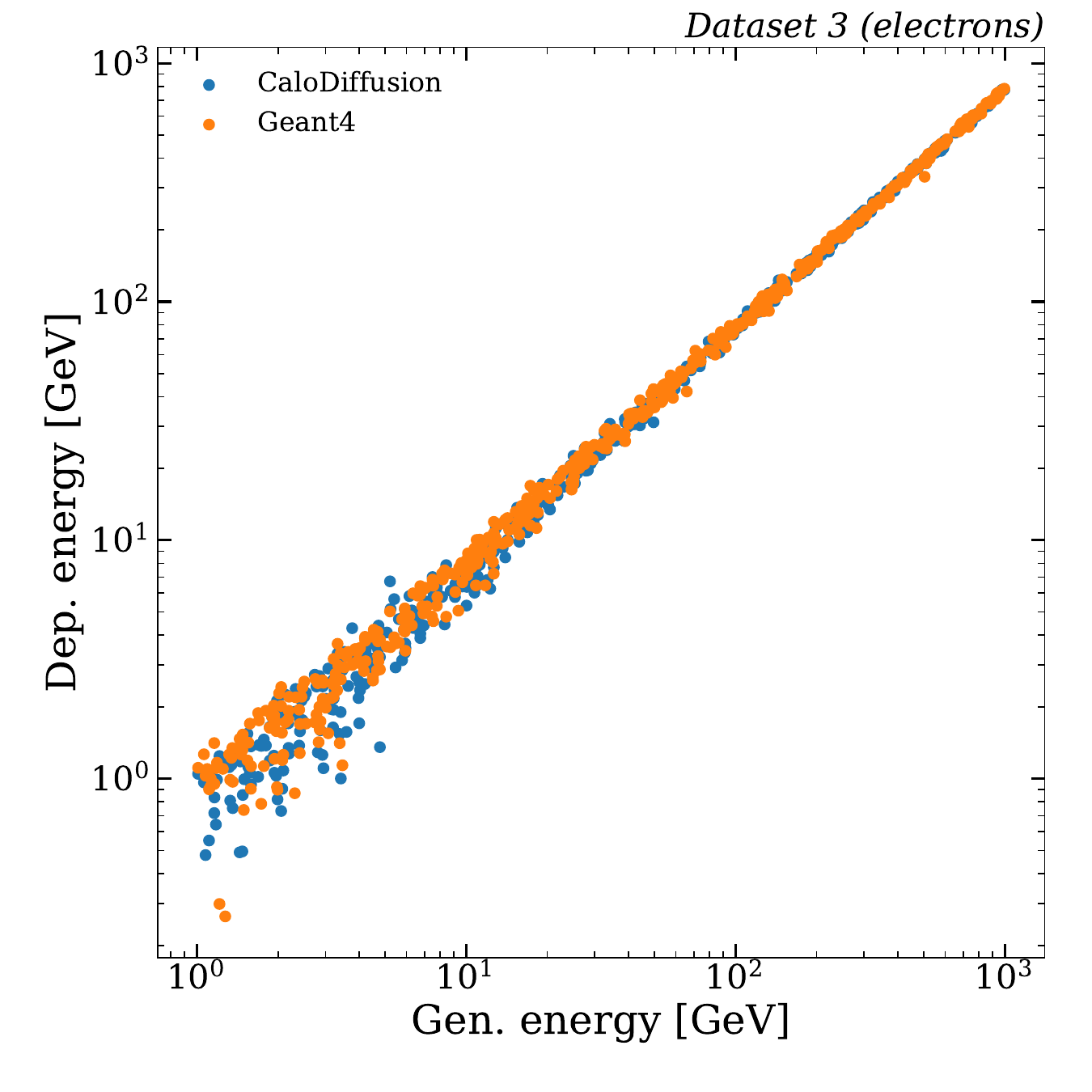}

         \caption{A comparison between \geant and \diffu showers on datasets 2 (top row) and 3 (bottom row).
         Shown are the distributions of the total shower energy (left) and the total shower energy divided by the incident particle energy (center), and a scatter plot of the two quantities (right).}

    \label{fig:energy_distribution}
\end{figure*}

For datasets 2 and 3, we examine the distribution of energy as a function of the layer of the calorimeter and as a function of the radial coordinate of the voxel.
We examine the total energy of the shower, the distribution of voxel energies, and the number of voxels with energy above 1 MeV.
The spatial properties of the shower are represented by the width of the shower in radial and angular dimensions (computed analogously to the Cartesian version defined above) separately for each layer of the calorimeter. 
The distributions for datasets 2 and 3 are shown in Figs.~\ref{fig:spatial_distribution},~\ref{fig:hits_distribution}, and~\ref{fig:energy_distribution}.

We generally find that \diffu is successful at modeling all the datasets considered.
The spatial distributions of the showers---the shower center and widths for dataset 1 and the layer/radial energy profile for datasets 2 and 3---are especially well reproduced.
We observe only very slight degradation in quality on dataset 3 compared to dataset 2, even though it features roughly a factor of 7 higher granularity. 
This underscores the advantage of the convolutional approach: because it is based on fully local operations, it can readily scale to higher-dimensional data. 

One of the most notable deficiencies is that \diffu produces a tail of low energy voxels for datasets 2 and 3, which is not seen in the \geant distributions.
The tail likely results from residual noise from the diffusion process that has not been fully removed by the model. 
The tail begins at approximately 10 keV and thus is not visible in dataset 1 because of the higher voxel energy threshold (10 MeV).  
The tail would likely be fully removed with a more realistic voxel minimum energy threshold applied to datasets 2 and 3. 
If not, such a low energy tail would still likely have minimal impact on the downstream reconstruction of the shower. 

Perhaps a more relevant deficiency of the model can be seen in the distribution of the shower response, the total shower energy divided by the incident particle energy. 
This is seen to be particularly discrepant in the photon sample of dataset 1, in which \geant exhibits a much narrower peak than \diffu, and mismodeling is visible in all datasets.
We have found distributions of such a `global' property of the shower to be among the hardest for the diffusion processes to capture, because most operations are done entirely locally. 
For such observables, it is not straightforward to add a dedicated loss term to the diffusion training because they are only well defined at the end of the diffusion process, but most of the training uses an intermediate step\footnote{We attempted to add a dedicated L2 loss term for the total shower energy, based on estimating the final shower from the intermediate noisy shower through a 1-step estimate of the denoised shower, but it did not produce any improvements, likely because of the amount of noise in this estimate.}.

In the future, a maximum mean discrepancy loss comparing the distributions from a large batch of events could be tried.
Another possibility would be to adopt a two-stage generation approach, as is done in Refs.~\cite{CaloClouds,L2LFlows,iCaloFlow}, in which the total energy of the shower, or the per-layer energy, is learned with a dedicated model and then used to normalize the output of the diffusion model.

There is also visible mismodeling of a peak in the energy distribution in layer 1 of the dataset 1 pion showers.
This peak comes from very low energy pions that deposit all of their energy in layers 0 and 1 of the calorimeter, producing very sparse showers. 
These showers are qualitatively different from the rest and perhaps could benefit from some dedicated training or optimization.

\subsection{Quantitative Metrics}
\label{subsec:metrics}

We compute several metrics sensitive to differences between the \geant and \diffu samples for quantitative assessment of our model's performance.

One proposed metric~\cite{CaloFlow,Gen_model_limits} is based on training a classifier to distinguish between the reference and synthetic samples.
An optimal classifier will learn a score proportional to the likelihood ratio between the two samples.
The closer the two samples are, the closer the likelihoods will be, and the classifier will struggle to distinguish between the two samples.
Performance can be quantified based on the area under the curve (AUC) from the receiver-operating characteristic (ROC) curve of this classifier evaluated on a statistically independent dataset.
An AUC of 1 would indicate there is a significant deficiency in the synthetic sample, such that the classifier is always able to distinguish it from a reference sample. 
An AUC of 0.5 would indicate the classifier cannot separate the two samples. 
Though Refs.~\cite{hep_eval_metrics,Gen_model_limits} showcased some limitations of the AUC in capturing subtle mismodelings, so far no ML-based calorimeter simulation has reported AUC scores very close to 0.5 on the \challenge or similar datasets.
Therefore, it is still a worthwhile metric to compare models. 

Following the setup of the \challenge, we employ two versions of this classifier test: 
one where the inputs to the classifier are the full showers themselves, along with the incident particle energy (low-level), and
one where the inputs are high-level, physics-informed features of the shower (high-level).
The high-level features are those used in the \challenge: the incident particle energy, the energy in each layer, and the center of energy and width of the shower in the $\eta$ and $\phi$ directions.
In both cases, the classifier is a fully connected network with 2 hidden layers, each with 2048 neurons. Dropout~\cite{dropout}, with a rate of 20\%, is used after each hidden layer.  

\begin{table}[hbtp]
    \centering
    \begin{tabular}{l c c c}
     & \multicolumn{2}{c}{Classifier AUC (low / high)} \\
    \multicolumn{1}{c}{Dataset} & \diffu & \caloflow & \caloscorev \\
    \hline
        1 (photons) & \textbf{0.62} / 0.62 & 0.70 / \textbf{0.55} & 0.76 / 0.59\\
        1 (pions) & \textbf{0.65} / \textbf{0.65} & 0.78 / 0.70 & - / - \\
        2 (electrons) & \textbf{0.56} / \textbf{0.56} & 0.80 / 0.80 & 0.60 / 0.62 \\
        3 (electrons) & \textbf{0.56} / \textbf{0.57} & 0.91 / 0.95 & 0.67 / 0.85 \\
    \end{tabular}
    \caption{The AUC values for a classifier trained to distinguish between \geant and synthetic showers.
    The first value listed is the AUC for the classifier trained on low-level features and the second is the AUC for the classifier trained on high-level features.
    The \diffu values are the average of 5 independent classifier trainings. In all cases, the variation in scores was observed to be 0.01 or less.
    In each row, the bold value is the best AUC value for each classifier type.} 
    \label{tab:aucs}
\end{table}

In Table~\ref{tab:aucs}, we compare the classifier AUC values for \diffu to those reported by \caloflow / \icaloflow~\cite{CaloFlow_dset1,iCaloFlow} (called here simply \caloflow), and \caloscorev, which are the only other models to have published quantitative results on the \challenge at the time of writing
\footnote{\caloflow actually reported results for a preliminary version of the pion sample of dataset 1 without a separate evaluation sample. Therefore, those classifier AUC values were computed from the same sample of showers as used in the training. \diffu results are based on the final version of the pion dataset, which includes separate training and evaluation sets. When training and testing on the older sample, \diffu has slightly improved AUC values, but we report here results on the final \challenge version for posterity.}.
\caloflow is actually a pair of models: the originally trained `teacher' model and a `student` model derived from the first model, optimized for inference speed.
\caloscorev is a score-based diffusion model and also features distilled versions based on progressive distillation. \caloscorev did not provide results for the pion version of dataset 1. 
As this version of \diffu was optimized for sample quality and has not used dedicated methods to improve sampling time, we compare to the teacher model of \caloflow and the undistilled version of \caloscorev\footnote{For dataset 3, the \caloscorev authors do not provide results on a model without distillation, so we compare to the 8-step distilled version.}, 
which have better performance and a similar generation time to \diffu. 
Future work will explore the development of a new version of \diffu with optimized generation speed, which would be more suitable for comparison to the faster versions of each model.

We find that \diffu produces classifier AUC values below 0.7 for all four datasets, indicating that the classifier struggles to distinguish between \diffu and \geant showers.
\diffu achieves better AUC scores than \caloflow and \texttt{CaloScore v2} for all cases except the photon showers of dataset 1 when using high-level features.
The performance gains of \diffu are especially prominent for the higher-dimensional datasets 2 and 3.

For \diffu, the classifiers trained on low-level features and high-level features have quite similar AUC values. 
This indicates that most of the discrimination power between \diffu and \geant showers is captured by these high-level features. 
We generally find that the low-level classifier overfits the training set significantly, and therefore an improved architecture would perhaps perform better.
However, we generally take this overfitting to be a positive sign, because it indicates that distinguishing between \geant and \diffu showers based on generalizable features is not easy.

\begin{table}[hbtp]
    \centering
    \begin{tabular}{l A A}
    \multicolumn{1}{c}{Dataset} & \multicolumn{2}{c}{FPD} & \multicolumn{2}{c}{KPD} \\
    \hline
        1 (photons) & 0&.014(1) & 0&.004(1) \\
        1 (pions) & 0&.029(1) & 0&.004(1) \\
        2 (electrons) & 0&.043(2) &  0&.0001(2) \\
        3 (electrons) & 0&.031(2) & 0&.0001(1) \\
        
    \end{tabular}
    \caption{Additional metrics comparing the agreement between showers generated with \geant and \diffu. The number in parentheses is the uncertainty in the last significant digit as evaluated with the \jetnet library.}
    \label{tab:additional_metrics}
\end{table}

We additionally report the Fr\'echet Particle Distance (FPD) and Kernel Particle Distance (KPD) metrics, suggested in Ref.~\cite{hep_eval_metrics} and implemented in the \jetnet library~\cite{JETNET}, interfaced to the \challenge evaluation code. 
We use the same high-level shower features as in the classifier test but omit the incident particle energy. 
We find that the FPD metric computed with these features is slightly biased; the reported value does not agree with zero within its uncertainty, even when comparing two samples of \geant showers.
We therefore normalize our reported values for FPD by subtracting the value
computed comparing two \geant samples\footnote{The FPD values computed comparing two \geant samples are 0.008, 0.0005, 0.008, and 0.011
for datasets 1 (photons), 1 (pions), 2 (electrons), and 3 (electrons), respectively.}.
We report these additional metrics in Table~\ref{tab:additional_metrics}.

Further quantitative comparisons with other approaches will be performed at the conclusion of the  \challenge. 
However, initial results from the \challenge~\cite{CaloChallenge_summaryTalk} indicated that a preliminary version of \diffu \footnote{The preliminary version did not use the attention layers and dimensionality reduction in $z$ that are included in the U-net architecture of the version in this paper.} was among the top submissions for every dataset. 

Ablation studies quantifying the performance improvements for various aspects of \diffu are discussed in Appendix \ref{sec:ablation}.

\subsection{Timing}
In Table~\ref{tab:timing}, we report the generation time of our model using different batch sizes on both CPUs and GPUs.
Results are based on a 2.6 GHz Intel E5-2650v2 ``Ivy Bridge'' 8-Core CPU and an NVIDIA V100 GPU.
The time required to generate a shower in \geant depends strongly on the incident energy of the particle. 
The average over the incident energies used in datasets 2 and 3 is $O(100\text{ s})$~\cite{iCaloFlow}.

\begin{table}[hbtp]
    \centering
    \begin{tabular}{l r@{\hskip 15pt} A A}
     & & \multicolumn{4}{c}{\hspace{-7pt}Time/Shower [s]} \\
    \multicolumn{1}{c}{Dataset} & \multicolumn{1}{c@{\hskip 15pt}}{Batch Size} & \multicolumn{2}{c}{CPU} & \multicolumn{2}{c}{GPU} \\
    \hline
        1 (photons)  & 1 & 9&.4 &  \hphantom{333}6&.3 \\
        (368 voxels) & 10 & 2&.0 & 0&.6 \\
                     & 100 & 1&.0 & 0&.1\\
        \hline
        1 (pions)    & 1 & 9&.8 & 6&.4 \\
        (533 voxels) & 10 & 2&.0 & 0&.6 \\
                     & 100 & 1&.0 & 0&.1 \\
        \hline
        2 (electrons) & 1 & 14&.8 & 6&.2 \\
        (6.5K voxels) & 10 & 4&.6 & 0&.6 \\
                      & 100 & 4&.0 & 0&.2 \\
        \hline
        3 (electrons)  & 1 & 52&.7 & 7&.1 \\
        (40.5K voxels) & 10 & 44&.1 & 2&.6 \\
                       & 100 & &- &  2&.0 \\
    \end{tabular}
    \caption{The shower generation time for \diffu on CPU and GPU for various batch sizes.}
    \label{tab:timing}
\end{table}

Because of the iterative denoising process during generation, diffusion models are usually slower than other ML approaches.
If limited to a batch size of one and running on a CPU, this version of \diffu may not satisfy the computation time requirement for a fast simulation.
Without any additional training or algorithmic changes, the \diffu generation time can be linearly improved by reducing the number of diffusion steps used in the sampling, with a cost to sample quality. We explore this tradeoff in Section \ref{subsec:sampling_steps}.

\subsection{Sampling Steps and Quality}
\label{subsec:sampling_steps}

\begin{figure}[htbp]
    \centering
    \includegraphics[width=0.48\textwidth]{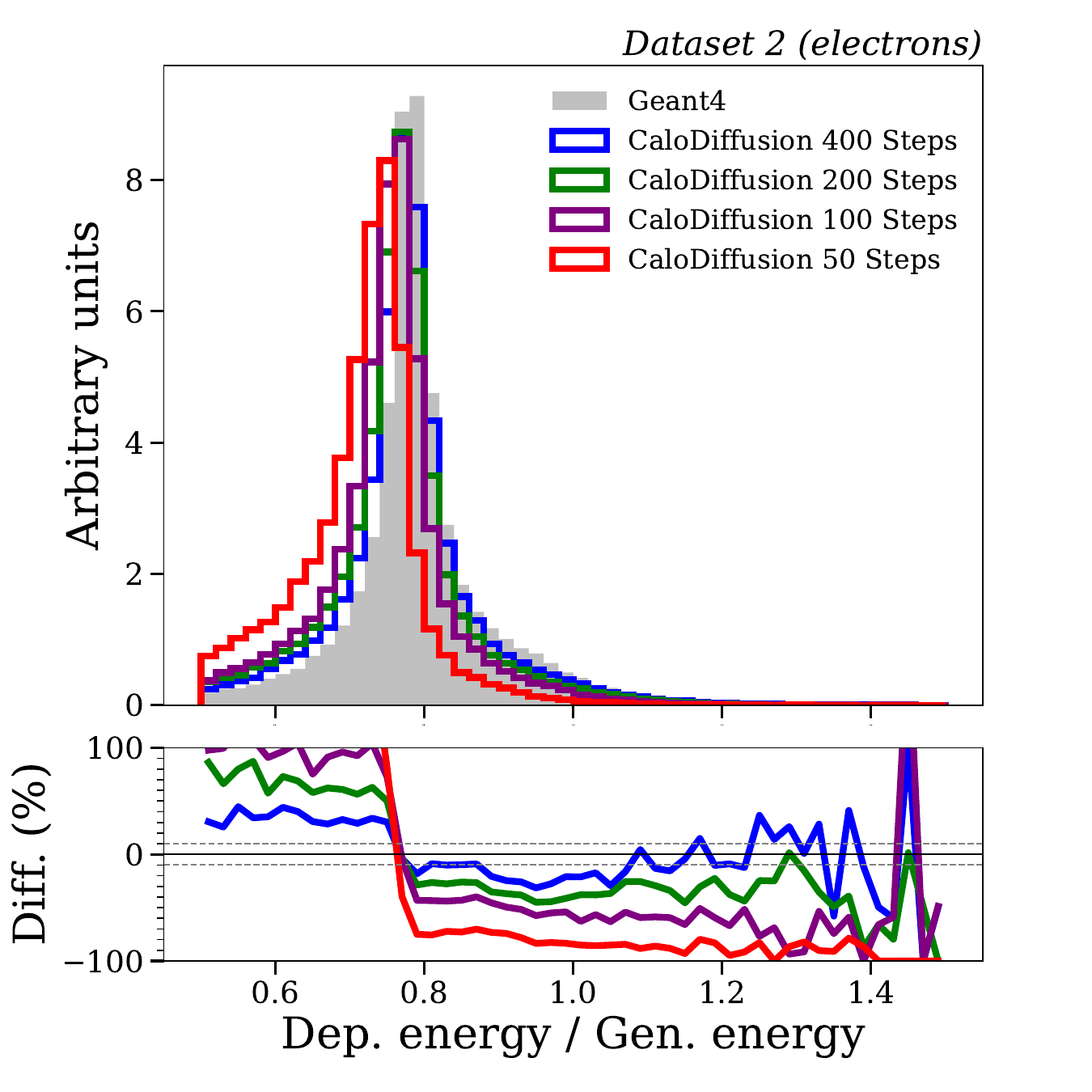}
    \caption{Distribution of the ratio of incident particle energy and total deposited energy of the shower comparing \diffu samples generated with different numbers of sampling steps to \geant.}
    \label{fig:Eratio_steps}
\end{figure}

\begin{sloppypar}In this work, our main goal was to demonstrate the fidelity achievable with the \diffu approach, rather than optimizing for generation speed.
We therefore chose the fewest diffusion steps that did not exhibit a significant decrease in sample quality.
However, significant reductions in the number of sampling steps can still result in high-quality samples.
This tradeoff between number of diffusion steps and sample quality was studied using dataset 2.
By changing the noise schedule, the model can be sampled using different numbers diffusion steps without retraining.
Inference time scales linearly with the number of diffusion steps regardless of batch size (using 200 steps generates samples twice as fast as 400 steps).
We find that one of the distributions most sensitive to the number of diffusion steps is the ratio of deposited to incident energy.
This seems to be one of the hardest features for the diffusion model to capture, and it degrades further with fewer steps.
A plot of this feature with different numbers of diffusion steps is shown in Fig. \ref{fig:Eratio_steps}.
In addition to the metrics reported in Sec.~\ref{subsec:metrics}, we report the separation power between \geant and \diffu on this 1D distribution.
The separation power is a modified $\chi^2$ metric proposed for calorimeter simulation in Ref.~\cite{Diefenbacher:2020rna} and implemented in the \challenge framework. 
Results are presented in Table~\ref{tab:sampling_steps}.\end{sloppypar}

\begin{table}[hbtp]
    \centering
    \begin{tabular}{r c c c}
    \multicolumn{1}{c}{Num. Steps} & \makecell{Classifier AUC \\ \small (low / high)} & FPD & E Ratio Sep. Power \\
    \hline
        400\hphantom{numbe} & 0.56 / 0.55 & 0.043(1) & 0.011 \\
        200\hphantom{numbe} & 0.61 / 0.56 & 0.046(1) & 0.036  \\
        100\hphantom{numbe} & 0.69 / 0.59 & 0.065(3) & 0.079 \\
        50\hphantom{numbe} & 0.83 / 0.67 & 0.110(4) & 0.251\\ 
    \end{tabular}
    \caption{Quantitative metrics comparing the agreement between showers generated with \geant and \diffu with different numbers of sampling steps for dataset 2 of the \challenge. The separation power is computed using the ratio of deposited to incident energy. See text for details.}
    \label{tab:sampling_steps}
\end{table}

Improving the generation time of diffusion models is an active area of research in the machine learning community.
Improved sampling algorithms have been proposed and shown to achieve higher sample quality for low numbers of diffusion steps~\cite{elucidating_design}.
Alternatively, once trained, the diffusion model can be `distilled' into a new model which requires an order of magnitude fewer diffusion steps~\cite{progressive_distill,consistency_models} with minimal loss in sample quality. 
This distillation approach was recently employed for the generation of particle jets using a point cloud representation in Refs.~\cite{Mikuni:2023dvk, PCDroid} and for detector simulation in Ref.~\cite{Mikuni:2023tqg}, still with some loss of quality.

An alternative approach would be to simplify the diffusion task of the network. 
The dimensionality of the data can be reduced by first compressing to a smaller latent space, running diffusion, and then decompressing back to the original space~\cite{LatentDiffusion}.
Alternatively, rather than starting the diffusion process from pure noise, it has been demonstrated that diffusion between two images is possible~\cite{ColdDiffu}.
One could therefore start the diffusion process from an approximate calorimeter simulation, generated by current non-ML fast simulation techniques. 
By providing input similar to the final result, the diffusion process would likely require fewer steps and some physical features may be learned more easily. 
This would be a similar approach to Refs.~\cite{denoising_CNNs}, in which CNNs were used to denoise a fast simulation to achieve higher quality results.
A conceptually related approach using diffusion with a Schr\"odinger bridge was recently demonstrated~\cite{Diefenbacher:2023flw}.
These techniques for refinement of low-level hits in calorimeter showers can complement regression-based refinement of high-level observables~\cite{Bein:2023ylt} by making the latter easier to learn and therefore even more precise.

\section{Conclusion}

In this work, we introduced \diffu, a new machine learning (ML) model that uses diffusion to generate calorimeter showers.
We employed several novel optimizations that exploit the underlying geometry of the calorimeter data.
We have also introduced the geometry latent mapping (\glam), a new approach to handle irregular geometrical structures in data.
\glam learns a lightweight embedding to transform the irregular data geometry into a regular shape, which can then be used in symmetry-preserving operations such as convolutions, and also learns the reverse transformation.
We have demonstrated that \diffu, combined with \glam, is able to generate high quality showers on a variety of datasets, some with high dimensionality.
We have set new benchmarks in quantitative performance metrics that demonstrate it is difficult to distinguish between \diffu and \geant showers.

Our work significantly advances the state of the art in the achievable physics performance from ML-based fast simulation techniques.
This is an important step to establish the viability of such techniques to resolve the simulation component of the computing challenges in the High Luminosity LHC era.
While the unoptimized generation time for diffusion models is slower than for some other ML architectures, producing showers in batches on GPUs is already noticeably faster than the \geant-based full detector simulation.
Future work will explore and compare a variety of approaches to improve the generation speed of \diffu and will apply \diffu with \glam to datasets with even more complicated geometries.

\section*{Code Availability}
\label{sec:code}

The code to reproduce the results in this paper, as well as the trained models, can be found at \url{https://github.com/OzAmram/CaloDiffusionPaper}.

\acknowledgments
We thank the organizers of the \challenge for providing the community datasets and evaluation code used in this work. We thank Raghav Kansal for assistance computing the KPD/FPD metrics on the \challenge datasets.

\section*{Funding Information}
O. Amram and K. Pedro are supported by Fermi Research Alliance,
LLC under Contract No. DE-AC02-07CH11359 with the U.S. Department of Energy, Office of
Science, Office of High Energy Physics. O. Amram is supported by the U.S. CMS Software and Computing Operations Program under the U.S. CMS HL-LHC R\&D Initiative.

\bibliographystyle{jhep}
\bibliography{main}

\appendix
\section{Ablation Studies}
\label{sec:ablation}
We report here ablation studies for several of the innovations and design choices used in this study.
We perform this ablation study on the pion sample of dataset 1 because it is the most difficult sample for \diffu to reproduce. 

The ablations we consider are: 
\begin{itemize}
    \item Not using the `layer' and `radial' images that allow for location-conditional convolutions. 
    \item Using regular Cartesian convolutions instead of cylindrical ones. 
    \item Using a fixed geometric embedding instead of the learnable \glam approach. The setup used to initialize \glam (Section~\ref{sec:glam}), based on the area overlap of cells, is employed for the fixed embedding. 
\end{itemize}
We attempted an additional ablation that replaced \glam with several fully connected dense layers, using no geometric information, but this model did not produce any reasonable results at the denoising task. 

For each choice under study, we retrained a different version of \diffu and generate samples to evaluate the impact. 
For the definitions of the metrics reported, see Tables~\ref{tab:aucs} and~\ref{tab:additional_metrics}.
We find that each ablation does lead to worse performance than the baseline, but no single change results in a substantial drop in performance.

\begin{table}[h!]
    \centering
    \renewcommand{\arraystretch}{1.8}
    \begin{tabular}{c c c c}
    Model & \makecell{Classifier AUC \\ \small (low / high)} & FPD & \makecell{E Ratio \\ Sep. Power} \\
    \hline
        Baseline Model & 0.65 / 0.65 & 0.029(1) & 0.0093 \\
        \makecell{Without layer and\\ radial images} & 0.67 / 0.69 & 0.038(1) & 0.0120 \\
        \makecell{Without cylindrical\\ convolutions} & 0.67 / 0.69 & 0.035(1) & 0.0110 \\
        \makecell{Fixed\\ geometric embedding} & 0.66 / 0.69 & 0.039(2) & 0.0118 \\

    \end{tabular}
    \caption{Quantitative metrics comparing the agreement between showers generated with \geant and different ablations of \diffu for the pion sample of dataset 1 of the \challenge.}
    \label{tab:ablation}
\end{table}

\end{document}